\documentclass[a4paper,10pt,fleqn]{article}

\usepackage{changepage}
\usepackage{amssymb}
\usepackage{slashed}
\usepackage{pifont}
\usepackage{geometry}
\usepackage{fleqn}
\usepackage{amsmath}
\usepackage{amsbsy}
\allowdisplaybreaks

\begin{document}

\newcommand\p{\partial}
\newcommand\Psibar{\overline{\Psi}}
\newcommand\psibar{\overline{\psi}}
\newcommand\chibar{\overline{\chi}}
\newcommand\tzeta{\widetilde{\zeta}}
\newcommand\Vbar{\overline{V}}
\newcommand\xbar{\overline{x}}
\newcommand\rme{\mathrm{e}}
\newcommand\rmi{\mathrm{i}}
\newcommand\rmc{\mathrm{c}}
\newcommand\rmd{\mathrm{d}}
\newcommand\rmexp{\mathrm{exp}}
\newcommand\rmln{\mathrm{ln}}
\newcommand\rmsin{\mathrm{sin}}
\newcommand\rmcos{\mathrm{cos}}
\newcommand\rmtan{\mathrm{tan}}
\newcommand\rmarcsin{\mathrm{arcsin}}
\newcommand\rmarccos{\mathrm{arccos}}
\newcommand\rmarctan{\mathrm{arctan}}
\newcommand\arcsec{\mathrm{arcsec}}
\newcommand\arcsinh{\mathrm{arcsinh}}
\newcommand\arccosh{\mathrm{arccosh}}
\newcommand\rmlim{\mathrm{lim}}
\newcommand\rmdet{\mathrm{det}}
\newcommand\rmker{\mathrm{ker}}
\newcommand\rmic{\mathrm{ic}}
\newcommand\rmt{\mathrm{t}}
\newcommand\rmT{\mathrm{T}}
\newcommand\rmRe{\mathrm{Re}}
\newcommand\rmIm{\mathrm{Im}}
\newcommand\scL{\mathcal{L}}
\newcommand\bux{\underline{\boldsymbol{x}}}
\newcommand\bua{\underline{\boldsymbol{a}}}
\newcommand\buu{\underline{\boldsymbol{u}}}
\newcommand\rhat{\hat{\boldsymbol{r}}}
\newcommand\xhat{\hat{\boldsymbol{x}}}
\newcommand\yhat{\hat{\boldsymbol{y}}}
\newcommand\zhat{\hat{\boldsymbol{z}}}
\newcommand\tV{\tilde{V}}
\newcommand\tT{\tilde{T}}
\newcommand\tW{\tilde{W}}
\newcommand\ta{\widetilde{a}}
\newcommand\tb{\widetilde{b}}
\newcommand\ba{\boldsymbol{a}}
\newcommand\bM{\boldsymbol{M}}
\newcommand\bW{\boldsymbol{W}}
\newcommand\bT{\boldsymbol{T}}
\newcommand\bv{\boldsymbol{v}}
\newcommand\bw{\boldsymbol{w}}
\newcommand\bxi{\boldsymbol{\xi}}
\newcommand\bep{\boldsymbol{\epsilon}}
\newcommand\sdual{{}^{*}\!s}
\newcommand\Sdual{{}^{*}\!S}
\newcommand\spart{\slashed{\partial}}
\newcommand\lspart{\overleftarrow{\slashed{\partial}}}




\section*{{\LARGE Fierz bilinear formulation of the Maxwell-Dirac equations and symmetry reductions}}
\vspace{0.5cm}



\begin{adjustwidth}{1cm}{}
{\large \textbf{S M Inglis and P D Jarvis}}

\vspace{0.2cm}
{\small\noindent School of Mathematics and Physics, University of Tasmania, Sandy Bay Campus, Private Bag 37, Hobart, Tasmania, 700}

\vspace{0.2cm}
{\small\noindent Email: {\tt sminglis@utas.edu.au, Peter.Jarvis@utas.edu.au}}

\vspace{0.5cm}
{\small
\noindent\textbf{Abstract.} We study the Maxwell-Dirac equations in a manifestly gauge invariant presentation using only the spinor bilinear scalar and pseudoscalar densities, and the vector and pseudovector currents, together with their quadratic Fierz relations. The internally produced vector potential is expressed via algebraic manipulation of the Dirac equation, as a rational function of the Fierz bilinears and first derivatives (valid on the support of the scalar density), which allows a gauge invariant vector potential to be defined. This leads to a Fierz bilinear formulation of the Maxwell tensor and of the Maxwell-Dirac equations, without any reference to gauge dependent quantities. We show how demanding invariance of tensor fields under the action of a fixed (but arbitrary) Lie subgroup of the Poincar\'{e} group leads to symmetry reduced equations. The procedure is illustrated, and the reduced equations worked out explicitly for standard spherical and cylindrical cases, which are coupled third order nonlinear PDEs. Spherical symmetry necessitates the existence of magnetic monopoles, which do not affect the coupled Maxwell-Dirac system due to magnetic terms canceling. In this paper we do not take up numerical computations. As a demonstration of the power of our approach, we also work out the symmetry reduced equations for two distinct classes of dimension 4 one-parameter families of Poincar\'{e} subgroups, one splitting and one non-splitting. The splitting class yields no solutions, whereas for the non-splitting class we find a family of formal exact solutions in closed form.}
\end{adjustwidth}







\section{Introduction}
The significance of the Dirac equation in physics is underlined by the immediate recognition it received following Dirac's seminal paper, even before its stunning experimental validation through its predictions of spectroscopic fine structure, and the discovery of the positron. It is hardly necessary to stress its subsequent leading role in the development of quantum electrodynamics as the paradigmatic interacting quantum field theory, which is also the exemplar for the matter sector of non-Abelian gauge theories and ultimately for the standard model of particle physics.  However, perturbative quantum field theory by construction starts with free particles and only then builds in interactions, and so effectively uses only trivial, free particle solutions to relativistic wave equations, specifically the Dirac equation for spin-1/2 fermions. It is surprising that, for such a major equation in mathematical physics, relatively little is known about classical solutions of the Dirac equation coupled to electromagnetism - that is, of the full, interacting Maxwell-Dirac equations - and what role they might play in non-perturbative effects in standard quantum electrodynamics, or indeed in generalizations of the second quantized theory itself.

In this article we develop a new formulation of symmetry-reduced Maxwell-Dirac equations. As we explain momentarily, we use a spinor bilinear approach, thereby realizing a long tradition of regarding the spinor wavefunction as a secondary entity, with its unphysical and unobservable phase and gauge ambiguities to be exorcised from calculations of measurable quantities. Although we develop the Fierz bilinear formulation for the general case, experience with other equations in physics suggests that, almost without exception, interesting solutions are those which possess special symmetry properties. Identifying symmetry constraints is therefore likely to be useful in filtering solution types. It is this strategy which is adopted for the present approach, exploiting the fact that the bilinear formulation is manifestly gauge invariant, thereby sidestepping questions of invariant gauge potentials. Moreover, in this setting, it is straightforward in principle to identify symmetry constraints arising from actions of \emph{any} admissible spacetime symmetry transformations. These we take to be associated with an \emph{arbitrary} (but fixed) Lie subgroup of the Poincar\'{e} group acting on four dimensional Minkowski space. In this paper we do not take up numerical computations, but we illustrate the setting up of symmetry reduced equations for some standard as well as nonstandard cases.

As mentioned, there is a substantial literature on the measurability of spinor wavefunctions, with historical origins in the ideas of de Broglie, Bohm, and Pauli. In the case of the Dirac theory, we cite Takabayasi \cite{Takabayasi-1957} as an exponent of the view that indeed the spinor bilinear quantities should be regarded as primary, coordinates of a type of relativistic fluid theory. More concretely, study of spinor bilinears involves the so-called Fierz algebra of their quadratic relations, intensively studied by, for example, Takahashi \cite{Takahashi-1982} and Zhelnorovich \cite{Zhelnorovich-1965} (see also Crawford \cite{Crawford-1985} and Kaempffer \cite{Kaempffer-1981}). It is this setting which underlies the `inversion' of the Dirac equation for the gauge potential, which is the starting point of the present work. In the original paper of Eliezer \cite{Eliezer-1958}, it was noted that the required manipulation involved finding the inverse of a certain $4\times 4$ complex matrix (the columns being simply the Dirac $\gamma$ matrices acting on the spinor wavefunction), and a straightforward calculation showed this matrix to be singular. Surprisingly however, imposing the additional constraint that the gauge potential should be real, does allow for a unique solution, and this was exploited by Radford and Booth \cite{Radford-1996}, \cite{Booth-Radford-1997} in two-component notation in their studies of the Maxwell-Dirac system (see below), and by Booth, Jarvis and Legg \cite{Booth-Legg-Jarvis-2001} in covariant form, with extensions to higher spacetime dimensions. In the latter paper, Eliezer's finding was resolved by pointing out that the the singular complex matrix in fact has full rank, when restricted to the relevant real subspace. In recent work \cite{Inglis-Jarvis-2012}, Inglis and Jarvis have provided an equivalent inversion formula for the Dirac equation with a non-Abelian gauge potential (with fermions in doublets of the gauge group $SU(2)$).

In this paper we point out that the algebraic inversion of the Dirac equation for the potential in fact allows for a complete gauge invariant reformulation of the Maxwell-Dirac equations. Firstly there is a natural gauge invariant vector field which acts as a proxy for the gauge potential, from which the latter can be recovered. Moreover, the Maxwell tensor can also be derived via the curl of this vector field, and hence ultimately given via the inversion formula in terms of the Fierz bilinears and derivatives. This leads to a Fierz current reformulation of the full Maxwell-Dirac equations, summarized in (\ref{Fierz inner product identity})-(\ref{Partial conservation of axial current}), which we believe to be new. 

Our aim is to apply this setting to investigate classical solutions of the resulting nonlinear coupled systems of partial differential equations. As with almost all significant equations in physics, interesting solutions are likely to be those with various special symmetry properties, and so we use this as a filter on their identification. However, in order not to prejudice the analysis, it is necessary to be quite general in the identification of such symmetric solutions. We formulate conditions for tensor fields to be invariant with respect to \emph{any} admissible space-time symmetry transformations. These we take to be associated with a fixed (but arbitrary) Lie subgroup of the relativistic Poincar\'{e} group acting in Minkowski space, which have been classified up to conjugacy class by Patera, Winternitz and Zassenhaus \cite{Patera-Winternitz-Zassenhaus-1975} and are 158 in number. Taking advantage of the gauge invariant presentation to sidestep issues of gauge equivalent fields, it is then straightforward in principle to set up the symmetry reduced Fierz-Maxwell-Dirac system of equations for any such symmetry type. In this work we content ourselves with working out the equations for selected standard cases, such as spherical and cylindrical symmetry, as well as two nonstandard cases $P_{11,2}$ and $\widetilde{P}_{13,10}$ obtained from \cite{Patera-Winternitz-Zassenhaus-1975}. Each of these has a free continuous parameter and hence represents an infinite family of symmetries. The reduced equations for these four cases are presented explicitly at the end of each subsection of section 5. The power of our systematic approach is illustrated by our ability to extend the study of the Fierz-Maxwell-Dirac system to these nonstandard symmetries.

In this paper we do not take up numerical computations, but to conclude these introductory remarks, it is worth reviewing what is known about solutions. Solitons, or highly localized solutions, have been investigated by Wakano \cite{Wakano-1966} and Lisi \cite{Lisi-1995}. Wakano obtained localized solutions to the Maxwell-Dirac equations when a dominant electrostatic potential $A^{0}$ was assumed, with no solutions existing for the case where $A^{0}$ was negligibly small compared with the three-vector potential, $A^{i}$. Lisi also numerically obtained an approximate localized solution by neglecting the three-vector potential, arguing that the angular dependence of $A^{i}$ would break the spherical symmetry. After reintroducing magnetic interaction via a perturbation and finding that it had a small effect on angular dependence, the full Maxwell-Dirac system was considered, and a normalized localized solution was found. Esteban et. al. \cite{Esteban-Georgiev-Sere-1996} employed a variational approach to finding localized solutions, and proved that stationary solutions do exist without making any approximations to the electromagnetic four vector potential.

Radford \cite{Radford-1996} performed a reduction and numerical analysis of the Maxwell-Dirac system under the assumption of a spherically symmetric, static Dirac spinor field, finding that localized compact objects with a shell like structure exist in this regime. At large distances, a shielding effect from the electrostatic charges dominates, and the field from the central charge distribution approaches a Coulombic form. Intriguingly, Radford also found that imposing static, spherical symmetry requires the existence of a magnetic monopole, a conclusion confirmed by the current study, albeit more generally. A major outcome of our study is that spherical symmetry \emph{alone} is enough to require the existence of a monopole, with magnetic charge $q_{m}=\mp2\pi/q$. However, we conclude that it has no effect on the coupled Maxwell-Dirac system, as all of the magnetic terms cancel out of the equations.

The requirement of magnetic monopoles is not shared by a subsequent study by Booth and Radford \cite{Booth-Radford-1997} on static cylindrically symmetric solutions. A solution describing a localized Dirac field with a concentric shell structure surrounding a charged axis, as well as finite linear charge density was obtained, and was compared to an equivalent ``linearized'' system, which lacked the self coupling between the Dirac and Maxwell equations. The relegation of the Maxwell field to an ``external'' potential resulted in both the localization of the Dirac field and boundedness of the charge density being destroyed.

Das and Kay \cite{Das-Kay-1989} investigated solutions to the Maxwell-Dirac system where the spinor field was assumed to be the form of a plane wave solution to the free Dirac equation. It was found that nontrivial solutions only exist when $m=0$, with the additional requirement that associated four vector fields be null. Plane wave solutions were also investigated by Bao and Li \cite{Bao-Li-2004} as a test case for their broad numerical scheme for the Maxwell-Dirac system, which yielded exact results. On the matter of the existence of general solutions, a very formidable global existence proof to the Cauchy problem for the Maxwell-Dirac equations has been provided by Flato et. al. \cite{Flato-Simon-Taflin-1997}, extending upon previous work on the matter by Gross \cite{Gross-1966}.

Lastly, we should mention the unpublished work of Legg \cite{Legg-2007}, upon which our current work is inspired, where Poincar\'{e} subgroup invariant solutions of the manifestly gauge invariant Maxwell-Dirac equations were investigated. Focusing on transitive Poincar\'{e} subgroups that have four-dimensional orbits, Legg found that the only one that resulted in a physically interesting solution was $\widetilde{P}_{13,10}$, one of the example subalgebras in the current work, and a reduction of the Maxwell-Dirac system and closed form solution was subsequently presented. Since the current work focuses on the reduction aspect, the investigation of the closed form solutions will appear in a later work.

The plan of the paper is as follows. Below, in section 2 we give a brief review of the algebraic manipulation of the Dirac equation, leading to its inversion for the electromagnetic vector potential $A^{\mu}$, as well as an alternative inverted form. This summarizes the work done in \cite{Booth-Legg-Jarvis-2001}, and serves as a guide to the inversion of the non-Abelian Dirac equation, described in \cite{Inglis-Jarvis-2012}. The section ends with a brief discussion on spinor bilinears and their corresponding Fierz identities.

In section 3, we make use of the Fierz expansion formula to derive the identities necessary to write the inverted Dirac equation in terms of bilinear tensors only, with no spinors explicitly visible. A manifestly gauge invariant vector potential $B^{\mu}$ is then defined by simply subtracting the gauge dependent bilinear term from the $A^{\mu}$ expression. A tetrad composed from the four mutually orthogonal four vector bilinear fields is then defined, and is used to convert the usual form of the field strength tensor into one which is manifestly gauge invariant, involving the sum of the four curl of $B^{\mu}$ and a rational term containing gauge invariant bilinears. The Maxwell-Dirac system is summarized at the end of the section.

Section 4 begins with a discussion of the Poincar\'{e} group, its generators and associated vector fields in the two representations applicable to this paper. The invariants and forms of scalar and vector fields invariant under four example Poincar\'{e} subgroups are then derived by constructing PDEs from the requirement that the Lie derivative of the fields under the transforming vector fields be zero, then solving them via the method of characteristics. The four subgroups considered are the standard spherical and cylindrical symmetries (subgroups $P_{3,4}$ and $P_{12,8}$ respectively), and two nonstandard subgroups $P_{11,2}$ and $\widetilde{P}_{13,10}$ obtained from \cite{Patera-Winternitz-Zassenhaus-1975}. In the case of the nonstandard subgroups, the invariant forms are calculated by cumulatively imposing the symmetry conditions required for each generator in the subgroup.

Section 5 involves the application of the subgroup symmetric forms calculated in section 4 to the Maxwell-Dirac system given at the end of section 3, in order to obtain a reduced system for each case. The reduction of the Fierz identities is considered first, then the gauge invariant vector potential, which is substituted into the field strength tensor, eliminating the vector potential from the system entirely. The final step is substitution of $F^{\mu\nu}$ into the inhomogeneous Maxwell equations, which generally results in a system of equations involving the two four vector fields $j^{\mu}$ and $k^{\mu}$, the two scalar fields $\sigma$ and $\omega$, and their derivatives up to third order. The reduced Fierz identities and consistency conditions are considered on a case by case basis. The spherical and cylindrical symmetries both yield very complicated systems of PDEs, so simplifying assumptions should be considered before seeking explicit solutions. The $P_{11,2}$ subgroup forces a trivial solution to the system, whereas the $\widetilde{P}_{13,10}$ case reduces (for one particular parameter choice) to the problem of solving the algebraic expression (\ref{P-13,10 reduced MD-Fierz equation}), which when taking account of (\ref{P-13,10 reduced orthogonality Fierz identity}) yields a four-parameter family of exact solutions in closed form.

Conclusions and prospects for further study are given in section 6, which is followed by four appendices. Appendix A contains a list of algebraic identities and definitions that are used throughout the paper. A list of tensor and vector potential coupling equations, consistency conditions and inversions resulting from the algebraic manipulation of the Dirac equation is summarized in appendix B. Appendix C contains a more detailed derivation of the tensor form of the vector potential than is given in section 3. Lastly, appendix D presents the explicit forms of the field strength tensor components for the cylindrically symmetric reduction.

\section{Dirac equation inversion and Fierz identities}
\subsection{Inversion}
The Dirac equation is the relativistic wave equation for spin-1/2 particles, such as electrons. For fermionic particles of charge $q$ interacting with an electromagnetic field, we require solutions to the Dirac equation form-invariant under a $U(1)$ Abelian gauge transformation, given by
\begin{equation}\label{U(1) covariant Dirac equation}
(\rmi\gamma^{\nu}\partial_{\nu}-q\gamma^{\nu}A_{\nu}-m)\psi=0.
\end{equation}
Conventions for Dirac algebra and spinor manipulations are given in appendix A. Our goal is to isolate the vector potential $A^{\mu}$. Rearranging gives us
\begin{equation}\label{U(1) covariant Dirac equation, rearranged for A}
\gamma^{\nu}\psi A_{\nu}=q^{-1}(\rmi\gamma^{\nu}\partial_{\nu}-m)\psi.
\end{equation}
We can form a bilinear spinor expression by multiplying by $\psibar\gamma^{\mu}$ from the left. Using the Dirac identity (\ref{gamma-gamma Dirac identity}), our expression becomes
\begin{equation}\label{Dirac equation after left-multiplication by psibar-gamma}
\psibar\psi A^{\mu}-\rmi\psibar\sigma^{\mu\nu}\psi A_{\nu}=q^{-1}[\rmi\psibar\gamma^{\mu}\gamma^{\nu}(\partial_{\nu}\psi)-m\psibar\gamma^{\mu}\psi].
\end{equation}
In order to eliminate the second term on the left-hand side, turn to the charge conjugate Dirac equation, which is similar in form to (\ref{U(1) covariant Dirac equation}), but with the sign of the charge reversed:
\begin{equation}
(\rmi\gamma^{\nu}\partial_{\nu}+q\gamma^{\nu}A_{\nu}-m)\psi^{\rmc}=0.
\end{equation}
The charge conjugate spinor is defined in terms of the regular spinor as \cite{Itzykson-Zuber-1980}
\begin{equation}
\psi^{\rmc}=C\psibar{}^{\rmT}=\rmi\gamma^{2}\gamma^{0}\psibar{}^{\rmT}.
\end{equation}
Similarly rearranging and left-multiplying by $\psibar{}^{\rmc}\gamma^{\mu}$, then applying the appropriate charge conjugation identities in appendix A.2, gives
\begin{equation}\label{Charge conjugate Dirac equation after left-multiplication by psibar-gamma}
-\psibar\psi A^{\mu}-\rmi\psibar\sigma^{\mu\nu}\psi A_{\nu}=q^{-1}[\rmi(\partial_{\nu}\psibar)\gamma^{\nu}\gamma^{\mu}\psi+m\psibar\gamma^{\mu}\psi].
\end{equation}
Subtracting (\ref{Charge conjugate Dirac equation after left-multiplication by psibar-gamma}) from (\ref{Dirac equation after left-multiplication by psibar-gamma}) and again using (\ref{gamma-gamma Dirac identity}), gives us the inverted form of the Dirac equation
\begin{equation}\label{Inverted Dirac equation}
A^{\mu}=\frac{1}{2q}\frac{\rmi[\psibar(\partial^{\mu}\psi)-(\partial^{\mu}\psibar)\psi]+\partial_{\nu}s^{\mu\nu}-2mj^{\mu}}{\sigma},
\end{equation}
where we have used the shorthand notation for Dirac bilinear tensors, listed in appendix A.3. There is an alternative inverted form for the Dirac equation, which involves left-multiplication of (\ref{U(1) covariant Dirac equation}) by $\psibar\gamma_{5}\gamma^{\mu}$ to form bilinears. Following the same steps as above yields the expression
\begin{equation}\label{Inverted dual Dirac equation}
A^{\mu}=\frac{1}{2q}\frac{\rmi[\psibar\gamma_{5}(\partial^{\mu}\psi)-(\partial^{\mu}\psibar)\gamma_{5}\psi]+\partial_{\nu}\sdual^{\mu\nu}}{\omega},
\end{equation}
which lacks a mass-dependent term. In addition to these inversions, we can derive other expressions by left-multiplying (\ref{U(1) covariant Dirac equation, rearranged for A}) and its charge conjugate analogue by $\psibar\Gamma$ and $\psibar{}^{\rmc}\Gamma$ respectively, for general elements $\Gamma$ of the Dirac algebra, then adding or subtracting the two equations. Among the resulting expressions are the continuity equation $\partial_{\mu}j^{\mu}=0$ and the current-field coupling $j^{\nu}A_{\nu}$. The full list of expressions obtained from ``bilinearizing'' the Dirac equation is given in appendix B.

\subsection{Fierz identities}
It is well known that quadratic relationships between Dirac bilinears of the form $\chibar\Gamma_{R}\psi$, where $\Gamma_{R}$ represents the sixteen basis elements of the Dirac algebra $\Gamma_{R}=\{I,\gamma^{\mu},\sigma^{\mu\nu},\gamma_{5}\gamma^{\mu},\gamma_{5}\}$, can be derived via a Fierz expansion of the product of two Dirac spinors in this basis
\begin{align}\label{Dirac spinor Fierz expansion}
\psi\chibar=\sum_{R=1}^{16}a_{R}\Gamma_{R}={}&(1/4)(\chibar\psi)I+(1/4)(\chibar\gamma_{\mu}\psi)\gamma^{\mu}+(1/8)(\chibar\sigma_{\mu\nu}\psi)\sigma^{\mu\nu} \nonumber \\
&-(1/4)(\chibar\gamma_{5}\gamma_{\mu}\psi)\gamma_{5}\gamma^{\mu}+(1/4)(\chibar\gamma_{5}\psi)\gamma_{5}.
\end{align}
Here, $a_{R}$ are the numerical coefficients multiplied by the Dirac bilinears. This expansion is inserted into products such as $j^{\mu}k^{\nu}\equiv\psibar\gamma^{\mu}(\psi\psibar)\gamma_{5}\gamma^{\nu}\psi$, for example. Experimenting with different combinations of bilinears, and combining the resulting equations yields many different interrelationships. Many Fierz identities are summarized in \cite{Takahashi-1982} and \cite{Crawford-1985}, but the most important for our purposes are
\begin{align}
&j_{\nu}j^{\nu}=-k_{\nu}k^{\nu}=-m^{\nu}m_{\nu}=-n^{\nu}n_{\nu}=\sigma^{2}-\omega^{2},\label{Inner product Fierz identities} \\
&j_{\nu}k^{\nu}=j_{\nu}m^{\nu}=j_{\nu}n^{\nu}=k_{\nu}m^{\nu}=k_{\nu}n^{\nu}=m_{\nu}n^{\nu}=0,\label{Orthogonality Fierz identities} \\
&\epsilon_{\mu\nu\rho\sigma}j^{\rho}k^{\sigma}=m_{\mu}n_{\nu}-m_{\nu}n_{\mu},\label{Spin-plane Fierz identity} \\
&s_{\mu\nu}=\frac{(\sigma\epsilon_{\mu\nu}{}^{\rho\sigma}-\omega\delta_{\mu\nu}{}^{\rho\sigma})j_{\rho}k_{\sigma}}{\sigma^{2}-\omega^{2}}, \label{s_mu-nu Fierz identity} \\
&\sdual_{\mu\nu}=\frac{(\omega\epsilon_{\mu\nu}{}^{\rho\sigma}-\sigma\delta_{\mu\nu}{}^{\rho\sigma})j_{\rho}k_{\sigma}}{\sigma^{2}-\omega^{2}}. \label{sdual_mu-nu Fierz identity}
\end{align}
This method can be extended to the $SU(2)$ spinor doublet case by building into (\ref{Dirac spinor Fierz expansion}) an expansion over the Pauli matrices, including the $2\times2$ identity matrix. Analogous non-Abelian expressions to (\ref{s_mu-nu Fierz identity}) and (\ref{sdual_mu-nu Fierz identity}) were derived in \cite{Inglis-Jarvis-2012} via the use of such expansions.

\section{Manifestly gauge invariant Maxwell-Dirac equations}
\subsection{Vector potential}
In order to avoid arbitrarily fixing the gauge, here we eliminate gauge dependent terms from our equations entirely, so that our Maxwell-Dirac system is \emph{manifestly} gauge invariant. We do this by reformulating (\ref{Inverted Dirac equation}) and (\ref{Inverted dual Dirac equation}) to be entirely in terms of the bilinear tensors listed in appendix A.3, getting rid of the incongruous $[\psibar(\partial_{\mu}\psi)-(\partial_{\mu}\psibar)\psi]$ terms. We can then easily pick off the gauge dependent parts ($m^{\mu}$ and $n^{\mu}$), to define a gauge-invariant vector potential, which we denote $B^{\mu}$. A Maxwell-Dirac formalism, completely in terms of manifestly gauge invariant tensors is then derived.

This approach is in the spirit of Takabayasi \cite{Takabayasi-1957}, whose philosophy regarded a relativistic quantum mechanical formalism strictly involving only ``observables'', such as tensors, as being preferable to one where somewhat ``unphysical'' objects such as spinors are explicitly included. A detailed derivation of $B^{\mu}$ is given in appendix C, but we give a brief overview here.

First, take the sum of the two versions of the inverted Dirac equation (\ref{Inverted Dirac equation}), (\ref{Inverted dual Dirac equation}), and divide by 2
\begin{align}\label{Inverted Dirac equation obtained via sum}
A_{\mu}={}&\frac{1}{4q}\left\{\frac{\rmi[\psibar(\partial_{\mu}\psi)-(\partial_{\mu}\psibar)\psi]\omega+\rmi[\psibar\gamma_{5}(\partial_{\mu}\psi)-(\partial_{\mu}\psibar)\gamma_{5}\psi]\sigma}{\sigma\omega}+\frac{\partial_{\nu}s_{\mu}{}^{\nu}}{\sigma}+\frac{\partial_{\nu}\sdual_{\mu}{}^{\nu}}{\omega}\right. \nonumber \\
&\left.\qquad-\frac{2mj_{\mu}}{\sigma}\right\}.
\end{align}
The appropriate tensor forms needed to replace the spinor terms are $j^{\nu}(\partial_{\mu}k_{\nu})$ and $m^{\nu}(\partial_{\mu}n_{\nu})$. Consider the first tensor:
\begin{equation}
j^{\nu}(\partial_{\mu}k_{\nu})=\psibar\gamma^{\nu}\psi\cdot(\partial_{\mu}\psibar)\gamma_{5}\gamma_{\nu}\psi+\psibar\gamma^{\nu}\psi\cdot\psibar\gamma_{5}\gamma_{\nu}(\partial_{\mu}\psi).
\end{equation}
Fierz expanding both terms and rearranging gives
\begin{equation}
 j^{\nu}(\partial_{\mu}k_{\nu})=(2/3)[\psibar(\partial_{\mu}\psi)-(\partial_{\mu}\psibar)\psi]\omega-(2/3)[\psibar\gamma_{5}(\partial_{\mu}\psi)-(\partial_{\mu}\psibar)\gamma_{5}\psi]\sigma-(1/3)k^{\nu}(\partial_{\mu}j_{\nu}).
\end{equation}
We must also consider the Fierz expansion of $k^{\nu}(\partial_{\mu}j_{\nu})$ in order to eliminate it from the expression, which after rearrangement is given by
\begin{equation}
 k^{\nu}(\partial_{\mu}j_{\nu})=(2/3)[\psibar\gamma_{5}(\partial_{\mu}\psi)-(\partial_{\mu}\psibar)\gamma_{5}\psi]\sigma-(2/3)[\psibar(\partial_{\mu}\psi)-(\partial_{\mu}\psibar)\psi]\omega-(1/3)j^{\nu}(\partial_{\mu}k_{\nu}).
\end{equation}
Using these two equations yields the new Fierz identity
\begin{equation}\label{Spinor-tensor identity for j-partial-k}
 j^{\nu}(\partial_{\mu}k_{\nu})=-k^{\nu}(\partial_{\mu}j_{\nu})=[\psibar(\partial_{\mu}\psi)-(\partial_{\mu}\psibar)\psi]\omega-[\psibar\gamma_{5}(\partial_{\mu}\psi)-(\partial_{\mu}\psibar)\gamma_{5}\psi]\sigma.
\end{equation}
In order to eliminate both of the bracketed spinor terms, we require another independent expression involving them. Such an expression is provided by $m^{\nu}(\partial_{\mu}n_{\nu})$, which in spinor form is
\begin{align}
 m^{\nu}(\partial_{\mu}n_{\nu})={}&(\rmi/4)[\psibar{}^{\rmc}\gamma^{\nu}\psi\!\cdot\!(\partial_{\mu}\psibar)\gamma_{\nu}\psi^{\rmc}+\psibar{}^{\rmc}\gamma^{\nu}\psi\!\cdot\!\psibar\gamma_{\nu}(\partial_{\mu}\psi^{\rmc})-\psibar{}^{\rmc}\gamma^{\nu}\psi\!\cdot\!(\partial_{\mu}\psibar{}^{\rmc})\gamma_{\nu}\psi \nonumber \\
&-\psibar{}^{\rmc}\gamma^{\nu}\psi\cdot\psibar{}^{\rmc}\gamma_{\nu}(\partial_{\mu}\psi)+\psibar\gamma^{\nu}\psi^{\rmc}\cdot(\partial_{\mu}\psibar)\gamma_{\nu}\psi^{\rmc}+\psibar\gamma^{\nu}\psi^{\rmc}\cdot\psibar\gamma_{\nu}(\partial_{\mu}\psi^{\rmc}) \nonumber \\
&-\psibar\gamma^{\nu}\psi^{\rmc}\cdot(\partial_{\mu}\psibar{}^{\rmc})\gamma_{\nu}\psi-\psibar\gamma^{\nu}\psi^{\rmc}\cdot\psibar{}^{\rmc}\gamma_{\nu}(\partial_{\mu}\psi)].
\end{align}
Fierz expanding the individual terms, and applying the appropriate charge conjugate and complex conjugate bilinear identities from appendix A, we obtain
\begin{align}
 m^{\nu}(\partial_{\mu}n_{\nu})={}&(\rmi/4)\{2[\psibar(\partial_{\mu}\psi)-(\partial_{\mu}\psibar)\psi]\sigma-2[\psibar\gamma_{5}(\partial_{\mu}\psi)-(\partial_{\mu}\psibar)\gamma_{5}\psi]\omega \nonumber \\
&+[\psibar\gamma_{\nu}(\partial_{\mu}\psi)-(\partial_{\mu}\psibar)\gamma_{\nu}\psi]j^{\nu}-[\psibar\gamma_{5}\gamma_{\nu}(\partial_{\mu}\psi)-(\partial_{\mu}\psibar)\gamma_{5}\gamma_{\nu}\psi]k^{\nu}\}.
\end{align}
Performing another round of Fierz expansions on the last two terms eventually provides us with another Fierz identity
\begin{align}
&[\psibar\gamma_{\nu}(\partial_{\mu}\psi)-(\partial_{\mu}\psibar)\gamma_{\nu}\psi]j^{\nu}=-[\psibar\gamma_{5}\gamma_{\nu}(\partial_{\mu}\psi)-(\partial_{\mu}\psibar)\gamma_{5}\gamma_{\nu}\psi]k^{\nu} \nonumber \\
&\qquad=[\psibar(\partial_{\mu}\psi)-(\partial_{\mu}\psibar)\psi]\sigma-[\psibar\gamma_{5}(\partial_{\mu}\psi)-(\partial_{\mu}\psibar)\gamma_{5}\psi]\omega,
\end{align}
which gives us our desired identity
\begin{equation}\label{Spinor-tensor identity for m-partial-n}
m^{\nu}(\partial_{\mu}n_{\nu})=\rmi[\psibar(\partial_{\mu}\psi)-(\partial_{\mu}\psibar)\psi]\sigma-\rmi[\psibar\gamma_{5}(\partial_{\mu}\psi)-(\partial_{\mu}\psibar)\gamma_{5}\psi]\omega.
\end{equation}
It can be shown via a similar process that
\begin{equation}\label{Tensor-tensor identity for m-partial-n}
m^{\nu}(\partial_{\mu}n_{\nu})=-n^{\nu}(\partial_{\mu}m_{\nu}).
\end{equation}
From substitution and rearrangement of (\ref{Spinor-tensor identity for j-partial-k}) and (\ref{Spinor-tensor identity for m-partial-n}), we get the expressions needed to eliminate spinors from the inverted Dirac equation entirely
\begin{align}
&[\psibar(\partial_{\mu}\psi)-(\partial_{\mu}\psibar)\psi]=-(\sigma^{2}-\omega^{2})^{-1}[j^{\nu}(\partial_{\mu}k_{\nu})\omega+\rmi m^{\nu}(\partial_{\mu}n_{\nu})\sigma],\label{psi-psi anti-product rule identity} \\
&[\psibar\gamma_{5}(\partial_{\mu}\psi)-(\partial_{\mu}\psibar)\gamma_{5}\psi]=-(\sigma^{2}-\omega^{2})^{-1}[j^{\nu}(\partial_{\mu}k_{\nu})\sigma+\rmi m^{\nu}(\partial_{\mu}n_{\nu})\omega].\label{psi-gamma5-psi anti-product rule identity}
\end{align}
Combining these two identities in the form they appear in (\ref{Inverted Dirac equation obtained via sum}), we get
\begin{align}
&\{\rmi[\psibar(\partial_{\mu}\psi)-(\partial_{\mu}\psibar)\psi]\omega+\rmi[\psibar\gamma_{5}(\partial_{\mu}\psi)-(\partial_{\mu}\psibar)\gamma_{5}\psi]\sigma\}(\sigma\omega)^{-1} \nonumber \\
&\qquad=\frac{2m^{\nu}(\partial_{\mu}n_{\nu})}{\sigma^{2}-\omega^{2}}-\frac{\rmi j^{\nu}(\partial_{\mu}k_{\nu})}{\sigma^{2}-\omega^{2}}\left[\frac{\sigma^{2}+\omega^{2}}{\sigma\omega}\right].
\end{align}
Substituting into (\ref{Inverted Dirac equation obtained via sum}), we obtain an expression for $A_{\mu}$ exclusively in tensor form
\begin{equation}\label{Prototype inverted tensorial Dirac equation}
 A_{\mu}=\frac{1}{4q}\left\{\frac{2m^{\nu}(\partial_{\mu}n_{\nu})}{\sigma^{2}-\omega^{2}}-\rmi j^{\nu}(\partial_{\mu}k_{\nu})\left[\frac{\sigma^{2}+\omega^{2}}{\sigma\omega(\sigma^{2}-\omega^{2})}\right]+\frac{\partial_{\nu}s_{\mu}{}^{\nu}}{\sigma}+\frac{\partial_{\nu}\sdual_{\mu}{}^{\nu}}{\omega}-\frac{2mj_{\mu}}{\sigma}\right\}.
\end{equation}
We can improve on this by substituting (\ref{psi-psi anti-product rule identity}) and (\ref{psi-gamma5-psi anti-product rule identity}) into (\ref{Inverted Dirac equation}) and (\ref{Inverted dual Dirac equation}) respectively, then subtracting and rearranging, obtaining a consistency condition for the Dirac equation in tensor form
\begin{equation}
\rmi j^{\nu}(\partial_{\mu}k_{\nu})=2m\omega j_{\mu}+\sigma\partial_{\nu}\sdual_{\mu}{}^{\nu}-\omega\partial_{\nu}s_{\mu}{}^{\nu}.
\end{equation}
Substituting this into (\ref{Prototype inverted tensorial Dirac equation}), we obtain after some algebraic manipulation, the final form of the inverted Dirac equation in tensor form
\begin{equation}
A_{\mu}=\frac{1}{2q}\frac{m^{\nu}(\partial_{\mu}n_{\nu})+\sigma\partial_{\nu}s_{\mu}{}^{\nu}-\omega\partial_{\nu}\sdual_{\mu}{}^{\nu}-2m\sigma j_{\mu}}{\sigma^{2}-\omega^{2}}.
\end{equation}
We define the gauge invariant vector potential simply by subtracting the only gauge dependent part from $A_{\mu}$
\begin{equation}\label{Gauge invariant vector potential in terms of s}
B_{\mu}=A_{\mu}-\frac{1}{2q}\frac{m^{\nu}(\partial_{\mu}n_{\nu})}{\sigma^{2}-\omega^{2}}=\frac{1}{2q}\frac{\sigma\partial_{\nu}s_{\mu}{}^{\nu}-\omega\partial_{\nu}\sdual_{\mu}{}^{\nu}-2m\sigma j_{\mu}}{\sigma^{2}-\omega^{2}}.
\end{equation}
The Fierz identities (\ref{s_mu-nu Fierz identity}) and (\ref{sdual_mu-nu Fierz identity}) can be used to eliminate the rank-2 tensors from the $B_{\mu}$ expression entirely. With a small amount of work, we find that
\begin{align}
& B_{\mu}=(1/2q)\{\epsilon_{\mu}{}^{\nu\rho\sigma}[(\sigma^{2}-\omega^{2})\partial_{\nu}(j_{\rho}k_{\sigma})-(1/2)j_{\rho}k_{\sigma}\partial_{\nu}(\sigma^{2}-\omega^{2})] \nonumber \\
&\qquad+\delta_{\mu}{}^{\nu\rho\sigma}[(\partial_{\nu}\sigma)\omega-\sigma(\partial_{\nu}\omega)]j_{\rho}k_{\sigma}\}(\sigma^{2}-\omega^{2})^{-2}-(1/q)m\sigma j_{\mu}(\sigma^{2}-\omega^{2})^{-1}.
\end{align}
It is apparent that $B_{\mu}$ is only finite when $\sigma^{2}-\omega^{2}\neq0$. It is perhaps appropriate to mention here that a common alternative definition of the pseudoscalar bilinear \cite{Crawford-1985}, \cite{Takabayasi-1957} is $\varpi=\psibar\rmi\gamma_{5}\psi$, so performing a change of variables, we would have in the denominator $\sigma^{2}+\varpi^{2}$. Since $\varpi$ is real \cite{Crawford-1985}, implying that $\omega$ is purely imaginary, the denominator only vanishes for $\sigma$ and $\omega$ vanishing independently. Additionally, we have the condition that $\sigma^{2}-\omega^{2}\geq0$.

\subsection{The tetrad of bilinears}
Here we make the claim based on (\ref{Inner product Fierz identities}) and (\ref{Orthogonality Fierz identities}) that the four mutually orthogonal four vector fields $j^{\mu}$, $m^{\mu}$, $n^{\mu}$ and $k^{\mu}$ constitute the columns of a tetrad \cite{Legg-2007}
\begin{equation}
\rmt^{\mu}{}_{\alpha}=(\sigma^{2}-\omega^{2})^{-1/2}[j^{\mu},m^{\mu},n^{\mu},k^{\mu}].
\end{equation}
$\mu=0,1,2,3$ is the spacetime index as usual, and $\alpha=0,1,2,3$ labels the columns, with $\alpha=0$ denoting the timelike field $j^{\mu}$ and $\alpha=1,2,3$ denoting the spacelike fields, $m^{\mu}$, $n^{\mu}$ and $k^{\mu}$ respectively. Gauge transformations can be thought of as rotations in the $m^{\mu}-n^{\mu}$ plane. The coefficient $(\sigma^{2}-\omega^{2})^{-1/2}$ behaves as a normalizing factor. Now consider the contraction of two tetrads via the $\mu$ index
\begin{equation}\label{Tetrad contraction via mu index}
\rmt^{\alpha}{}_{\mu}\rmt^{\mu}{}_{\beta}=(\sigma^{2}-\omega^{2})^{-1}\left(\begin{array}{cccc}j_{\mu}j^{\mu} & j_{\mu}m^{\mu} & j_{\mu}n^{\mu} & j_{\mu}k^{\mu} \\
-m_{\mu}j^{\mu} & -m_{\mu}m^{\mu} & -m_{\mu}n^{\mu} & -m_{\mu}k^{\mu} \\
-n_{\mu}j^{\mu} & -n_{\mu}m^{\mu} & -n_{\mu}n^{\mu} & -n_{\mu}k^{\mu} \\
-k_{\mu}j^{\mu} & -k_{\mu}m^{\mu} & -k_{\mu}n^{\mu} & -k_{\mu}k^{\mu}\end{array}\right)=\delta^{\alpha}{}_{\beta},
\end{equation}
which in matrix notation is simply $(\eta\rmt^{\rmT}\eta)\rmt=I$, implying that $(\eta\rmt^{\rmT}\eta)=\rmt^{-1}$. Putting the inverse tetrad on the right, and labeling the indices appropriately gives us
\begin{equation}\label{Tetrad contraction via alpha index}
\rmt^{\mu}{}_{\alpha}\rmt^{\alpha}{}_{\nu}=(\sigma^{2}-\omega^{2})^{-1}(j^{\mu}j_{\nu}-m^{\mu}m_{\nu}-n^{\mu}n_{\nu}-k^{\mu}k_{\nu})=\delta^{\mu}{}_{\nu},
\end{equation}
an identity which we will find useful in the next section. Taking the derivative of (\ref{Tetrad contraction via mu index}) and rearranging gives
\begin{equation}\label{Tetrad derivative identity}
\rmt^{\nu}{}_{\alpha}(\partial_{\mu}\rmt_{\nu\beta})=-\rmt^{\nu}{}_{\beta}(\partial_{\mu}\rmt_{\nu\alpha}),
\end{equation}
which is antisymmetric in $\alpha$ and $\beta$. In fact, this is a generalization of (\ref{Spinor-tensor identity for j-partial-k}) and (\ref{Tensor-tensor identity for m-partial-n}), which were originally derived via the Fierz expansion method. A result of the antisymmetry is that if $\alpha=\beta$, the term vanishes, which tells us that the four vector fields multiplied by the normalizing factor $(\sigma^{2}-\omega^{2})^{-1}$ are orthogonal to their own partial four derivatives. Substituting the components of the tetrad for $\alpha=\beta$ into (\ref{Tetrad derivative identity}) gives an identity in terms of the unnormalized vectors
\begin{equation}\label{x-partial-x contracted identity}
 j^{\nu}(\partial_{\mu}j_{\nu})=-m^{\nu}(\partial_{\mu}m_{\nu})=-n^{\nu}(\partial_{\mu}n_{\nu})=-k^{\nu}(\partial_{\mu}k_{\nu})=\sigma(\partial_{\mu}\sigma)-\omega(\partial_{\mu}\omega),
\end{equation}
which is just the derivative of (\ref{Inner product Fierz identities}).

\subsection{Field strength tensor}
The electromagnetic field strength tensor is defined as
\begin{equation}
F_{\mu\nu}=\partial_{\mu}A_{\nu}-\partial_{\nu}A_{\mu}.
\end{equation}
This is a gauge invariant tensor, but in this form it is not \emph{manifestly} gauge invariant because it explicitly contains the gauge dependent term $A_{\mu}$. The manifestly gauge invariant $F_{\mu\nu}$ was originally obtained by Takabayasi \cite{Takabayasi-1957}, then again by Legg \cite{Legg-2007}, this derivation mirroring that of the latter. Replace $A_{\mu}$ using (\ref{Gauge invariant vector potential in terms of s})
\begin{equation}
F_{\mu\nu}=\partial_{\mu}B_{\nu}-\partial_{\nu}B_{\mu}+\frac{1}{2q}\left\{\partial_{\mu}\left[\frac{m^{\rho}(\partial_{\nu}n_{\rho})}{\sigma^{2}-\omega^{2}}\right]-\partial_{\nu}\left[\frac{m^{\rho}(\partial_{\mu}n_{\rho})}{\sigma^{2}-\omega^{2}}\right]\right\}.
\end{equation}
Our goal is to eliminate the gauge dependent terms $m^{\mu}$ and $n^{\mu}$ from the expression entirely. Expanding the derivatives in the bracketed term gives us
\begin{align}
&\partial_{\mu}\left[\frac{m^{\rho}(\partial_{\nu}n_{\rho})}{\sigma^{2}-\omega^{2}}\right]-\partial_{\nu}\left[\frac{m^{\rho}(\partial_{\mu}n_{\rho})}{\sigma^{2}-\omega^{2}}\right]=\frac{(\partial_{\mu}m^{\rho})(\partial_{\nu}n_{\rho})-(\partial_{\nu}m^{\rho})(\partial_{\mu}n_{\rho})}{\sigma^{2}-\omega^{2}} \nonumber \\
&\qquad+\frac{m^{\rho}(\partial_{\mu}n_{\rho})\partial_{\nu}(\sigma^{2}-\omega^{2})-m^{\rho}(\partial_{\nu}n_{\rho})\partial_{\mu}(\sigma^{2}-\omega^{2})}{(\sigma^{2}-\omega^{2})^{2}}.\label{Expanded bracket term in F-mu-nu}
\end{align}
Focusing on the left-hand term, if we insert the identity $\delta_{\sigma}{}^{\rho}$ between each four vector derivative in the numerator, then expand using (\ref{Tetrad contraction via alpha index}), we get
\begin{align}
&[(\partial_{\mu}m^{\sigma})\delta_{\sigma}{}^{\rho}(\partial_{\nu}n_{\rho})-(\partial_{\nu}m^{\sigma})\delta_{\sigma}{}^{\rho}(\partial_{\mu}n_{\rho})](\sigma^{2}-\omega^{2})^{-1} \nonumber \\
&=[(\partial_{\mu}m^{\sigma})j_{\sigma}j^{\rho}(\partial_{\nu}n_{\rho})-(\partial_{\mu}m^{\sigma})k_{\sigma}k^{\rho}(\partial_{\nu}n_{\rho})-(\partial_{\mu}m^{\sigma})m_{\sigma}m^{\rho}(\partial_{\nu}n_{\rho}) \nonumber \\
&\qquad-(\partial_{\mu}m^{\sigma})n_{\sigma}n^{\rho}(\partial_{\nu}n_{\rho})-(\partial_{\nu}m^{\sigma})j_{\sigma}j^{\rho}(\partial_{\mu}n_{\rho})+(\partial_{\nu}m^{\sigma})k_{\sigma}k^{\rho}(\partial_{\mu}n_{\rho}) \nonumber \\
&\qquad+(\partial_{\nu}m^{\sigma})m_{\sigma}m^{\rho}(\partial_{\mu}n_{\rho})+(\partial_{\nu}m^{\sigma})n_{\sigma}n^{\rho}(\partial_{\mu}n_{\rho})](\sigma^{2}-\omega^{2})^{-2} \nonumber \\
&=(m^{\sigma}n^{\rho}-m^{\rho}n^{\sigma})[(\partial_{\mu}j_{\sigma})(\partial_{\nu}j_{\rho})-(\partial_{\mu}k_{\sigma})(\partial_{\nu}k_{\rho})](\sigma^{2}-\omega^{2})^{-2} \nonumber \\
&\qquad-[m^{\sigma}(\partial_{\mu}n_{\sigma})\partial_{\nu}(\sigma^{2}-\omega^{2})-m^{\sigma}(\partial_{\nu}n_{\sigma})\partial_{\mu}(\sigma^{2}-\omega^{2})](\sigma^{2}-\omega^{2})^{-2}.\label{Left-hand term expanded from tetrad identity}
\end{align}
To get to the first term in the last step, we used the tetrad identity (\ref{Tetrad derivative identity}) to switch the partial derivatives onto the gauge independent tensors, then factorized. The second term in the last step follows from using (\ref{x-partial-x contracted identity}) to replace terms like $m^{\rho}(\partial_{\mu}m_{\rho})$ with $-(1/2)\partial_{\mu}(\sigma^{2}-\omega^{2})$, and using and (\ref{Tetrad derivative identity}) to place all the derivatives onto the $n_{\sigma}$ vectors. Substituting (\ref{Left-hand term expanded from tetrad identity}) into (\ref{Expanded bracket term in F-mu-nu}), the right-hand term in (\ref{Expanded bracket term in F-mu-nu}) cancels out, leaving us with
\begin{equation}
\partial_{\mu}\left[\frac{m^{\rho}(\partial_{\nu}n_{\rho})}{\sigma^{2}-\omega^{2}}\right]-\partial_{\nu}\left[\frac{m^{\rho}(\partial_{\mu}n_{\rho})}{\sigma^{2}-\omega^{2}}\right]=\frac{(m^{\sigma}n^{\rho}-m^{\rho}n^{\sigma})[(\partial_{\mu}j_{\sigma})(\partial_{\nu}j_{\rho})-(\partial_{\mu}k_{\sigma})(\partial_{\nu}k_{\rho})]}{(\sigma^{2}-\omega^{2})^{2}}.
\end{equation}
Applying the Fierz identity (\ref{Spin-plane Fierz identity}), we can eliminate the gauge dependent tensors entirely, giving us the desired expression
\begin{equation}
F_{\mu\nu}=\partial_{\mu}B_{\nu}-\partial_{\nu}B_{\mu}+\frac{1}{2q}\frac{\epsilon^{\sigma\rho\kappa\tau}j_{\kappa}k_{\tau}[(\partial_{\mu}j_{\sigma})(\partial_{\nu}j_{\rho})-(\partial_{\mu}k_{\sigma})(\partial_{\nu}k_{\rho})]}{(\sigma^{2}-\omega^{2})^{2}},
\end{equation}
the manifestly gauge invariant electromagnetic field strength tensor.

\subsection{Maxwell-Dirac equations}
In summary, our Maxwell-Dirac system consists of the Fierz identities
\begin{align}
&j_{\mu}j^{\mu}=-k_{\mu}k^{\mu}=\sigma^{2}-\omega^{2},\label{Fierz inner product identity} \\
&j_{\mu}k^{\mu}=0,\label{Fierz orthogonality identity}
\end{align}
the gauge invariant form of the inverted Dirac equation
\begin{align}\label{Gauge invariant vector potential}
& B_{\mu}=(1/2q)\{\epsilon_{\mu}{}^{\nu\rho\sigma}[(\sigma^{2}-\omega^{2})\partial_{\nu}(j_{\rho}k_{\sigma})-(1/2)j_{\rho}k_{\sigma}\partial_{\nu}(\sigma^{2}-\omega^{2})] \nonumber \\
&\qquad+\delta_{\mu}{}^{\nu\rho\sigma}[(\partial_{\nu}\sigma)\omega-\sigma(\partial_{\nu}\omega)]j_{\rho}k_{\sigma}\}(\sigma^{2}-\omega^{2})^{-2}-(1/q)m\sigma j_{\mu}(\sigma^{2}-\omega^{2})^{-1},
\end{align}
the manifestly gauge invariant field strength tensor
\begin{equation}\label{Manifestly gauge invariant field strength tensor}
F_{\mu\nu}=\partial_{\mu}B_{\nu}-\partial_{\nu}B_{\mu}+\frac{1}{2q}\frac{\epsilon^{\sigma\rho\kappa\tau}j_{\kappa}k_{\tau}[(\partial_{\mu}j_{\sigma})(\partial_{\nu}j_{\rho})-(\partial_{\mu}k_{\sigma})(\partial_{\nu}k_{\rho})]}{(\sigma^{2}-\omega^{2})^{2}},
\end{equation}
and the inhomogeneous Maxwell equations
\begin{equation}\label{Maxwell equations}
\partial_{\nu}F^{\nu\mu}=qj^{\mu}.
\end{equation}
We can also use two of the physical constraints obtained from manipulating the Dirac equation listed in appendix B, the continuity equation and its pseudovector analogue
\begin{align}
&\partial_{\mu}j^{\mu}=0,\label{Continuity equation} \\
&\partial_{\mu}k^{\mu}=-2\rmi m\omega.\label{Partial conservation of axial current}
\end{align}
Because we are using the inverted Dirac equation in the field strength tensor instead of an ``external'' electromagnetic field, we are demanding that the charged fermionic field itself be the source of the field. Substituting into the Maxwell equations yields a self-consistent set of PDEs describing the behaviour of the fermion field under its own electromagnetic field. In the next sections, we will consider how the imposition of symmetry under select subgroups of the Poincar\'{e} group affects this system, and as we shall see, depending on the subgroup we choose, the complexity of the Maxwell-Dirac system varies dramatically.

\section{Symmetry of tensor fields under Poincar\'{e} subgroups}
Now that we have obtained our Maxwell-Dirac system, we must consider how we are to go about finding symmetry reductions. Observing equations (\ref{Gauge invariant vector potential})-(\ref{Maxwell equations}), we can see that we have a third-order non-linear coupled set of PDEs; a very formidable system indeed. It is fortunate that most physically interesting situations have symmetry under a certain subgroup of the Poincar\'{e} group, two prime examples being the spherical and cylindrical symmetries. These particular cases were studied by Radford and Booth \cite{Radford-1996}, \cite{Booth-Radford-1997}, with the additional constraint of having a static Dirac field, which assumes that there must exist a Lorentz frame in which there is no current flow, $j^{\mu}=\delta_{0}{}^{\mu}j^{0}$. Since their work was done in the gauge dependent 2-spinor formalism, a specific gauge was chosen to remove gauge ambiguity. In this study, we will apply these same symmetries to the Maxwell-Dirac system, but since we are working with inherently gauge invariant tensor fields only, we have the advantage of not having to choose any specific gauge arbitrarily, which could result in a loss of generality.

The situations of spherical and cylindrical symmetry are but two of many possibilities for analyzing the structure of the reduced Maxwell-Dirac system in the presence of symmetries. Given that we are dealing with relativistic wave equations compatible with the underlying action of the Poincar\'{e} group of transformations on Minkowski space, appropriate symmetries are therefore subgroups of the Poincar\'{e} group. The comprehensive classification by Patera, Winternitz and Zassenhaus \cite{Patera-Winternitz-Zassenhaus-1975} (hereafter PWZ), identifies all 158 continuous subgroups of the Poincar\'{e} group up to conjugacy, and the methods we develop are in principle able to give Maxwell-Dirac symmetry reductions for \emph{any} of these subgroups. At the Lie algebra level, the PWZ scheme uses the known list $F_{i}$, $i=1,2,...,15$ of distinct subalgebras of the Lorentz group Lie algebra, to establish a corresponding classification $P_{i,j}$ of Poincar\'{e} subalgebras, where the Lorentz part $F_{i}$ is extended by an ideal $N_{i,j}$ containing translation generators for some $j=1,2,...,n_{i}$. In addition, there exists a further exceptional set denoted $\widetilde{P}_{i,j}$, for certain $i,j$. Whereas the $P_{i,j}=F_{i}+N_{i,j}$ split over the translation generators, the $\widetilde{P}_{i,j}=\widetilde{F}_{i}+N_{i,j}$ do not. Although each $\widetilde{F}_{i}$ is isomorphic to its counterpart $F_{i}$ as a Lie algebra, it is \emph{non-conjugate} to $F_{i}$ within the Poincar\'{e} Lie algebra, as its generators are irrevocably ``tied up'' in linear combinations with the translation generators.

In this paper we work in the context of all admissible symmetry reductions of the Maxwell-Dirac system, but we illustrate the method with a small selection of test cases. The standard limits of spherical and cylindrical symmetry (subgroups $P_{3,4}$ and $P_{12,8}$ in the PWZ list) exemplify subgroups arising from three-dimensional geometry, biased towards a particular reference frame. A subgroup not explicitly covered in this paper, but an interesting extension of the spherical case is that of the hyperbolic symmetry subgroup $SO(2,1)$, represented in the PWZ list by $P_{4,4}$. This illustrates the case of a simple, but non-compact Lie algebra, and would make an interesting comparison case to more involved analyses of the $SO(3)$ Maxwell-Dirac symmetry reduction, due to its algebraic similarity. Finally, we take up two cases (with solvable Lie algebras) which specify one-parameter families of symmetries. The first, $P_{11,2}$, features an unusual ``screw'' generator, which is a parametric combination of a translation and a rotation about the corresponding axis. The second, $\widetilde{P}_{13,10}$ is a non-splitting subalgebra, with a parameter fixing the amount of translation generator entrained in the definition of a certain Lorentz generator in a minimal presentation.

In this section, we will calculate the scalar and vector field forms invariant under each subgroup, covering the actual Maxwell-Dirac reductions in the next section. As mentioned, the spherical and cylindrical cases reduce to a complicated system of nonlinear PDEs, there is no solution for $P_{11,2}$, and for $P_{13,10}$ the Maxwell-Dirac system reduces to a set of algebraic relations. Initially, we use the method described by Olver \cite{Olver-1986} to obtain a reduced set of independent variables, ``invariants'' henceforth, jointly invariant under the action of all the generators of the subgroup. It follows that arbitrary functions of the invariants will also be invariant under subgroup transformations, that is, they constitute solutions to the PDEs corresponding to a symmetric infinitesimal group action. Components of the invariant four-vector field must also be solutions to the PDEs corresponding to invariance under the group action. Such a set of differential equations is provided by the Lie derivative \cite{Wald-1984}, which defines the directional derivative of a tensor field of rank $(k,l)$ along the infinitesimal transformation vector field $\bxi\equiv\xi^{\sigma}\partial_{\sigma}$:
\begin{align}\label{Lie Derivative of a general tensor field}
&\mathcal{L}_{\bxi}T^{\mu_{1}\mu_{2}...\mu_{k}}{}_{\nu_{1}\nu_{2}...\nu_{l}}=\xi^{\sigma}\partial_{\sigma}T^{\mu_{1}\mu_{2}...\mu_{k}}{}_{\nu_{1}\nu_{2}...\nu_{l}}-(\partial_{\sigma}\xi^{\mu_{1}})T^{\sigma\mu_{2}...\mu_{k}}{}_{\nu_{1}\nu_{2}...\nu_{l}} \nonumber \\
&\qquad-(\partial_{\sigma}\xi^{\mu_{2}})T^{\mu_{1}\sigma...\mu_{k}}{}_{\nu_{1}\nu_{2}...\nu_{l}}-\ ...\ +(\partial_{\nu_{1}}\xi^{\sigma})T^{\mu_{1}\mu_{2}...\mu_{k}}{}_{\sigma\nu_{2}...\nu_{l}} \nonumber \\
&\qquad+(\partial_{\nu_{2}}\xi^{\sigma})T^{\mu_{1}\mu_{2}...\mu_{k}}{}_{\nu_{1}\sigma...\nu_{l}}+\ ... ,
\end{align}
with invariance under $\bxi$ imposed by setting
\begin{equation}\label{Zero Lie derivative}
\mathcal{L}_{\bxi}T^{\mu_{1}\mu_{2}...\mu_{k}}{}_{\nu_{1}\nu_{2}...\nu_{l}}=0.
\end{equation}
A scalar field $\phi$ is a rank $(0,0)$ tensor field, and the vector field with upper index $\Phi^{\mu}$ is of rank $(1,0)$. Note that ``rank'' $(k,l)$ in this context refers to the transformation properties as Lorentz tensor type objects, where $k$ is the number of contravariant (upstairs) indices, and $l$ is the number of covariant (downstairs) indices. Therefore, a tensor field such as $\Phi^{\mu}$ is of rank $(1,0)$, and transforms as a contravariant vector field under Lorentz transformations. The field strength tensor $F_{\mu\nu}$ is of rank $(0,2)$, transforming as a tensor with two covariant Lorentz indices. We do not need to consider tensor fields with $k+l>1$ however, since we have shown that $F_{\mu\nu}$ can be described in terms of scalar and four-vector fields of rank $(0,0)$ and $(0,1)$ respectively. Scalar and vector fields invariant under the transformation vector field $\bxi$ must solve the respective PDEs
\begin{align}
&\mathcal{L}_{\bxi}\phi=\xi^{\sigma}\partial_{\sigma}\phi=0 \label{Invariant scalar field PDE} \\
&\mathcal{L}_{\bxi}\Phi^{\mu}=\xi^{\sigma}\partial_{\sigma}\Phi^{\mu}-(\partial_{\sigma}\xi^{\mu})\Phi^{\sigma}=0. \label{Invariant vector field PDE}
\end{align}
Solutions to (\ref{Invariant scalar field PDE}) are calculated by using the method of characteristics to obtain the characteristic trajectories, which are the group invariants. Arbitrary scalar functions of these invariants solve (\ref{Invariant scalar field PDE}), as can be confirmed via substitution. Solutions to (\ref{Invariant vector field PDE}) are obtained along the same lines, but the situation is complicated by the second term, which mixes some of the vector components. We can obtain a characteristic system of ODEs involving the $\Phi^{\mu}$ components, from which we get algebraic expressions that allow us to make an accurate guess as to what forms components should take for invariance. The guessed solutions are confirmed by substituting into the equations generated by (\ref{Invariant vector field PDE}).

\subsection{The Poincar\'{e} generators}
One of the most important groups in special relativity is the Poincar\'{e} group $\mathcal{P}$, which consists of the Lorentz group of rotations and boosts, $SO(1,3)$, as well as the Abelian group of translations in four dimensions, $T(4)$. Since $\mathcal{P}$ is a Lie group, we can take infinitesimal translations and rotations, building finite transformations from infinitesimal ones through exponentiation. This allows us to work with the mathematically simpler Lie algebra of the Poincar\'{e} group, $L(\mathcal{P})$, which forms a vector space with the generators as the basis. Subalgebras of $L(\mathcal{P})$ are described in terms of their constituent generators, and are by definition closed under a Lie bracket operation. The six infinitesimal generators of the Lorentz group are defined by
\begin{equation}\label{Lorentz group generator equation}
(l_{\alpha\beta})^{\mu}{}_{\nu}=\delta_{\alpha}{}^{\mu}\eta_{\beta\nu}-\delta_{\beta}{}^{\mu}\eta_{\alpha\nu}\qquad(\alpha,\beta=0,1,2,3),
\end{equation}
with $l_{\alpha\beta}=-l_{\beta\alpha}$, and an arbitrary infinitesimal Lorentz transformation on the coordinate frame is
\begin{equation}
\Lambda^{\mu}{}_{\nu}x^{\nu}=[I+(1/2)\omega^{\alpha\beta}l_{\alpha\beta}]^{\mu}{}_{\nu}x^{\nu},
\end{equation}
where $\omega^{\alpha\beta}=-\omega^{\beta\alpha}$ are the six infinitesimal parameters associated with each generator. In explicit matrix form, the Lorentz generators are
\begin{align}
& l_{01}=\left(\!\!\!\begin{array}{cccc}0 & -1 & 0 & 0 \\
-1 & 0 & 0 & 0 \\
0 & 0 & 0 & 0 \\
0 & 0 & 0 & 0\end{array}\!\!\right),\ l_{02}=\left(\!\!\!\begin{array}{cccc}0 & 0 & -1 & 0 \\
0 & 0 & 0 & 0 \\
-1 & 0 & 0 & 0 \\
0 & 0 & 0 & 0\end{array}\!\!\right),\ l_{03}=\left(\!\!\!\begin{array}{cccc}0 & 0 & 0 & -1 \\
0 & 0 & 0 & 0 \\
0 & 0 & 0 & 0 \\
-1 & 0 & 0 & 0\end{array}\!\!\!\right), \nonumber \\
& l_{12}=\left(\begin{array}{cccc}0 & 0 & 0 & 0 \\
0 & 0 & -1 & 0 \\
0 & 1 & 0 & 0 \\
0 & 0 & 0 & 0\end{array}\right),\ l_{13}=\left(\begin{array}{cccc}0 & 0 & 0 & 0 \\
0 & 0 & 0 & -1 \\
0 & 0 & 0 & 0 \\
0 & 1 & 0 & 0\end{array}\!\!\right),\ l_{23}=\left(\begin{array}{cccc}0 & 0 & 0 & 0 \\
0 & 0 & 0 & 0 \\
0 & 0 & 0 & -1 \\
0 & 0 & 1 & 0\end{array}\!\!\right).
\end{align}
The first three generators are boosts along the $x$, $y$ and $z$-axes respectively and the bottom three generators correspond to rotations in the $x-y$, $x-z$ and $y-z$ planes. Is is easy to show using (\ref{Lorentz group generator equation}) that the Lie bracket of the Lorentz algebra is
\begin{equation}
[l_{\alpha\beta},l_{\gamma\delta}]=\eta_{\alpha\delta}l_{\beta\gamma}+\eta_{\beta\gamma}l_{\alpha\delta}-\eta_{\alpha\gamma}l_{\beta\delta}-\eta_{\beta\delta}l_{\alpha\gamma}.
\end{equation}
A commonly used representation of these generators is
\begin{equation}
 K_{1}=-l_{01},\ \ K_{2}=-l_{02},\ \ K_{3}=-l_{03},\ \ L_{1}=l_{23},\ \ L_{2}=-l_{13},\ \ L_{3}=l_{12}.
\end{equation}
We define the components of the transformation vector field $\bxi_{X}$ corresponding to generator $X$, to be $\xi_{X}{}^{\mu}\equiv X^{\mu}{}_{\nu}x^{\nu}$. In the $K-L$ representation, these vector fields are
\begin{align}
&\xi_{K_{1}}{}^{\mu}=\left(\begin{array}{c}x \\
t \\
0 \\
0\end{array}\right),\ \xi_{K_{2}}{}^{\mu}=\left(\begin{array}{c}y \\
0 \\
t \\
0\end{array}\right),\ \xi_{K_{3}}{}^{\mu}=\left(\begin{array}{c}z \\
0 \\
0 \\
t\end{array}\right), \nonumber \\
&\xi_{L_{1}}{}^{\mu}=\left(\!\!\begin{array}{c}0 \\
0 \\
-z \\
y\end{array}\!\!\right),\ \xi_{L_{2}}{}^{\mu}=\left(\!\!\begin{array}{c}0 \\
z \\
0 \\
-x\end{array}\!\!\right),\ \xi_{L_{3}}{}^{\mu}=\left(\!\!\begin{array}{c}0 \\
-y \\
x \\
0\end{array}\!\!\right)
\end{align}
The vector fields $\xi_{P_{\nu}}{}^{\mu}$ corresponding to the infinitesimal translation operators $P_{\nu}$ are simply four vectors with components $\delta_{\nu}^{\mu}$ that act on the coordinate space additively
\begin{equation}
P_{\nu}\cdot x:\ x^{\mu}\rightarrow x^{\mu}+\varepsilon\xi_{P_{\nu}}{}^{\mu}=x^{\mu}+\varepsilon\delta_{\nu}^{\mu}.
\end{equation}
The generators of all of the Poincar\'{e} subalgebras are listed by PWZ \cite{Patera-Winternitz-Zassenhaus-1975} using an alternative representation to that defined above, but which allows convenient extension to larger subgroups of the conformal group of spacetime transformations, such as the similitude (Weyl) group \cite{Patera-Winternitz-Zassenhaus-1975}, \cite{Patera-Winternitz-Zassenhaus-1975(II)}. It is important to note that if any extensions of the Maxwell-Dirac symmetry reduction to the conformal group are undertaken, the physical system must be restricted to \emph{massless} particles only. The PWZ Lorentz generators are
\begin{align}
& B_{1}=2L_{3}=2l_{12},\ \ B_{2}=-2K_{3}=2l_{03},\ \ B_{3}=-L_{2}-K_{1}=l_{13}+l_{01}, \nonumber \\
& B_{4}=L_{1}-K_{2}=l_{23}+l_{02}, \ \ B_{5}=L_{2}-K_{1}=-l_{13}+l_{01}, \nonumber \\
&B_{6}=L_{1}+K_{2}=l_{23}-l_{02},
\end{align}
with explicit matrix form
\begin{align}
&B_{1}=\left(\!\!\begin{array}{cccc}0 & 0 & 0 & 0 \\
0 & 0 & -2 & 0 \\
0 & 2 & 0 & 0 \\
0 & 0 & 0 & 0\end{array}\!\!\right),\ B_{2}=\left(\!\!\!\begin{array}{cccc}0 & 0 & 0 & -2 \\
0 & 0 & 0 & 0 \\
0 & 0 & 0 & 0 \\
-2 & 0 & 0 & 0\end{array}\!\!\right),\ B_{3}=\left(\!\!\!\begin{array}{cccc}0 & -1 & 0 & 0 \\
-1 & 0 & 0 & -1 \\
0 & 0 & 0 & 0 \\
0 & 1 & 0 & 0\end{array}\!\!\right), \nonumber \\
&B_{4}=\left(\!\!\!\begin{array}{cccc}0 & 0 & -1 & 0 \\
0 & 0 & 0 & 0 \\
-1 & 0 & 0 & -1 \\
0 & 0 & 1 & 0\end{array}\!\!\!\right),\ B_{5}=\left(\!\!\begin{array}{cccc}0 & -1 & 0 & 0 \\
-1 & 0 & 0 & 1 \\
0 & 0 & 0 & 0 \\
0 & -1 & 0 & 0\end{array}\!\!\right),\ B_{6}=\left(\!\!\begin{array}{cccc}0 & 0 & 1 & 0 \\
0 & 0 & 0 & 0 \\
1 & 0 & 0 & -1 \\
0 & 0 & 1 & 0\end{array}\!\!\right).
\end{align}
In this representation, it is appropriate to replace $t$ and $z$ with the light cone coordinates $l_{+}\equiv t+z$ and $l_{-}\equiv t-z$. The transformation vector fields for each generator are
\begin{align}
&\xi_{B_{1}}{}^{\mu}=\left(\!\!\begin{array}{c}0 \\
-2y \\
2x \\
0\end{array}\!\!\right),\ \xi_{B_{2}}{}^{\mu}=\left(\!\!\begin{array}{c}-l_{+}+l_{-} \\
0 \\
0 \\
-l_{+}-l_{-}\end{array}\!\!\right),\ \xi_{B_{3}}{}^{\mu}=\left(\!\!\begin{array}{c}-x \\
-l_{+} \\
0 \\
x\end{array}\!\!\right), \nonumber \\
&\xi_{B_{4}}{}^{\mu}=\left(\begin{array}{c}-y \\
0 \\
-l_{+} \\
y\end{array}\right),\ \xi_{B_{5}}{}^{\mu}=\left(\!\!\begin{array}{c}-x \\
-l_{-} \\
0 \\
-x\end{array}\!\!\right),\ \xi_{B_{6}}{}^{\mu}=\left(\begin{array}{c}y \\
0 \\
l_{-} \\
y\end{array}\right).
\end{align}
The translation generators in the PWZ representation are
\begin{equation}
 X_{1}=(1/2)(P_{0}-P_{3}),\ \ X_{2}=P_{2},\ \ X_{3}=-P_{1},\ \ X_{4}=(1/2)(P_{0}+P_{3}),
\end{equation}
where $X_{1}$ and $X_{4}$ correspond to translations along the $l_{-}$ and $l_{+}$ axes respectively. The corresponding (constant) vector fields are
\begin{equation}\label{PWZ translation vector fields}
 \xi_{X_{1}}{}^{\mu}=\left(\!\!\begin{array}{c}1/2 \\
0 \\
0 \\
-1/2\end{array}\!\!\right),\ \xi_{X_{2}}{}^{\mu}=\left(\begin{array}{c}0 \\
0 \\
1 \\
0\end{array}\right),\ \xi_{X_{3}}{}^{\mu}=\left(\!\!\begin{array}{c}0 \\
-1 \\
0 \\
0\end{array}\!\!\right),\ \xi_{X_{4}}{}^{\mu}=\left(\!\!\begin{array}{c}1/2 \\
0 \\
0 \\
1/2\end{array}\!\!\right).
\end{equation}
There is another $B$-generator, a composite of $B_{1}$ and $B_{2}$, which is present in the PWZ $F_{5}$ and $F_{11}$ subalgebras of the Lorentz group, as well as the $P_{5,i}$ and $P_{11,i}$ subalgebras of the Poincar\'{e} group
\begin{equation}
B_{\varphi}=\cos\varphi B_{1}+\sin\varphi B_{2},\ \ \ 0<\varphi<\pi,\ \ \varphi\neq\pi/2.
\end{equation}
This generator corresponds to a simultaneous rotation around, and boost along the $z$-axis, the so-called ``screw'' group, $S(1)$. The continuous parameter varies the generator from being almost a pure rotation $(\varphi\approx0)$, to an almost pure boost $(\varphi\approx\pi/2)$. The extreme cases when $\varphi\rightarrow0^{+}$ and $\varphi\rightarrow\pi/2^{-}$, meaning $B_{\varphi}\rightarrow B_{1}$ and $B_{\varphi}\rightarrow B_{2}$, are actually Lorentz subalgebras in their own right, and are given the PWZ labels $F_{12}$ and $F_{13}$ respectively. The explicit matrix form and vector field for $B_{\varphi}$ are
\begin{equation}\label{B_phi matrix and vector field}
 B_{\varphi}=\left(\!\!\begin{array}{cccc}0 & 0 & 0 & -2\sin\varphi \\
0 & 0 & -2\cos\varphi & 0 \\
0 & 2\cos\varphi & 0 & 0 \\
-2\sin\varphi & 0 & 0 & 0\end{array}\!\!\right),\ \ \ \xi_{B_{\varphi}}{}^{\mu}=\left(\!\!\begin{array}{c}-2z\sin\varphi \\
-2y\cos\varphi \\
2x\cos\varphi \\
-2t\sin\varphi\end{array}\!\!\right).\
\end{equation}

\subsection{Spherical symmetry (subgroup $P_{3,4}$)}
The condition for scalar and vector fields to be spherically symmetric is that they be invariant under the action of the $SO(3)$ group, which consists of the three rotation generators $L_{1}$, $L_{2}$ and $L_{3}$. The Lie derivative of a scalar field invariant under $L_{1}$ is
\begin{equation}
\mathcal{L}_{L_{1}}\phi=\xi_{L_{1}}{}^{\sigma}\partial_{\sigma}\phi=-z\partial_{y}\phi+y\partial_{z}\phi=0,
\end{equation}
which yields the characteristic system
\begin{equation}
\frac{\rmd y}{-z}=\frac{\rmd z}{y},
\end{equation}
giving us the $L_{1}$ invariant $\rho=\sqrt{y^{2}+z^{2}}$. The requirement for invariance under $L_{2}$ is
\begin{equation}\label{L_{2} invariant scalar field PDE}
\mathcal{L}_{L_{2}}\phi=\xi_{L_{2}}{}^{\sigma}\partial_{\sigma}\phi=z\partial_{x}\phi-x\partial_{z}\phi=0.
\end{equation}
To impose that $\phi$ to be jointly invariant under $L_{1}$ and $L_{2}$, we require that the solution to (\ref{L_{2} invariant scalar field PDE}) be a function of $t$, $x$ and $\rho$. Using the chain rule, we find that the PDE becomes independent from $z$ explicitly
\begin{equation}
\rho\partial_{x}\phi-x\partial_{\rho}\phi=0.
\end{equation}
The characteristic equation is
\begin{equation}
\frac{\rmd x}{\rho}=\frac{\rmd\rho}{-x},
\end{equation}
from which we obtain the joint invariant $r=\sqrt{x^{2}+y^{2}+z^{2}}$. Lastly, we require that
\begin{equation}
\mathcal{L}_{L_{3}}\phi=\xi_{L_{3}}{}^{\sigma}\partial_{\sigma}\phi=-y\partial_{x}\phi+x\partial_{y}\phi=0,
\end{equation}
which is automatically satisfied by $\phi(t,r)$. Now consider the four vector field $\Phi^{\mu}$ invariant under $L_{1}$. From (\ref{Invariant vector field PDE}) for $\mu=0-3$, we obtain the following PDEs
\begin{subequations}
\begin{align}
&\mathcal{L}_{L_{1}}\Phi^{0}=-z\partial_{y}\Phi^{0}+y\partial_{z}\Phi^{0}=0, \\
&\mathcal{L}_{L_{1}}\Phi^{1}=-z\partial_{y}\Phi^{1}+y\partial_{z}\Phi^{1}=0, \\
&\mathcal{L}_{L_{1}}\Phi^{2}=-z\partial_{y}\Phi^{2}+y\partial_{z}\Phi^{2}+\Phi^{3}=0, \\
&\mathcal{L}_{L_{1}}\Phi^{3}=-z\partial_{y}\Phi^{3}+y\partial_{z}\Phi^{3}-\Phi^{2}=0.
\end{align}
\end{subequations}
Likewise, for $L_{2}$ we obtain
\begin{subequations}
\begin{align}
&\mathcal{L}_{L_{2}}\Phi^{0}=z\partial_{x}\Phi^{0}-x\partial_{z}\Phi^{0}=0, \\
&\mathcal{L}_{L_{2}}\Phi^{1}=z\partial_{x}\Phi^{1}-x\partial_{z}\Phi^{1}-\Phi^{3}=0, \\
&\mathcal{L}_{L_{2}}\Phi^{2}=z\partial_{x}\Phi^{2}-x\partial_{z}\Phi^{2}=0, \\
&\mathcal{L}_{L_{2}}\Phi^{3}=z\partial_{x}\Phi^{3}-x\partial_{z}\Phi^{3}+\Phi^{1}=0,
\end{align}
\end{subequations}
and for $L_{3}$ we get
\begin{subequations}
\begin{align}
&\mathcal{L}_{L_{3}}\Phi^{0}=-y\partial_{x}\Phi^{0}+x\partial_{y}\Phi^{0}=0,\label{Vector field L3 symmetry, mu=0} \\
&\mathcal{L}_{L_{3}}\Phi^{1}=-y\partial_{x}\Phi^{1}+x\partial_{y}\Phi^{1}+\Phi^{2}=0,\label{Vector field L3 symmetry, mu=1} \\
&\mathcal{L}_{L_{3}}\Phi^{2}=-y\partial_{x}\Phi^{2}+x\partial_{y}\Phi^{2}-\Phi^{1}=0,\label{Vector field L3 symmetry, mu=2} \\
&\mathcal{L}_{L_{3}}\Phi^{3}=-y\partial_{x}\Phi^{3}+x\partial_{y}\Phi^{3}=0.\label{Vector field L3 symmetry, mu=3}
\end{align}
\end{subequations}
Noticing that $\Phi^{0}$ obeys the same set of PDEs as $\phi$, we can immediately conclude that $\Phi^{0}=a(t,r)$. By taking the combination $x\mathcal{L}_{L_{1}}\Phi^{i}+y\mathcal{L}_{L_{2}}\Phi^{i}+z\mathcal{L}_{L_{3}}\Phi^{i}$ for $i=1,2,3$, we simplify the other PDEs to the algebraic set
\begin{subequations}
\begin{align}
&z\Phi^{2}-y\Phi^{3}=0, \\
&x\Phi^{3}-z\Phi^{1}=0, \\
&y\Phi^{1}-x\Phi^{2}=0,
\end{align}
\end{subequations}
giving us the solution $\Phi^{i}=x^{i}b(t,r)$. So $SO(3)$ invariant four vector fields must have the general form
\begin{equation}\label{Spherically symmetric four vector field}
\Phi^{\mu}=\left(\begin{array}{c}a(t,r) \\
xb(t,r) \\
yb(t,r) \\
zb(t,r)\end{array}\right).
\end{equation}

\subsection{Cylindrical symmetry (subgroup $P_{12,8}$)}
For tensor fields to be cylindrically symmetric, they must be invariant under rotation around, and translation along, a single axis. Choosing the rotation plane to be $x-y$, the axis must be $z$, so the infinitesimal invariance generators are $L_{3}$ and $P_{3}$. The scalar field must satisfy the relatively trivial PDEs
\begin{align}
&\mathcal{L}_{P_{3}}\phi=\xi_{P_{3}}{}^{\sigma}\partial_{\sigma}\phi=\partial_{z}\phi=0, \\
&\mathcal{L}_{L_{3}}\phi=\xi_{L_{3}}{}^{\sigma}\partial_{\sigma}\phi=-y\partial_{x}\phi+x\partial_{y}\phi=0.
\end{align}
The first equation tells us that $\phi$ is independent of $z$, and from the second we obtain the invariant $\rho=\sqrt{x^{2}+y^{2}}$. Cylindrical symmetry requires that scalar fields be functions of $t$ and $\rho$ only. The vector fields must satisfy
\begin{equation}
\mathcal{L}_{P_{3}}\Phi^{\mu}=\xi_{P_{3}}{}^{\sigma}\partial_{\sigma}\Phi^{\mu}=0,
\end{equation}
as well as equations (\ref{Vector field L3 symmetry, mu=0}) to (\ref{Vector field L3 symmetry, mu=3}). We can immediately conclude that $\Phi^{0}=a(t,\rho)$ and $\Phi^{3}=d(t,\rho)$, since they solve the same equations as $\phi$. We can also say that $\Phi^{1}$ and $\Phi^{2}$ are independent of $z$, but due to components mixing, they are not pure functions of $t$ and $\rho$. From (\ref{Vector field L3 symmetry, mu=1}) and (\ref{Vector field L3 symmetry, mu=2}), we obtain the respective characteristic systems
\begin{subequations}
\begin{align}
&\frac{\rmd x}{-y}=\frac{\rmd y}{x}=\frac{\rmd\Phi^{1}}{-\Phi^{2}}, \\
&\frac{\rmd x}{-y}=\frac{\rmd y}{x}=\frac{\rmd\Phi^{2}}{\Phi^{1}},
\end{align}
\end{subequations}
obviously implying that
\begin{equation}
\frac{\rmd\Phi^{1}}{-\Phi^{2}}=\frac{\rmd\Phi^{2}}{\Phi^{1}}.
\end{equation}
Integrating and taking into account the fact that arbitrary functions of $t$ and $\rho$ are constant along characteristic curves, we obtain the algebraic constraint
\begin{equation}\label{Cylindrical symmetry vector field component algebraic constraint}
(\Phi^{1})^{2}+(\Phi^{2})^{2}=f(t,\rho),
\end{equation}
which accepts solutions of the form $\Phi^{1}=xb(t,\rho)-yc(t,\rho)$ and $\Phi^{2}=yb(t,\rho)+xc(t,\rho)$. This gives us the form of the cylindrically symmetric four vector field
\begin{equation}\label{Cylindrically symmetric four vector field}
\Phi^{\mu}=\left(\begin{array}{c}a(t,\rho) \\
xb(t,\rho)-yc(t,\rho) \\
yb(t,\rho)+xc(t,\rho) \\
d(t,\rho)\end{array}\right).
\end{equation}

\subsection{$P_{11,2}$ symmetry (``screw'' subgroup)}
The Poincar\'{e} subalgebra $P_{11,2}$ as defined by PWZ consists of the single Lorentz generator $B_{\varphi}$, and the three translation generators $X_{1}$, $X_{2}$ and $X_{3}$. In this subsection, we will find the symmetric form of the fields for the $B_{\varphi}$ generator first, then proceed through the translation generators in numerical order. The process is a cumulative one, in that once we have derived the form of the $B_{\varphi}$ invariant fields, we apply the $X_{1}$ invariance condition to them in this form, resulting in a more restricted form, and so on.

\subsubsection{$B_{\varphi}$ invariant fields}
From (\ref{B_phi matrix and vector field}), we can see that the $B_{\varphi}$ invariance condition for a scalar field is
\begin{equation}
\mathcal{L}_{B_{\varphi}}\phi=\xi_{B_{\varphi}}{}^{\sigma}\partial_{\sigma}\phi=-2z\sin\varphi\;\partial_{t}\phi-2y\cos\varphi\;\partial_{x}\phi+2x\cos\varphi\;\partial_{y}\phi-2t\sin\varphi\;\partial_{z}\phi=0.
\end{equation}
Since we are using the light cone coordinates, we must use the chain rule to rewrite the derivatives
\begin{align}
&\partial_{t}\phi=\partial_{+}\phi+\partial_{-}\phi,\label{Chain rule to convert dt to d+ and d-} \\
&\partial_{z}\phi=\partial_{+}\phi-\partial_{-}\phi,\label{Chain rule to convert dz to d+ and d-}
\end{align}
resulting in the PDE
\begin{equation}\label{PDE for B-varphi invariant scalar field}
-l_{+}\sin\varphi\;\partial_{+}\phi+l_{-}\sin\varphi\;\partial_{-}\phi-y\cos\varphi\;\partial_{x}\phi+x\cos\varphi\;\partial_{y}\phi=0,
\end{equation}
where for simplicity, we have defined $\partial_{+}\equiv\partial/\partial l_{+}$ and $\partial_{-}\equiv\partial/\partial l_{-}$. From the method of characteristics, we get the system of six ODEs
\begin{equation}\label{B-varphi scalar field characteristic system}
\frac{\rmd l_{+}}{-l_{+}\sin\varphi}=\frac{\rmd l_{-}}{l_{-}\sin\varphi}=\frac{\rmd x}{-y\cos\varphi}=\frac{\rmd y}{x\cos\varphi},
\end{equation}
from which we obtain the six invariants
\begin{subequations}
\begin{align}
&|L|=|l_{+}l_{-}|=|t^{2}-z^{2}|, \\
&\rho=\sqrt{x^{2}+y^{2}}, \\
&\alpha=\cos\varphi\;\ln|l_{+}|-\sin\varphi\;\arcsin(x/\rho), \\
&\beta=\cos\varphi\;\ln|l_{+}|+\sin\varphi\;\arcsin(y/\rho), \\
&\gamma=\cos\varphi\;\ln|l_{-}|+\sin\varphi\;\arcsin(x/\rho), \\
&\delta=\cos\varphi\;\ln|l_{-}|-\sin\varphi\;\arcsin(y/\rho),
\end{align}
\end{subequations}
where we obtain the last four invariants via integration by recognizing that $\rho$ is a constant in the characteristic system. Not all of these invariants are independent. For example, adding $\alpha$ and $\gamma$, then rearranging gives
\begin{equation}
|L|=\exp\left(\frac{\alpha+\gamma}{\cos\varphi}\right).
\end{equation}
In addition to $|L|$ and $\rho$, we can construct a neat form for a third invariant from the list $\alpha$, $\beta$, $\gamma$, $\delta$. Consider the combination
\begin{equation}
\alpha-\delta=\beta-\gamma=\cos\varphi\;(\ln|l_{+}|-\ln|l_{-}|)+\sin\varphi\;[\arcsin(y/\rho)-\arcsin(x/\rho)].
\end{equation}
After some manipulation, we find that
\begin{equation}
\exp[(\alpha-\delta)/\cos\varphi]=(l_{+}/l_{-})\exp\{\tau\arctan[(y^{2}-x^{2})/2xy]\},
\end{equation}
where $\tau\equiv\tan\varphi$. Using the logarithmic form of arctan, and introducing polar coordinates in the $x-y$ plane
\begin{equation}
y+\rmi x=\rho\rme^{\rmi\chi},
\end{equation}
where $\chi\equiv\arctan(x/y)$, we find that
\begin{equation}
\exp[(\alpha-\delta)/\cos\varphi]=-(l_{+}/l_{-})\rme^{-2\tau\chi}.
\end{equation}
Taking the negative reciprocal of this, we arrive at the form for the new invariant
\begin{equation}
\zeta_{\varphi}=(l_{-}/l_{+})\rme^{2\tau\chi},
\end{equation}
with the $\varphi$ subscript indicating that the invariant is dependent on the value of the free group parameter. In summary, by imposing $B_{\varphi}$ invariance, we have gone from the independent variable set $(l_{+},l_{-},x,y)$ to the reduced set $(|L|,\rho,\zeta_{\varphi})$. Arbitrary scalar functions of the latter set are solutions to (\ref{PDE for B-varphi invariant scalar field}), which can be checked via substituting the partial derivatives of $\phi(|L|,\rho,\zeta_{\varphi})$
\begin{subequations}
\begin{align}
&\partial_{+}\phi=l_{-}(L/|L|)\partial_{|L|}\phi-(\zeta_{\varphi}/l_{+})\partial_{\zeta}\phi,\label{Scalar function of B-varphi invariants chain rule for partial-l_+} \\
&\partial_{-}\phi=l_{+}(L/|L|)\partial_{|L|}\phi+(\zeta_{\varphi}/l_{-})\partial_{\zeta}\phi,\label{Scalar function of B-varphi invariants chain rule for partial-l_-} \\
&\partial_{x}\phi=(x/\rho)\partial_{\rho}\phi+(2\tau\zeta_{\varphi}y/\rho^{2})\partial_{\zeta}\phi,\label{Scalar function of B-varphi invariants chain rule for partial-x} \\
&\partial_{y}\phi=(y/\rho)\partial_{\rho}\phi-(2\tau\zeta_{\varphi}x/\rho^{2})\partial_{\zeta}\phi.\label{Scalar function of B-varphi invariants chain rule for partial-y}
\end{align}
\end{subequations}
Now the vector field components must satisfy
\begin{equation}
\mathcal{L}_{B_{\varphi}}\Phi^{\mu}=\xi_{B_{\varphi}}{}^{\sigma}\partial_{\sigma}\Phi^{\mu}-(\partial_{\sigma}\xi_{B_{\varphi}}{}^{\mu})\Phi^{\sigma}=0,
\end{equation}
which for $\mu=0-3$ gives us
\begin{subequations}
\begin{align}
&-l_{+}\sin\varphi\;\partial_{+}\Phi^{0}+l_{-}\sin\varphi\;\partial_{-}\Phi^{0}-y\cos\varphi\;\partial_{x}\Phi^{0}+x\cos\varphi\;\partial_{y}\Phi^{0}+\sin\varphi\;\Phi^{3}=0,\label{Vector field B-varphi symmetry, mu=0} \\
&-l_{+}\sin\varphi\;\partial_{+}\Phi^{1}+l_{-}\sin\varphi\;\partial_{-}\Phi^{1}-y\cos\varphi\;\partial_{x}\Phi^{1}+x\cos\varphi\;\partial_{y}\Phi^{1}+\cos\varphi\;\Phi^{2}=0, \\
&-l_{+}\sin\varphi\;\partial_{+}\Phi^{2}+l_{-}\sin\varphi\;\partial_{-}\Phi^{2}-y\cos\varphi\;\partial_{x}\Phi^{2}+x\cos\varphi\;\partial_{y}\Phi^{2}-\cos\varphi\;\Phi^{1}=0, \\
&-l_{+}\sin\varphi\;\partial_{+}\Phi^{3}+l_{-}\sin\varphi\;\partial_{-}\Phi^{3}-y\cos\varphi\;\partial_{x}\Phi^{3}+x\cos\varphi\;\partial_{y}\Phi^{3}+\sin\varphi\;\Phi^{0}=0.\label{Vector field B-varphi symmetry, mu=3}
\end{align}
\end{subequations}
The method of characteristics gives us a system of ODEs like (\ref{B-varphi scalar field characteristic system}) for each equation, but with an extra $\rmd\Phi^{\mu}$ part. Equating these parts gives us an ODE system involving only the vector field components
\begin{equation}
\frac{\rmd\Phi^{0}}{-\sin\varphi\;\Phi^{3}}=\frac{\rmd\Phi^{1}}{-\cos\varphi\;\Phi^{2}}=\frac{\rmd\Phi^{2}}{\cos\varphi\;\Phi^{1}}=\frac{\rmd\Phi^{3}}{-\sin\varphi\;\Phi^{0}}.
\end{equation}
Integrating this ODE system, we obtain a list of algebraic constraints on the vector field components
\begin{subequations}
\begin{align}
&A^{2}(|L|,\rho,\zeta_{\varphi})=(\Phi^{0})^{2}-(\Phi^{3})^{2},\label{B-varphi quadratic algebraic constraint for mu=0,3} \\
&B^{2}(|L|,\rho,\zeta_{\varphi})=(\Phi^{1})^{2}+(\Phi^{2})^{2},\label{B-varphi quadratic algebraic constraint for mu=1,2} \\
&f(|L|,\rho,\zeta_{\varphi})=\tau\arcsin(\Phi^{1}/B)-\arccosh(\Phi^{0}/A), \\
&g(|L|,\rho,\zeta_{\varphi})=\tau\arcsin(\Phi^{2}/B)+\arccosh(\Phi^{0}/A), \\
&h(|L|,\rho,\zeta_{\varphi})=\tau\arcsin(\Phi^{1}/B)-\arcsinh(\Phi^{3}/A), \\
&k(|L|,\rho,\zeta_{\varphi})=\tau\arcsin(\Phi^{2}/B)+\arcsinh(\Phi^{3}/A),
\end{align}
\end{subequations}
where we have used the fact that arbitrary functions of the invariants are constants in the characteristic ODE system. By introducing angular coordinates in the $\Phi^{0}-\Phi^{3}$ and $\Phi^{1}-\Phi^{2}$ planes, we find that $f=h$ and $g=k$, so we effectively have only four independent constraints. After some investigation, we find that the $f$ and $g$ angular constraints provide a superfluous level of detail that can be eliminated by a simple relabeling of functions. Focusing on the quadratic constraints (\ref{B-varphi quadratic algebraic constraint for mu=0,3}) and (\ref{B-varphi quadratic algebraic constraint for mu=1,2}), we find that an appropriate generic form for the four vector field is
\begin{equation}\label{B-varphi invariant four vector field form}
\Phi^{\mu}=\left(\begin{array}{c}ta(|L|,\rho,\zeta_{\varphi})+zb(|L|,\rho,\zeta_{\varphi}) \\
xc(|L|,\rho,\zeta_{\varphi})-yd(|L|,\rho,\zeta_{\varphi}) \\
yc(|L|,\rho,\zeta_{\varphi})+xd(|L|,\rho,\zeta_{\varphi}) \\
za(|L|,\rho,\zeta_{\varphi})+tb(|L|,\rho,\zeta_{\varphi})\end{array}\right).
\end{equation}
Verification can be obtained by substituting into (\ref{Vector field B-varphi symmetry, mu=0})-(\ref{Vector field B-varphi symmetry, mu=3}), (\ref{B-varphi quadratic algebraic constraint for mu=0,3}) and (\ref{B-varphi quadratic algebraic constraint for mu=1,2}). Note that since the Poincar\'{e} subgroup $P_{11,6}$ consists of the single Lorentz generator $B_{\varphi}$, this is the invariant four vector field form for that subgroup. $P_{11,2}$ consists of $B_{\varphi}$, as well as the three translation generators $X_{1}$, $X_{2}$ and $X_{3}$, the corresponding vector fields being listed in (\ref{PWZ translation vector fields}).

\subsubsection{$B_{\varphi}$, $X_{1}$ invariant fields}
The condition that scalar fields are simultaneously invariant under $B_{\varphi}$ and $X_{1}$ is
\begin{equation}
\xi_{X_{1}}{}^{\sigma}\partial_{\sigma}\phi(|L|,\rho,\zeta_{\varphi})=\partial_{-}\phi(|L|,\rho,\zeta_{\varphi})=0,\label{B-varphi, X1 invariant scalar field Lie derivative}
\end{equation}
which upon applying the chain rule and multiplying through by $l_{-}$, gives the PDE
\begin{equation}\label{B-varphi, X1 scalar field PDE}
|L|\partial_{|L|}\phi+\zeta_{\varphi}\partial_{\zeta}\phi=0.
\end{equation}
The corresponding characteristic ODE for this equation is
\begin{equation}
\frac{\rmd|L|}{|L|}=\frac{\rmd\zeta_{\varphi}}{\zeta_{\varphi}},
\end{equation}
which can be integrated to yield the $\{B_{\varphi},X_{1}\}$ joint invariant
\begin{equation}
\tzeta_{\varphi}\equiv L/\zeta_{\varphi}=l_{+}^{2}\rme^{-2\tau\chi}.
\end{equation}
So in order to be jointly invariant along the characteristics generated by both $B_{\varphi}$ and $X_{1}$, the scalar field $\phi$ must be a function of $(\rho,\tzeta_{\varphi})$ only. Since neither $\rho$, nor $\tzeta_{\varphi}$ are dependent on $l_{-}$, it is obvious that $\phi(\rho,\tzeta_{\varphi})$ solves (\ref{B-varphi, X1 invariant scalar field Lie derivative}).

For four vector fields to be simultaneously invariant under $B_{\varphi}$ and $X_{1}$, we require that
\begin{equation}
\mathcal{L}_{X_{1}}\Phi^{\mu}=\xi_{X_{1}}{}^{\sigma}\partial_{\sigma}\Phi^{\mu}-(\partial_{\sigma}\xi_{X_{1}}{}^{\mu})\Phi^{\sigma}=\partial_{-}\Phi^{\mu}=0,
\end{equation}
where $\Phi^{\mu}$ are the components of (\ref{B-varphi invariant four vector field form}). The second term is zero because $\xi_{X_{1}}{}^{\mu}$ is a constant vector field. Carrying out the derivatives yields the four PDEs
\begin{subequations}
\begin{align}
&l_{-}(a-b)/2+t(|L|\partial_{|L|}a+\zeta_{\varphi}\partial_{\zeta}a)+z(|L|\partial_{|L|}b+\zeta_{\varphi}\partial_{\zeta}b)=0,\label{B-varphi, X1 four vector PDE mu=0} \\
&x(|L|\partial_{|L|}c+\zeta_{\varphi}\partial_{\zeta}c)-y(|L|\partial_{|L|}d+\zeta_{\varphi}\partial_{\zeta}d)=0,\label{B-varphi, X1 four vector PDE mu=1} \\
&y(|L|\partial_{|L|}c+\zeta_{\varphi}\partial_{\zeta}c)+x(|L|\partial_{|L|}d+\zeta_{\varphi}\partial_{\zeta}d)=0,\label{B-varphi, X1 four vector PDE mu=2} \\
&-l_{-}(a-b)/2+z(|L|\partial_{|L|}a+\zeta_{\varphi}\partial_{\zeta}a)+t(|L|\partial_{|L|}b+\zeta_{\varphi}\partial_{\zeta}b)=0,\label{B-varphi, X1 four vector PDE mu=3}
\end{align}
\end{subequations}
where we have taken the additional step of multiplying through by $l_{-}$. Adding (\ref{B-varphi, X1 four vector PDE mu=0}) to (\ref{B-varphi, X1 four vector PDE mu=3}) and subtracting (\ref{B-varphi, X1 four vector PDE mu=3}) from (\ref{B-varphi, X1 four vector PDE mu=0}) gives the two respective equations
\begin{subequations}
\begin{align}
&|L|\partial_{|L|}a+\zeta_{\varphi}\partial_{\zeta}a=-(|L|\partial_{|L|}b+\zeta_{\varphi}\partial_{\zeta}b), \\
&b-a=|L|\partial_{|L|}a+\zeta_{\varphi}\partial_{\zeta}a-(|L|\partial_{|L|}b+\zeta_{\varphi}\partial_{\zeta}b).
\end{align}
\end{subequations}
The first equation can be substituted into the second to eliminate either the $a$ or $b$ derivatives from the right hand side, giving us non-homogeneous PDEs for $a$ and $b$:
\begin{subequations}
\begin{align}
&(b-a)/2=|L|\partial_{|L|}a+\zeta_{\varphi}\partial_{\zeta}a, \\
&(a-b)/2=|L|\partial_{|L|}b+\zeta_{\varphi}\partial_{\zeta}b,
\end{align}
\end{subequations}
to which we can apply the method of characteristics, giving us the respective ODE systems
\begin{subequations}
\begin{align}
&\frac{\rmd|L|}{|L|}=\frac{\rmd\zeta_{\varphi}}{\zeta_{\varphi}}=\frac{2\rmd a}{b-a}, \\
&\frac{\rmd|L|}{|L|}=\frac{\rmd\zeta_{\varphi}}{\zeta_{\varphi}}=\frac{2\rmd b}{a-b}.
\end{align}
\end{subequations}
The first two parts of each ODE system again tell us that characteristics lie along curves of constant $\tzeta_{\varphi}=l_{+}^{2}\rme^{-2\tau\chi}$. We can equate the right hand sides of both ODE systems and integrate to yield the algebraic constraint on the forms of $a$ and $b$
\begin{equation}\label{B-varphi, X1 a,b algebraic constraint}
a(|L|,\rho,\zeta_{\varphi})+b(|L|,\rho,\zeta_{\varphi})=f(\rho,\tzeta_{\varphi}),
\end{equation}
where $f$ is a constant in this ODE system. This constraint is satisfied if we simply impose that $a$ and $b$ now be functions of $\rho$ and $\tzeta_{\varphi}$. Substituting these into the $\mu=0$ Lie derivative gives
\begin{equation}
\partial_{-}\Phi^{0}=\partial_{-}[ta(\rho,\tzeta_{\varphi})+zb(\rho,\tzeta_{\varphi})]=a/2-b/2=0,
\end{equation}
which is satisfied only if $a=b$. The $\mu=3$ equation gives the same result, so we conclude that the form of the $\Phi^{0}$ and $\Phi^{3}$ components is
\begin{equation}
\Phi^{0}=\Phi^{3}=l_{+}a(\rho,\tzeta_{\varphi}).
\end{equation}
Moving on to the $\mu=1,2$ PDEs, taking the combinations $x$(\ref{B-varphi, X1 four vector PDE mu=1})$+y$(\ref{B-varphi, X1 four vector PDE mu=2}) as well as $x$(\ref{B-varphi, X1 four vector PDE mu=2}) $-y$(\ref{B-varphi, X1 four vector PDE mu=1}) yields the two respective PDEs
\begin{subequations}
\begin{align}
&|L|\partial_{|L|}c+\zeta_{\varphi}\partial_{\zeta}c=0, \\
&|L|\partial_{|L|}d+\zeta_{\varphi}\partial_{\zeta}d=0.
\end{align}
\end{subequations}
Since these equations are of exactly the same form as the scalar field PDE (\ref{B-varphi, X1 scalar field PDE}), we conclude that $c$ and $d$ must be functions of $(\rho,\tzeta_{\varphi})$. Our $\{B_{\varphi},X_{1}\}$ invariant four vector field is therefore of the form
\begin{equation}\label{B-varphi, X1 invariant four vector field}
\Phi^{\mu}=\left(\begin{array}{c}l_{+}a(\rho,\tzeta_{\varphi}) \\
xb(\rho,\tzeta_{\varphi})-yc(\rho,\tzeta_{\varphi}) \\
yb(\rho,\tzeta_{\varphi})+xc(\rho,\tzeta_{\varphi}) \\
l_{+}a(\rho,\tzeta_{\varphi})\end{array}\right).
\end{equation}
Note that this invariant form corresponds to the $P_{11,5}$ Poincar\'{e} subalgebra.

\subsubsection{$B_{\varphi}$, $X_{1}$, $X_{2}$ invariant fields}
Scalar fields simultaneously invariant under $B_{\varphi}$, $X_{1}$ and $X_{2}$ must satisfy
\begin{equation}
\xi_{X_{2}}{}^{\sigma}\partial_{\sigma}\phi(\rho,\tzeta_{\varphi})=\partial_{y}\phi(\rho,\tzeta_{\varphi})=0,
\end{equation}
which when applying the chain rule, gives us the PDE
\begin{equation}\label{B-varphi, X1, X2 invariant scalar field PDE}
y(1/\rho)\partial_{\rho}\phi+x(2\tau\tzeta_{\varphi}/\rho^{2})\partial_{\tzeta}\phi=0.
\end{equation}
For this equation to hold for all $x$, $y$, it must be that the derivative terms are zero, implying that $\phi$ is a constant.

For four vector fields invariant under these generators, we require
\begin{equation}
\xi_{X_{2}}{}^{\sigma}\partial_{\sigma}\Phi^{\mu}-(\partial_{\sigma}\xi_{X_{2}}{}^{\mu})\Phi^{\sigma}=\partial_{y}\Phi^{\mu}=0,
\end{equation}
with $\Phi^{\mu}$ components given by (\ref{B-varphi, X1 invariant four vector field}). The equation for the $\mu=0$ and $\mu=3$ components is
\begin{equation}
y(1/\rho)\partial_{\rho}a+x(2\tau\tzeta_{\varphi}/\rho^{2})\partial_{\tzeta}a=0,
\end{equation}
which has exactly the same form as (\ref{B-varphi, X1, X2 invariant scalar field PDE}), so $a$ is constant. After applying the chain rule, the $\mu=1$ and $\mu=2$ equations respectively are
\begin{subequations}
\begin{align}
& c=x^{2}(2\tau\tzeta_{\varphi}/\rho^{2})\partial_{\tzeta}b+xy(1/\rho)\partial_{\rho}b-xy(2\tau\tzeta_{\varphi}/\rho^{2})\partial_{\tzeta}c-y^{2}(1/\rho)\partial_{\rho}c, \\
& b=-x^{2}(2\tau\tzeta_{\varphi}/\rho^{2})\partial_{\tzeta}c-xy(1/\rho)\partial_{\rho}c-xy(2\tau\tzeta_{\varphi}/\rho^{2})\partial_{\tzeta}b-y^{2}(1/\rho)\partial_{\rho}b.
\end{align}
\end{subequations}
Since it is established that $c$ and $b$ are both functions of $\rho$ and $\tzeta_{\varphi}$, for these expressions to hold for all $x$, $y$, it must be that $b=c=0$. Intuitively speaking, if we imagine the form of a vector field symmetric under rotations around the $z-$axis, with $x$ and $y$ components $\Phi^{1}$ and $\Phi^{2}$, including the additional requirement of symmetry along $y-$axis translations forces both components to be zero. We have determined the form of the $\{B_{\varphi},X_{1},X_{2}\}$ invariant four vector field to be
\begin{equation}\label{P-11,2 invariant four vector field}
\Phi^{\mu}=\left(\begin{array}{c}l_{+}a \\
0 \\
0 \\
l_{+}a\end{array}\right),
\end{equation}
where $a$ is a constant.

\subsubsection{$B_{\varphi}$, $X_{1}$, $X_{2}$, $X_{3}$ invariant fields}
Finally, for full $P_{11,2}$ symmetry, we require that globally constant scalars satisfy
\begin{equation}
\xi_{X_{3}}{}^{\sigma}\partial_{\sigma}\phi=-\partial_{x}\phi=0,
\end{equation}
which they obviously do. Four vector fields must satisfy
\begin{equation}\label{X3 invariance Lie derivative}
\xi_{X_{3}}{}^{\sigma}\partial_{\sigma}\Phi^{\mu}-(\partial_{\sigma}\xi_{X_{3}}{}^{\mu})\Phi^{\sigma}=-\partial_{x}\Phi^{\mu}=0.
\end{equation}
Since $\Phi^{0}$ and $\Phi^{3}$ in (\ref{P-11,2 invariant four vector field}) are already independent of $x$, (\ref{X3 invariance Lie derivative}) is automatically satisfied. We therefore conclude that (\ref{P-11,2 invariant four vector field}) is the full $P_{11,2}$ invariant four vector field form.

\subsection{$\widetilde{P}_{13,10}$ symmetry}
The final example we cover in this paper is the non-splitting $\widetilde{P}_{13,10}$ Poincar\'{e} subalgebra, consisting of the generators $B_{2}+\lambda X_{2}$ $(\lambda>0)$, $X_{1}$, $X_{3}$ and $X_{4}$. The non-splitting aspect is manifested in the fact that the pure Lorentz generator $B_{2}$ is ``tied up'' with the translation generator $\lambda X_{2}$, where the non-zero parameter $\lambda$ determines the relative weight of the translation part. To condense the notation, we define the abbreviated form of the non-split generator, $\widetilde{B}_{\lambda}\equiv B_{2}+\lambda X_{2}$, where the tilde is to emphasize that it is not a pure Lorentz generator. As in the previous subsection, we will consider the Lorentz-translation generator $\widetilde{B}_{\lambda}$ first, then sequentially apply the translation generators in numerical order.

\subsubsection{$\widetilde{B}_{\lambda}$ invariant fields}
$B_{2}$ generates hyperbolic rotations in the $t-z$ plane, and $X_{2}$ generates translations along the $y-$axis. Since the individual vector fields point in orthogonal directions, the vector field corresponding to $\widetilde{B}_{\lambda}$ is simply the linear combination
\begin{equation}
\xi_{B_{2}}{}^{\mu}+\lambda\xi_{X_{2}}{}^{\mu}\equiv\xi_{\widetilde{B}_{\lambda}}{}^{\mu}=\left(\begin{array}{c}-l_{+}+l_{-} \\
0 \\
\lambda \\
-l_{+}-l_{-}\end{array}\right).
\end{equation}
A scalar field $\phi$ invariant under the action of $\widetilde{B}_{\lambda}$ solves
\begin{equation}
\mathcal{L}_{\widetilde{B}_{\lambda}}\phi=\xi_{\widetilde{B}_{\lambda}}{}^{\sigma}\partial_{\sigma}\phi=(-l_{+}+l_{-})\partial_{t}\phi+\lambda\partial_{y}\phi+(-l_{+}-l_{-})\partial_{z}\phi=0.
\end{equation}
Applying the chain rule (\ref{Chain rule to convert dt to d+ and d-}), (\ref{Chain rule to convert dz to d+ and d-}), this PDE becomes
\begin{equation}\label{PDE for B-lambda invariant scalar field}
-2l_{+}\partial_{+}\phi+2l_{-}\partial_{-}\phi+\lambda\partial_{y}\phi=0,
\end{equation}
which has the set of characteristic ODEs
\begin{equation}\label{B_lambda invariant scalar field characteristic ODEs}
\frac{\rmd l_{+}}{-2l_{+}}=\frac{\rmd l_{-}}{2l_{-}}=\frac{\rmd y}{\lambda}.
\end{equation}
Integrating these three equations yields the set of invariants
\begin{subequations}
\begin{align}
&|L|=|l_{+}l_{-}|=|t^{2}-z^{2}|, \\
&\alpha=|l_{+}|\rme^{2y/\lambda}, \\
&\beta=|l_{-}|\rme^{-2y/\lambda},
\end{align}
\end{subequations}
If $\alpha$ and $\beta$ are invariants, then so is
\begin{equation}
\alpha\beta=|l_{+}||l_{-}|\rme^{2y/\lambda}\rme^{-2y/\lambda}=|L|,
\end{equation}
implying that $|L|$ is not independent from the other two. Choosing $\phi$ to be an arbitrary function of $\alpha$, $\beta$ and $x$, we find that (\ref{PDE for B-lambda invariant scalar field}) is satisfied. For future reference, the partial derivatives of $\phi(\alpha,\beta,x)$ are
\begin{subequations}
\begin{align}
&\partial_{t}\phi=\partial_{\alpha}\phi\cdot\partial_{+}\alpha+\partial_{\beta}\phi\cdot\partial_{-}\beta=(l_{+}/|l_{+}|)\rme^{2y/\lambda}\partial_{\alpha}\phi+(l_{-}/|l_{-}|)\rme^{-2y/\lambda}\partial_{\beta}\phi,\label{B-lambda invariant dt chain rule} \\
&\partial_{y}\phi=\partial_{\alpha}\phi\cdot\partial_{y}\alpha+\partial_{\beta}\phi\cdot\partial_{y}\beta=(2|l_{+}|/\lambda)\rme^{2y/\lambda}\partial_{\alpha}\phi-(2|l_{-}|/\lambda)\rme^{-2y/\lambda}\partial_{\beta}\phi,\label{B-lambda invariant dy chain rule} \\
&\partial_{z}\phi=\partial_{\alpha}\phi\cdot\partial_{+}\alpha-\partial_{\beta}\phi\cdot\partial_{-}\beta=(l_{+}/|l_{+}|)\rme^{2y/\lambda}\partial_{\alpha}\phi-(l_{-}/|l_{-}|)\rme^{-2y/\lambda}\partial_{\beta}\phi,\label{B-lambda invariant dz chain rule}
\end{align}
\end{subequations}
where $\partial_{t}l_{+}=\partial_{z}l_{+}=\partial_{t}l_{-}=1$ and $\partial_{z}l_{-}=-1$ are implicit. Invariant vector fields must solve
\begin{equation}
\mathcal{L}_{\widetilde{B}_{\lambda}}\Phi^{\mu}=\xi_{\widetilde{B}_{\lambda}}{}^{\sigma}\partial_{\sigma}\Phi^{\mu}-(\partial_{\sigma}\xi_{\widetilde{B}_{\lambda}}{}^{\mu})\Phi^{\sigma}=0,
\end{equation}
which for $\mu=0-3$ gives us the set of PDEs
\begin{subequations}
\begin{align}
&-2l_{+}\partial_{+}\Phi^{0}+2l_{-}\partial_{-}\Phi^{0}+\lambda\partial_{y}\Phi^{0}+2\Phi^{3}=0,\label{Vector field B-lambda symmetry, mu=0} \\
&-2l_{+}\partial_{+}\Phi^{1}+2l_{-}\partial_{-}\Phi^{1}+\lambda\partial_{y}\Phi^{1}=0,\label{Vector field B-lambda symmetry, mu=1} \\
&-2l_{+}\partial_{+}\Phi^{2}+2l_{-}\partial_{-}\Phi^{2}+\lambda\partial_{y}\Phi^{2}=0,\label{Vector field B-lambda symmetry, mu=2} \\
&-2l_{+}\partial_{+}\Phi^{3}+2l_{-}\partial_{-}\Phi^{3}+\lambda\partial_{y}\Phi^{3}+2\Phi^{0}=0.\label{Vector field B-lambda symmetry, mu=3}
\end{align}
\end{subequations}
Since (\ref{Vector field B-lambda symmetry, mu=1}) and (\ref{Vector field B-lambda symmetry, mu=2}) are the same form as (\ref{PDE for B-lambda invariant scalar field}), they have the same characteristic solution as the scalar field: $\Phi^{1}=c(\alpha,\beta,x)$ and $\Phi^{2}=d(\alpha,\beta,x)$. Applying the method of characteristics to the non-homogeneous PDEs (\ref{Vector field B-lambda symmetry, mu=0}) and (\ref{Vector field B-lambda symmetry, mu=3}), we obtain two systems of ODEs exactly like (\ref{B_lambda invariant scalar field characteristic ODEs}), but with extra $\rmd\Phi^{\mu}$ parts. Equating these parts gives
\begin{equation}
\frac{\rmd\Phi^{0}}{2\Phi^{3}}=\frac{\rmd\Phi^{3}}{2\Phi^{0}},
\end{equation}
which when integrated yields the algebraic constraint
\begin{equation}
(\Phi^{0})^{2}-(\Phi^{3})^{2}=f(\alpha,\beta,x).\label{B-lambda algebraic constraint for mu=0,3}
\end{equation}
This has the solution
\begin{subequations}
\begin{align}
&\Phi^{0}=l_{+}a(\alpha,\beta,x)+l_{-}b(\alpha,\beta,x), \\
&\Phi^{3}=l_{+}a(\alpha,\beta,x)-l_{-}b(\alpha,\beta,x),
\end{align}
\end{subequations}
which can be checked via substitution. We have determined the form of the $\widetilde{B}_{\lambda}$ invariant four vector field to be
\begin{equation}\label{B-lambda invariant four-vector field}
\Phi^{\mu}=\left(\begin{array}{c}l_{+}a+l_{-}b \\
c \\
d \\
l_{+}a-l_{-}b\end{array}\right),
\end{equation}
where $a$, $b$, $c$ and $d$ are functions of $\alpha$, $\beta$ and $x$. This form can be verified by inserting the appropriate components into (\ref{Vector field B-lambda symmetry, mu=0})-(\ref{Vector field B-lambda symmetry, mu=3}), as well as (\ref{B-lambda algebraic constraint for mu=0,3}).

\subsubsection{$\widetilde{B}_{\lambda}$, $X_{1}$ invariant fields}
Scalar fields invariant under both $\widetilde{B}_{\lambda}$ and $X_{1}$ must solve
\begin{equation}
\mathcal{L}_{X_{1}}\phi(\alpha,\beta,x)=\xi_{X_{1}}{}^{\sigma}\partial_{\sigma}\phi(\alpha,\beta,x)=(1/2)(\partial_{t}\phi-\partial_{z}\phi)=0.
\end{equation}
Applying (\ref{B-lambda invariant dt chain rule}) and (\ref{B-lambda invariant dz chain rule}) gives us
\begin{equation}
(l_{-}/|l_{-}|)\rme^{-2y/\lambda}\partial_{\beta}\phi=0,
\end{equation}
but since $l_{-}/|l_{-}|=\pm1$ ($l_{-}\neq0$) and $\rme^{-2y/\lambda}$ are positive definite, we require
\begin{equation}
\partial_{\beta}\phi=0,
\end{equation}
implying that $\phi$ is independent of $\beta$. For the four vector field to be invariant under $\widetilde{B}_{\lambda}$ and $X_{1}$, we must have
\begin{equation}
\mathcal{L}_{X_{1}}\Phi^{\mu}=\xi_{X_{1}}{}^{\sigma}\partial_{\sigma}\Phi^{\mu}-(\partial_{\sigma}\xi_{X_{1}}{}^{\mu})\Phi^{\sigma}=(1/2)(\partial_{t}\Phi^{\mu}-\partial_{z}\Phi^{\mu})=0,
\end{equation}
where the components of $\Phi^{\mu}$ are given by (\ref{B-lambda invariant four-vector field}). The four PDEs obtained after carrying out the derivatives are
\begin{subequations}
\begin{align}
&b+l_{+}(l_{-}/|l_{-}|)\rme^{-2y/\lambda}\partial_{\beta}a+l_{-}(l_{-}/|l_{-}|)\rme^{-2y/\lambda}\partial_{\beta}b=0,\label{B-lambda, X1 invariance PDE for mu=0} \\
&\partial_{t}c-\partial_{z}c=0,\label{B-lambda, X1 invariance PDE for mu=1} \\
&\partial_{t}d-\partial_{z}d=0,\label{B-lambda, X1 invariance PDE for mu=2} \\
&-b+l_{+}(l_{-}/|l_{-}|)\rme^{-2y/\lambda}\partial_{\beta}a-l_{-}(l_{-}/|l_{-}|)\rme^{-2y/\lambda}\partial_{\beta}b=0.\label{B-lambda, X1 invariance PDE for mu=3}
\end{align}
\end{subequations}
Since (\ref{B-lambda, X1 invariance PDE for mu=1}) and (\ref{B-lambda, X1 invariance PDE for mu=2}) are the same PDEs as in the scalar field case, $c$ and $d$ must both be independent of $\beta$. We can easily decouple (\ref{B-lambda, X1 invariance PDE for mu=0}) and (\ref{B-lambda, X1 invariance PDE for mu=3}) by adding or subtracting them. Adding and discarding the non-zero terms gives
\begin{equation}
l_{+}\partial_{\beta}a=0,
\end{equation}
which requires $\partial_{\beta}a=0$ everywhere, except possibly at $l_{+}=0$. Ignoring this technicality, we can say that $a$ must be a function of $\alpha$ and $x$ only. Subtracting the two PDEs and rearranging gives
\begin{equation}
\partial_{\beta}b(\alpha,\beta,x)=-b(\alpha,\beta,x)/\beta,
\end{equation}
which has the solution
\begin{equation}
b(\alpha,\beta,x)=b(\alpha,x)/\beta.
\end{equation}
Our $\widetilde{B}_{\lambda}$, $X_{1}$ invariant four vector field is therefore
\begin{equation}\label{B-lambda, X1 invariant four vector field}
\Phi^{\mu}=\left(\begin{array}{c}l_{+}a+(l_{-}/\beta)b \\
c \\
d \\
l_{+}a-(l_{-}/\beta)b\end{array}\right),
\end{equation}
where $a$, $b$, $c$ and $d$ are functions of $\alpha$ and $x$.

\subsubsection{$\widetilde{B}_{\lambda}$, $X_{1}$, $X_{3}$ invariant fields}
The invariance condition for scalar fields is now
\begin{equation}
\mathcal{L}_{X_{3}}\phi(\alpha,x)=\xi_{X_{3}}{}^{\sigma}\partial_{\sigma}\phi(\alpha,x)=\partial_{x}\phi=0,
\end{equation}
implying that $\phi$ must be independent of $x$, a function of the single variable $\alpha$. The invariance condition for four vector fields is
\begin{equation}
\mathcal{L}_{X_{3}}\Phi^{\mu}=\xi_{X_{3}}{}^{\sigma}\partial_{\sigma}\Phi^{\mu}-(\partial_{\sigma}\xi_{X_{3}}{}^{\mu})\Phi^{\sigma}=\partial_{x}\Phi^{\mu}=0,
\end{equation}
where the components $\Phi^{\mu}$ are given in (\ref{B-lambda, X1 invariant four vector field}). The four PDEs for each component are
\begin{subequations}
\begin{align}
&l_{+}\partial_{x}a+(l_{-}/\beta)\partial_{x}b=0, \\
&\partial_{x}c=0, \\
&\partial_{x}d=0, \\
&l_{+}\partial_{x}a-(l_{-}/\beta)\partial_{x}b=0.
\end{align}
\end{subequations}
The middle two equations obviously imply that $c$ and $d$ are functions of $\alpha$ only. Adding the first and last equations gives
\begin{equation}
l_{+}\partial_{x}a=0.
\end{equation}
Again ignoring the ambiguity at the $l_{+}=0$ point, we conclude that $a$ is a function of $\alpha$ only. Subtracting the last PDE from the first gives us
\begin{equation}
(l_{-}/\beta)\partial_{x}b=(l_{-}/|l_{-}|)\rme^{2y/\lambda}\partial_{x}b=0,
\end{equation}
implying that that $b$ is a function of $\alpha$ only. The $\widetilde{B}_{\lambda}$, $X_{1}$, $X_{3}$ invariant form of the four vector field is
\begin{equation}\label{B-lambda, X1, X3 invariant four-vector field}
\Phi^{\mu}=\left(\begin{array}{c}l_{+}a+(l_{-}/\beta)b \\
c \\
d \\
l_{+}a-(l_{-}/\beta)b\end{array}\right),
\end{equation}
where $a$, $b$, $c$ and $d$ are functions of $\alpha$ only.

\subsubsection{$\widetilde{B}_{\lambda}$, $X_{1}$, $X_{3}$, $X_{4}$ invariant fields}
Lastly, for full $\widetilde{P}_{13,10}$ invariance, the scalar field must satisfy
\begin{equation}
\mathcal{L}_{X_{4}}\phi(\alpha)=\xi_{X_{4}}{}^{\sigma}\partial_{\sigma}\phi(\alpha)=(1/2)(\partial_{t}\phi+\partial_{z}\phi)=0.
\end{equation}
After applying the chain rule and disregarding the non-zero terms, we get
\begin{equation}
\partial_{\alpha}\phi=0,
\end{equation}
so $\phi$ is a constant. $\widetilde{P}_{13,10}$ invariant vector fields must satisfy
\begin{equation}
\mathcal{L}_{X_{4}}\Phi^{\mu}=\xi_{X_{4}}{}^{\sigma}\partial_{\sigma}\Phi^{\mu}-(\partial_{\sigma}\xi_{X_{4}}{}^{\mu})\Phi^{\sigma}=(1/2)(\partial_{t}\Phi^{\mu}+\partial_{z}\Phi^{\mu})=0,
\end{equation}
where $\Phi^{\mu}$ has components (\ref{B-lambda, X1, X3 invariant four-vector field}). The PDEs for each component are
\begin{subequations}
\begin{align}
&a+l_{+}(l_{+}/|l_{+}|)\rme^{2y/\lambda}\partial_{\alpha}a+(L/|L|)\rme^{4y/\lambda}\partial_{\alpha}b=0, \\
&\partial_{t}c+\partial_{z}c=0, \\
&\partial_{t}d+\partial_{z}d=0, \\
&a+l_{+}(l_{+}/|l_{+}|)\rme^{2y/\lambda}\partial_{\alpha}a-(L/|L|)\rme^{4y/\lambda}\partial_{\alpha}b=0.
\end{align}
\end{subequations}
The middle two PDEs are the same as the scalar field case, so $c$ and $d$ are both constants. Adding the first and last equations and rearranging gives
\begin{equation}
\partial_{\alpha}a(\alpha)=-a(\alpha)/\alpha,
\end{equation}
which has the solution
\begin{equation}
a(\alpha)=a/\alpha,
\end{equation}
where the $a$ on the right-hand side is a constant. Finally, subtracting the last PDE from the first, we find that
\begin{equation}
\partial_{\alpha}b=0,
\end{equation}
so $b$ is a constant. The $\widetilde{P}_{13,10}$ invariant form of the four vector field is
\begin{equation}\label{P_13,10 invariant four vector field}
\Phi^{\mu}=\left(\begin{array}{c}(l_{+}/|l_{+}|)\rme^{-2y/\lambda}a+(l_{-}/|l_{-}|)\rme^{2y/\lambda}b \\
c \\
d \\
(l_{+}/|l_{+}|)\rme^{-2y/\lambda}a-(l_{-}/|l_{-}|)\rme^{2y/\lambda}b\end{array}\right).
\end{equation}

\section{Maxwell-Dirac symmetry reductions}
In this section, we apply the Poincar\'{e} symmetry subgroups from section 4 to the Maxwell-Dirac equations (\ref{Fierz inner product identity})-(\ref{Maxwell equations}), and observe how the system reduces under these constraints. By substituting (\ref{Gauge invariant vector potential}) into (\ref{Manifestly gauge invariant field strength tensor}), and subsequently substituting (\ref{Manifestly gauge invariant field strength tensor}) into (\ref{Maxwell equations}), we find that the system depends only on the two four vector fields $j^{\mu}$ and $k^{\mu}$, and the two scalar fields $\sigma$ and $\omega$. Strictly speaking, $k^{\mu}$ is a pseudovector, and $\omega$ is a pseudoscalar, which means their sign under a Lorentz transformation depends on the determinant of the transforming matrix. The sign is negative for improper transformations, but since we are only dealing with symmetry under Lorentz transformations connected to the identity, we shall treat $k^{\mu}$ as a regular four vector field and $\omega$ as a scalar. It can be shown, at least in the spherical symmetry case, that if one interprets $k^{\mu}$ in terms of a rank-3 tensor $T_{\nu\sigma\rho}$ fully contracted with a rank-4 Levi-Civita symbol $\epsilon^{\mu\nu\sigma\rho}$, and then imposes the symmetric form of $T_{\nu\sigma\rho}$, $k^{\mu}$ has the correct four-vector form for that symmetry group. For groups other than $SO(3)$, we are assuming that this is the case. The Fierz identities (\ref{Fierz inner product identity}) and (\ref{Fierz orthogonality identity}) can be used to eliminate two dependent variables, which in our analysis we choose to be from the $k^{\mu}$ four vector field.

Our analysis will proceed as follows. For each symmetry subgroup, we will restate the forms that the fields must take, then use the Fierz identities to obtain expressions for two of the dependent functions in $k^{\mu}$ in terms of other dependent functions. These expressions are used to eliminate the two functions from the system entirely, except in the cylindrical case, where this would unnecessary complicate things. Next, we reduce $B^{\mu}$ (\ref{Gauge invariant vector potential}) for $\mu=0-3$ by substituting in the subgroup-invariant forms for the fields. Due to the length of the calculations in the spherical and cylindrical cases, we will just state the results, where we find that $B^{\mu}$ has the correct form for a subgroup-invariant four vector field, with the dependent functions in terms of $j^{\mu}$, $k^{\mu}$, $\sigma$ and $\omega$. In order to save space, we introduce the more abbreviated derivative notation $\partial_{t}\sigma\equiv\sigma_{t}$ and $\partial_{r}j_{a}\equiv j_{a,r}$.

Following this, we reduce $F^{\mu\nu}$ (\ref{Manifestly gauge invariant field strength tensor}) by substituting the subgroup-invariant forms of the fields, as well as our previously obtained $B^{\mu}$. In the spherically and cylindrically symmetric cases, the $B^{\mu}$ components are quite long, so we enlist the aid of Mathematica to carry out the derivatives and factorization, along with some further manual manipulation. The results of the manual manipulation can be checked for errors by comparing with the original Mathematica output. Once the form of $F^{\mu\nu}$ has been obtained, we can substitute it into the Maxwell equations (\ref{Maxwell equations}), yielding up to four equations, purely in terms of $j^{\mu}$, $k^{\mu}$, $\sigma$ and $\omega$ only; the Maxwell-Dirac equations. We shall see that the Maxwell-Dirac system varies wildly in complexity depending on what the chosen symmetry group is, from a simple algebraic system in the $P_{11,2}$ case, to the cylindrical case, which yields a coupled set of third-order, non-linear PDEs too long to easily write in closed form.

\subsection{Spherical symmetry (subgroup $P_{3,4}$)}
\subsubsection{Fierz identities}
From (\ref{Spherically symmetric four vector field}) in section 4.2, $j^{\mu}$ and $k^{\mu}$ have the form
\begin{equation}
j^{\mu}=\left(\begin{array}{c}j_{a} \\
xj_{b} \\
yj_{b} \\
zj_{b}\end{array}\right),\qquad k^{\mu}=\left(\begin{array}{c}k_{a} \\
xk_{b} \\
yk_{b} \\
zk_{b}\end{array}\right),
\end{equation}
where $j_{a}$, $j_{b}$, $k_{a}$ and $k_{b}$ are all functions of $t$ and $r=\sqrt{x^{2}+y^{2}+z^{2}}$, as are $\sigma$ and $\omega$. Lowering the index of the four vectors turns the column into a row, and changes the sign of the $\mu=1,2,3$ components. Contracting $j^{\mu}$ and $k^{\mu}$ with themselves, and using the Fierz identity (\ref{Fierz inner product identity}), we get
\begin{equation}\label{Spherical inner product Fierz identity}
j_{a}^{2}-r^{2}j_{b}^{2}=-k_{a}^{2}+r^{2}k_{b}^{2}=\sigma^{2}-\omega^{2}.
\end{equation}
We can rearrange this to solve for $k_{a}$
\begin{equation}\label{Spherical inner product identity rearranged for ka}
k_{a}=\pm\sqrt{r^{2}(j_{b}^{2}+k_{b}^{2})-j_{a}^{2}}.
\end{equation}
Contracting $j^{\mu}$ with $k^{\mu}$ and using the orthogonality condition (\ref{Fierz orthogonality identity})
\begin{equation}\label{Spherical orthogonality condition}
j_{a}k_{a}-r^{2}j_{b}k_{b}=0,
\end{equation}
then substituting (\ref{Spherical inner product identity rearranged for ka}) and performing some simple algebraic manipulation gives us the identity
\begin{equation}
k_{b}=\pm j_{a}/r.
\end{equation}
Substituting this back into (\ref{Spherical inner product identity rearranged for ka}) gives the other identity
\begin{equation}
k_{a}=\pm rj_{b}.
\end{equation}
We have determined that $k^{\mu}$ can be expressed entirely in terms of the dependent functions of $j^{\mu}$
\begin{equation}
k^{\mu}=\pm\left(\begin{array}{c}rj_{b} \\
(x/r)j_{a} \\
(y/r)j_{a} \\
(z/r)j_{a}\end{array}\right).
\end{equation}

\subsubsection{Vector potential}
Our next step is to substitute our symmetric fields into (\ref{Gauge invariant vector potential}), and simplify for each $\mu$. When performing the calculations, dealing with each term in the numerator of $B^{\mu}$ separately makes them much easier to handle. We will briefly outline the calculations for the first two components, then skip to the symmetric form of $B^{\mu}$. When $\mu=0$, the first component is
\begin{align}
&\epsilon^{0\nu\rho\sigma}(\sigma^{2}-\omega^{2})\partial_{\nu}(j_{\rho}k_{\sigma}) \nonumber \\
&\qquad=(\sigma^{2}-\omega^{2})[\partial_{1}(j_{2}k_{3})\!-\!\partial_{1}(j_{3}k_{2})\!-\!\partial_{2}(j_{1}k_{3})\!+\!\partial_{2}(j_{3}k_{1})\!+\!\partial_{3}(j_{1}k_{2})\!-\!\partial_{3}(j_{2}k_{1})] \nonumber \\
&\qquad=\pm(\sigma^{2}-\omega^{2})\{\partial_{x}[(yz/r)j_{a}j_{b}]-\partial_{x}[(yz/r)j_{a}j_{b}]-\partial_{y}[(xz/r)j_{a}j_{b}] \nonumber \\
&\qquad\qquad+\partial_{y}[(xz/r)j_{a}j_{b}]+\partial_{z}[(xy/r)j_{a}j_{b}]-\partial_{z}[(xy/r)j_{a}j_{b}]\} \nonumber \\
&\qquad=0.
\end{align}
The second term in $B^{0}$ expands in exactly the same way, except the partial derivative operator acts on $\sigma^{2}-\omega^{2}$, so we get
\begin{equation}
-(1/2)\epsilon^{0\nu\rho\sigma}j_{\rho}k_{\sigma}\partial_{\nu}(\sigma^{2}-\omega^{2})=0.
\end{equation}
The third term in $B^{0}$ is
\begin{align}
&\delta^{0\nu\rho\sigma}[(\partial_{\nu}\sigma)\omega-\sigma(\partial_{\nu}\omega)]j_{\rho}k_{\sigma}=\rmi(j^{0}k^{\nu}-j^{\nu}k^{0})[(\partial_{\nu}\sigma)\omega-\sigma(\partial_{\nu}\omega)] \nonumber \\
&\qquad=\pm\rmi[rj_{a}j_{b}(\partial_{t}\sigma)\omega+(x/r)j_{a}^{2}(\partial_{x}\sigma)\omega+(y/r)j_{a}^{2}(\partial_{y}\sigma)\omega+(z/r)j_{a}^{2}(\partial_{z}\sigma)\omega \nonumber \\
&\qquad\qquad-rj_{a}j_{b}\sigma(\partial_{t}\omega)-(x/r)j_{a}^{2}\sigma(\partial_{x}\omega)-(y/r)j_{a}^{2}\sigma(\partial_{y}\omega)-(z/r)j_{a}^{2}\sigma(\partial_{z}\omega) \nonumber \\
&\qquad\qquad-rj_{a}j_{b}(\partial_{t}\sigma)\omega-rxj_{b}^{2}(\partial_{x}\sigma)\omega-ryj_{b}^{2}(\partial_{y}\sigma)\omega-rzj_{b}^{2}(\partial_{z}\sigma)\omega \nonumber \\
&\qquad\qquad+rj_{a}j_{b}\sigma(\partial_{t}\omega)+rxj_{b}^{2}\sigma(\partial_{x}\omega)+ryj_{b}^{2}\sigma(\partial_{y}\omega)+rzj_{b}^{2}\sigma(\partial_{z}\omega)] \nonumber \\
&\qquad=\pm\rmi[(j_{a}^{2}/r)-rj_{b}^{2}](x^{2}/r+y^{2}/r+z^{2}/r)[(\partial_{r}\sigma)\omega-\sigma(\partial_{r}\omega)] \nonumber \\
&\qquad=\pm\rmi(\sigma^{2}-\omega^{2})[(\partial_{r}\sigma)\omega-\sigma(\partial_{r}\omega)].
\end{align}
Substituting these terms back into $B^{0}$ gives, after canceling terms and applying our abbreviated notation
\begin{equation}
B^{0}=[\pm(\rmi/2)(\sigma_{r}\omega-\sigma\omega_{r})-m\sigma j_{a}][q(\sigma^{2}-\omega^{2})]^{-1}.
\end{equation}
Now take the $\mu=1$ case. The first term in $B^{1}$ is
\begin{align}
&\epsilon^{1\nu\rho\sigma}(\sigma^{2}-\omega^{2})\partial_{\nu}(j_{\rho}k_{\sigma}) \nonumber \\
&\qquad=(\sigma^{2}-\omega^{2})[-\partial_{0}(j_{2}k_{3})+\partial_{0}(j_{3}k_{2})+\partial_{2}(j_{0}k_{3})-\partial_{2}(j_{3}k_{0})-\partial_{3}(j_{0}k_{2}) \nonumber \\
&\qquad\qquad+\partial_{3}(j_{2}k_{0})] \nonumber \\
&\qquad=\pm(\sigma^{2}-\omega^{2})[(yz/r^{3})j_{a}^{2}-(2yz/r^{2})j_{a}(\partial_{r}j_{a})+(yz/r)j_{b}^{2}+2yzj_{b}(\partial_{r}j_{b}) \nonumber \\
&\qquad\qquad-(yz/r^{3})j_{a}^{2}+(2yz/r^{2})j_{a}(\partial_{r}j_{a})-(yz/r)j_{b}^{2}-2yzj_{b}(\partial_{r}j_{b})] \nonumber \\
&\qquad=0.
\end{align}
Since we had to apply the derivative operators in this case, we must be more careful about the second term, which is
\begin{align}
&-(1/2)\epsilon^{1\nu\rho\sigma}j_{\rho}k_{\sigma}\partial_{\nu}(\sigma^{2}-\omega^{2}) \nonumber \\
&\qquad=\pm(1/2)[(yz/r)j_{a}j_{b}\partial_{t}-(yz/r)j_{a}j_{b}\partial_{t}+(z/r)j_{a}^{2}\partial_{y}-zrj_{b}^{2}\partial_{y}-(y/r)j_{a}^{2}\partial_{z} \nonumber \\
&\qquad\qquad+yrj_{b}^{2}\partial_{z}](\sigma^{2}-\omega^{2}) \nonumber \\
&\qquad=\pm(1/2)[(2yz/r^{2})j_{a}^{2}-2yzj_{b}^{2}-(2yz/r^{2})j_{a}^{2}+2yzj_{b}^{2}][\sigma(\partial_{r}\sigma)-\omega(\partial_{r}\omega)] \nonumber \\
&\qquad=0.
\end{align}
Taking advantage of the abbreviated derivative notation, the third term in $B^{1}$ is
\begin{align}
&\delta^{1\nu\rho\sigma}[(\partial_{\nu}\sigma)\omega-\sigma(\partial_{\nu}\omega)]j_{\rho}k_{\sigma}=\rmi(j^{1}k^{\nu}-j^{\nu}k^{1})[(\partial_{\nu}\sigma)\omega-\sigma(\partial_{\nu}\omega)] \nonumber \\
&\qquad=\pm\rmi[xrj_{b}^{2}(\sigma_{t}\omega-\sigma\omega_{t})+(xy/r)j_{a}j_{b}(\sigma_{y}\omega-\sigma\omega_{y}) \nonumber \\
&\qquad\qquad+(xz/r)j_{a}j_{b}(\sigma_{z}\omega-\sigma\omega_{z})-(x/r)j_{a}^{2}(\sigma_{t}\omega-\sigma\omega_{t}) \nonumber \\
&\qquad\qquad-(xy/r)j_{a}j_{b}(\sigma_{y}\omega-\sigma\omega_{y})-(xz/r)j_{a}j_{b}(\sigma_{z}\omega-\sigma\omega_{z})] \nonumber \\
&\qquad=\mp\rmi(x/r)(\sigma^{2}-\omega^{2})(\sigma_{t}\omega-\sigma\omega_{t}).
\end{align}
Substituting into $B^{1}$ and making appropriate cancellations gives
\begin{equation}
B^{1}=x[\mp(\rmi/2r)(\sigma_{t}\omega-\sigma\omega_{t})-m\sigma j_{b}][q(\sigma^{2}-\omega^{2})]^{-1}.
\end{equation}
There is a similar result for $\mu=2$ and $\mu=3$, but instead of an $x$, there is a $y$ and $z$ respectively. We have found that $B^{\mu}$ assumes the form required for a spherically symmetric four vector field
\begin{equation}
B^{\mu}=\left(\begin{array}{c}B_{a} \\
xB_{b} \\
yB_{b} \\
zB_{b}\end{array}\right),
\end{equation}
where $B_{a}$ and $B_{b}$ are functions of the invariants, given by
\begin{align}
&B_{a}=[\pm(\rmi/2)(\sigma_{r}\omega-\sigma\omega_{r})-m\sigma j_{a}][q(\sigma^{2}-\omega^{2})]^{-1},\label{Spherical symmetry, Ba} \\
&B_{b}=[\mp(\rmi/2r)(\sigma_{t}\omega-\sigma\omega_{t})-m\sigma j_{b}][q(\sigma^{2}-\omega^{2})]^{-1}.\label{Spherical symmetry, Bb}
\end{align}

\subsubsection{Field strength tensor} Now we turn our attention to $F_{\mu\nu}$ (\ref{Manifestly gauge invariant field strength tensor}), which is antisymmetric with six independent components. Since the calculations are quite lengthy, we enlist the computational aid of Mathematica. Considering the $\mu=0$, $\nu=i=1,2,3$ components of $F_{\mu\nu}$ first, we find that
\begin{equation}
\epsilon^{\sigma\rho\kappa\tau}j_{\kappa}k_{\tau}[(\partial_{0}j_{\sigma})(\partial_{i}j_{\rho})-(\partial_{0}k_{\sigma})(\partial_{i}k_{\rho})]=0,
\end{equation}
so the rational term vanishes. We are left with the four-curl term
\begin{equation}
F_{0i}=\partial_{0}B_{i}-\partial_{i}B_{0}=x_{i}[-\partial_{t}B_{b}-(1/r)\partial_{r}B_{a}],
\end{equation}
which is of the form
\begin{equation}
F_{0i}=x_{i}F_{a}(t,r),
\end{equation}
where the function of the invariants in terms of the bilinear fields is
\begin{align}
& F_{a}(t,r)=(1/qr)(\sigma^{2}-\omega^{2})^{-2}\{-2m[\sigma j_{a}(\sigma\sigma_{r}-\omega\omega_{r})+r\sigma j_{b}(\sigma\sigma_{t}-\omega\omega_{t})] \nonumber \\
&\qquad\pm\rmi[\sigma\omega(\sigma_{r}^{2}-\sigma_{t}^{2}+\omega_{r}^{2}-\omega_{t}^{2})+(\sigma^{2}+\omega^{2})(\sigma_{t}\omega_{t}-\sigma_{r}\omega_{r})]\} \nonumber \\
&\qquad+(1/qr)(\sigma^{2}-\omega^{2})^{-1}[m(\sigma_{r}j_{a}+\sigma j_{a,r}+r\sigma_{t}j_{b}+r\sigma j_{b,t}) \nonumber \\
&\qquad\pm(\rmi/2)(\sigma_{tt}\omega-\sigma\omega_{tt}-\sigma_{rr}\omega+\sigma\omega_{rr})].
\end{align}
Now consider the purely spatial components, $F_{ij}$. The four-curl term in this case is
\begin{equation}
\partial_{i}B_{j}-\partial_{j}B_{i}=-(x_{j}x_{i}/r)\partial_{r}B_{b}+(x_{i}x_{j}/r)\partial_{r}B_{b}=0,
\end{equation}
leaving us with the rational term only. Expanding the three independent $F_{ij}$ using Mathematica, we find that
\begin{subequations}
\begin{align}
&F_{12}=zF_{b}(t,r), \\
&F_{13}=-yF_{b}(t,r), \\
&F_{23}=xF_{b}(t,r),
\end{align}
\end{subequations}
where
\begin{equation}
F_{b}(t,r)=\pm\frac{1}{2q}\left(\frac{j_{a}^{4}}{r^{3}}-\frac{2j_{a}^{2}j_{b}^{2}}{r}+rj_{b}^{4}\right)(\sigma^{2}-\omega^{2})^{-2}.
\end{equation}
Factorizing, and using the inner product Fierz identity (\ref{Spherical inner product Fierz identity}), we find that this simplifies to
\begin{equation}
F_{b}(t,r)=\pm\frac{1}{2qr^{3}}.
\end{equation}
The spatial components of the field strength tensor become
\begin{subequations}
\begin{align}
&F_{12}=-M_{z}=\pm\frac{1}{2q}\frac{z}{r^{3}}, \\
&F_{13}=M_{y}=\mp\frac{1}{2q}\frac{y}{r^{3}}, \\
&F_{23}=-M_{x}=\pm\frac{1}{2q}\frac{x}{r^{3}},
\end{align}
\end{subequations}
implying a magnetic field of the form
\begin{equation}\label{Magnetic monopole field}
\boldsymbol{M}=\mp\frac{1}{2q}\frac{\rhat}{r^{2}},
\end{equation}
where we have unconventionally denoted the magnetic field by $\bM=M_{x}\xhat+M_{y}\yhat+M_{z}\zhat$ to avoid confusion with the gauge invariant vector potential. We can see that the $\boldsymbol{M}$-field is radially pointing and obeys an inverse square law, implying that there is a magnetic monopole at the origin. Taking the divergence of (\ref{Magnetic monopole field}), we find that the only non-zero point is located at $r=0$. Calling the magnetic charge $q_{m}$ and the associated magnetic charge density $\rho_{m}$, we must have
\begin{equation}\label{Divergence of monopole field}
\nabla\cdot\bM=\rho_{m}=q_{m}\delta(r),
\end{equation}
the volume integral of the right-hand side being equal to $q_{m}$. Applying the divergence theorem to the left-hand side of the volume integral of (\ref{Divergence of monopole field}), we find that
\begin{equation}
\int\nabla\cdot\bM\rmd V=\oint\bM\cdot\rmd\ba=\mp\frac{2\pi}{q}=q_{m},
\end{equation}
which is in agreement with the Dirac quantization condition \cite{Dirac-1948}
\begin{equation}
q_{m}q/4\pi=n/2,
\end{equation}
for the special case where $n=\mp1$.

This result follows simply from imposing spherical symmetry on our manifestly gauge invariant Dirac and Fierz formalism, and is a generalization of Radford's formalism \cite{Radford-1996}, in that the monopole field appears in both the static \emph{and} non-static cases. We shall see in the next section that the $F_{b}$ dependent terms, and hence the magnetic monopole aspect, cancel out of the Maxwell equations, so monopoles have no effect on the physics of the coupled Maxwell-Dirac system. In summary, our spherically symmetric field strength tensor is of the form
\begin{equation}
F_{\mu\nu}=\left(\begin{array}{cccc}0 & xF_{a} & yF_{a} & zF_{a} \\
-xF_{a} & 0 & zF_{b} & -yF_{b} \\
-yF_{a} & -zF_{b} & 0 & xF_{b} \\
-zF_{a} & yF_{b} &-xF_{b} & 0\end{array}\right).
\end{equation}

\subsubsection{Maxwell equations}
Combining the reduced field strength tensor (which was obtained from the inverted Dirac equation) with Maxwell's equations (\ref{Maxwell equations}) results in the full Maxwell-Dirac system. It is easy to show that for the current symmetry group $SO(3)$, there are only two independent equations
\begin{align}
&3F_{a}+r\partial_{r}F_{a}=qj_{a} \\
&\partial_{t}F_{a}=-qj_{b},
\end{align}
for $\mu=0$ and $\mu=i$ respectively. The $F_{b}$ terms appear in the $\mu=i$ equations, but cancel out, indicating that despite magnetic monopoles entering into the field strength tensor, they do not affect the physics of the coupled Maxwell-Dirac system. Carrying out the derivatives with Mathematica and factorizing manually gives the two Maxwell-Dirac equations in terms of $\sigma$, $\omega$, $j_{a}$, $j_{b}$ and their $t$ and $r$ derivatives, up to third order. The first equation is
\begin{align}\label{Spherical symmetry, first MD equation}
& q^{2}j_{a}=-(\sigma^{2}-\omega^{2})^{-3}4(\sigma\sigma_{r}-\omega\omega_{r})\{-2m[\sigma j_{a}(\sigma\sigma_{r}-\omega\omega_{r})+r\sigma j_{b}(\sigma\sigma_{t}-\omega\omega_{t})] \nonumber \\
&\qquad\pm\rmi[(\sigma^{2}+\omega^{2})(\sigma_{t}\omega_{t}-\sigma_{r}\omega_{r})+\sigma\omega(\sigma_{r}^{2}-\sigma_{t}^{2}+\omega_{r}^{2}-\omega_{t}^{2})]\} \nonumber \\
&\qquad+(\sigma^{2}-\omega^{2})^{-2}\{-2m[(3\sigma j_{b}+r\sigma_{r}j_{b}+r\sigma j_{b,r})(\sigma\sigma_{t}-\omega\omega_{t})+(2\sigma_{r}j_{a}+2\sigma j_{a,r} \nonumber \\
&\qquad+2\sigma j_{a}/r+r\sigma_{t}j_{b}+r\sigma j_{b,t})(\sigma\sigma_{r}-\omega\omega_{r})+\sigma j_{a}(\sigma_{r}^{2}+\sigma\sigma_{rr}-\omega_{r}^{2}-\omega\omega_{rr}) \nonumber \\
&\qquad+r\sigma j_{b}(\sigma\sigma_{tr}+\sigma_{t}\sigma_{r}-\omega\omega_{tr}-\omega_{t}\omega_{r})]\pm\rmi[2(\sigma\sigma_{r}+\omega\omega_{r}+\sigma^{2}/r+\omega^{2}/r)(\sigma_{t}\omega_{t} \nonumber \\
&\qquad-\sigma_{r}\omega_{r})+(\sigma\omega_{tt}-\sigma_{tt}\omega-\sigma\omega_{rr}+\sigma_{rr}\omega)(\sigma\sigma_{r}-\omega\omega_{r})+(\sigma\omega_{r}+\sigma_{r}\omega \nonumber \\
&\qquad+2\sigma\omega/r)(\sigma_{r}^{2}-\sigma_{t}^{2}+\omega_{r}^{2}-\omega_{t}^{2})+(\sigma^{2}+\omega^{2})(\sigma_{t}\omega_{tr}+\sigma_{tr}\omega_{t}-\sigma_{r}\omega_{rr}-\sigma_{rr}\omega_{r}) \nonumber \\
&\qquad+2\sigma\omega(\sigma_{r}\sigma_{rr}-\sigma_{t}\sigma_{tr}+\omega_{r}\omega_{rr}-\omega_{t}\omega_{tr})]\}+(\sigma^{2}-\omega^{2})^{-1}\{m(\sigma_{rr}j_{a}+2\sigma_{r}j_{a,r} \nonumber \\
&\qquad+\sigma j_{a,rr}+r\sigma_{tr}j_{b}+r\sigma_{t}j_{b,r}+r\sigma_{r}j_{b,t}+r\sigma j_{b,tr}+3\sigma_{t}j_{b}+3\sigma j_{b,t}+2\sigma_{r}j_{a}/r \nonumber \\
&\qquad+2\sigma j_{a,r}/r)\pm\rmi[(1/2)(\sigma\omega_{rrr}+\sigma_{r}\omega_{rr}-\sigma_{rr}\omega_{r}-\sigma_{rrr}\omega-\sigma\omega_{ttr}-\sigma_{r}\omega_{tt} \nonumber \\
&\qquad+\sigma_{tt}\omega_{r}+\sigma_{ttr}\omega)-(1/r)(\sigma\omega_{tt}-\sigma_{tt}\omega-\sigma\omega_{rr}+\sigma_{rr}\omega)]\},
\end{align}
and the second is
\begin{align}\label{Spherical symmetry, second MD equation}
& q^{2}rj_{b}=(\sigma^{2}-\omega^{2})^{-3}4(\sigma\sigma_{t}-\omega\omega_{t})\{-2m[\sigma j_{a}(\sigma\sigma_{r}-\omega\omega_{r})+r\sigma j_{b}(\sigma\sigma_{t}-\omega\omega_{t})] \nonumber \\
&\qquad\pm\rmi[(\sigma^{2}+\omega^{2})(\sigma_{t}\omega_{t}-\sigma_{r}\omega_{r})+\sigma\omega(\sigma_{r}^{2}-\sigma_{t}^{2}+\omega_{r}^{2}-\omega_{t}^{2})]\} \nonumber \\
&\qquad-(\sigma^{2}-\omega^{2})^{-2}\{-2m[(\sigma j_{a,r}+\sigma_{r}j_{a}+2r\sigma_{t}j_{b}+2r\sigma j_{b,t})(\sigma\sigma_{t}-\omega\omega_{t}) \nonumber \\
&\qquad+(\sigma j_{a,t}+\sigma_{t}j_{a})(\sigma\sigma_{r}-\omega\omega_{r})+\sigma j_{a}(\sigma\sigma_{tr}+\sigma_{t}\sigma_{r}-\omega\omega_{tr}-\omega_{t}\omega_{r})+r\sigma j_{b}(\sigma_{t}^{2} \nonumber \\
&\qquad+\sigma\sigma_{tt}-\omega_{t}^{2}-\omega\omega_{tt})]\pm\rmi[(\sigma_{rr}\omega-\sigma\omega_{rr}-\sigma_{tt}\omega+\sigma\omega_{tt})(\sigma\sigma_{t}-\omega\omega_{t}) \nonumber \\
&\qquad+2(\sigma\sigma_{t}+\omega\omega_{t})(\sigma_{t}\omega_{t}-\sigma_{r}\omega_{r})+(\sigma_{t}\omega+\sigma\omega_{t})(\sigma_{r}^{2}-\sigma_{t}^{2}+\omega_{r}^{2}-\omega_{t}^{2}) \nonumber \\
&\qquad+(\sigma^{2}+\omega^{2})(\sigma_{tt}\omega_{t}+\sigma_{t}\omega_{tt}-\sigma_{tr}\omega_{r}-\sigma_{r}\omega_{tr})+2\sigma\omega(\sigma_{tr}\sigma_{r}-\sigma_{tt}\sigma_{t}+\omega_{tr}\omega_{r} \nonumber \\
&\qquad-\omega_{tt}\omega_{t})]\}-(\sigma^{2}-\omega^{2})^{-1}[m(\sigma_{tr}j_{a}+\sigma_{t}j_{a,r}+\sigma_{r}j_{a,t}+\sigma j_{a,tr}+r\sigma_{tt}j_{b} \nonumber \\
&\qquad+2r\sigma_{t}j_{b,t}+r\sigma j_{b,tt})\pm(\rmi/2)(\sigma_{ttt}\omega+\sigma_{tt}\omega_{t}-\sigma_{t}\omega_{tt}-\sigma\omega_{ttt}-\sigma_{trr}\omega-\sigma_{rr}\omega_{t} \nonumber \\
&\qquad+\sigma_{t}\omega_{rr}+\sigma\omega_{trr})].
\end{align}
Note that we still have the freedom to eliminate another field, by applying the Fierz identity
\begin{equation}\label{Spherical symmetry, unused Fierz identity}
j_{a}^{2}-r^{2}j_{b}^{2}=\sigma^{2}-\omega^{2}.
\end{equation}
Additionally, applying spherical symmetry to (\ref{Continuity equation}) and (\ref{Partial conservation of axial current}) we have the two respective physical constraint equations
\begin{align}
&j_{a,t}+3j_{b}+rj_{b,r}=0,\label{Spherical symmetry, continuity equation} \\
&rj_{b,t}+(2/r)j_{a}+j_{a,r}=\mp2\rmi m\omega.\label{Spherical symmetry, partial conservation of axial current equation}
\end{align}

\subsection{Cylindrical symmetry (subgroup $P_{12,8}$)}
\subsubsection{Fierz identities} From (\ref{Cylindrically symmetric four vector field}), $j^{\mu}$ and $k^{\mu}$ take the form
\begin{equation}\label{Cylindrically symmetric j-mu}
j^{\mu}=\left(\begin{array}{c}j_{a} \\
xj_{b}-yj_{c} \\
yj_{b}+xj_{c} \\
j_{d}\end{array}\right),\ \ \ k^{\mu}=\left(\begin{array}{c}k_{a} \\
xk_{b}-yk_{c} \\
yk_{b}+xk_{c} \\
k_{d}\end{array}\right),
\end{equation}
where $j_{a}$, $j_{b}$, etc. are functions of $t$ and $\rho=\sqrt{x^{2}+y^{2}}$. From (\ref{Fierz inner product identity}), these forms imply
\begin{equation}\label{Cylindrical inner product Fierz identity}
j_{a}^{2}-\rho^{2}(j_{b}^{2}+j_{c}^{2})-j_{d}^{2}=-k_{a}^{2}+\rho^{2}(k_{b}^{2}+k_{c}^{2})+k_{d}^{2}=\sigma^{2}-\omega^{2},
\end{equation}
where $\sigma$ and $\omega$ are both functions of $t$ and $\rho$. Additionally, from (\ref{Fierz orthogonality identity}) we have
\begin{equation}\label{Cylindrical orthogonality Fierz identity}
j_{a}k_{a}-\rho^{2}(j_{b}k_{b}+j_{c}k_{c})-j_{d}k_{d}=0.
\end{equation}
In the cylindrical case, there are four dependent functions in each four vector field (as opposed to two in the spherical case), so we can arbitrarily choose to eliminate two of them using the Fierz identities. Let us choose to solve for $k_{a}$ and $k_{d}$ as a single example. Rearranging (\ref{Cylindrical orthogonality Fierz identity}) gives
\begin{equation}\label{Cylindrical orthogonality Fierz identity rearranged for ka}
k_{a}=[\rho^{2}(j_{b}k_{b}+j_{c}k_{c})+j_{d}k_{d}]j_{a}^{-1}.
\end{equation}
Substituting this into (\ref{Cylindrical inner product Fierz identity}) and rearranging, we obtain the quadratic expression for $k_{d}$
\begin{align}
&(j_{d}^{2}-j_{a}^{2})k_{d}^{2}+2\rho^{2}j_{d}(j_{b}k_{b}+j_{c}k_{c})k_{d}+[j_{a}^{4}-\rho^{2}j_{a}^{2}(j_{b}^{2}+j_{c}^{2}+k_{b}^{2}+k_{c}^{2}) \nonumber \\
&\qquad+\rho^{4}(j_{b}k_{b}+j_{c}k_{c})^{2}-j_{a}^{2}j_{d}^{2}]=0,
\end{align}
which has the solution according to the quadratic formula
\begin{align}
& k_{d}=(-\rho^{2}j_{d}(j_{b}k_{b}+j_{c}k_{c})\pm\{\rho^{4}j_{d}^{2}(j_{b}k_{b}+j_{c}k_{c})^{2}-(j_{d}^{2}-j_{a}^{2})[j_{a}^{4}-\rho^{2}j_{a}^{2}(j_{b}^{2}+j_{c}^{2} \nonumber \\
&\qquad+k_{b}^{2}+k_{c}^{2})+\rho^{4}(j_{b}k_{b}+j_{c}k_{c})^{2}-j_{a}^{2}j_{d}^{2}]\}^{1/2})(j_{d}^{2}-j_{a}^{2})^{-1},
\end{align}
after canceling out the factor of 2. Algebraic manipulation of the square root argument yields the simpler form
\begin{align}
& k_{d}=\{-\rho^{2}j_{d}(j_{b}k_{b}+j_{c}k_{c})\pm j_{a}[(j_{a}^{2}-j_{d}^{2})^{2}-\rho^{2}(j_{a}^{2}-j_{d}^{2})(j_{b}^{2}+j_{c}^{2}+k_{b}^{2}+k_{c}^{2}) \nonumber \\
&\qquad+\rho^{4}(j_{b}k_{b}+j_{c}k_{c})^{2}]^{1/2}\}(j_{d}^{2}-j_{a}^{2})^{-1},
\end{align}
which can be substituted into (\ref{Cylindrical orthogonality Fierz identity rearranged for ka}) to give
\begin{align}
& k_{a}=\{-\rho^{2}j_{a}(j_{b}k_{b}+j_{c}k_{c})\pm j_{d}[(j_{a}^{2}-j_{d}^{2})^{2}-\rho^{2}(j_{a}^{2}-j_{d}^{2})(j_{b}^{2}+j_{c}^{2}+k_{b}^{2}+k_{c}^{2})\nonumber \\
&\qquad+\rho^{4}(j_{b}k_{b}+j_{c}k_{c})^{2}]^{1/2}\}(j_{d}^{2}-j_{a}^{2})^{-1}.
\end{align}
Unlike the spherical case, the Fierz identities do not provide a tidy replacement of the components of $k^{\mu}$, so in our calculation of the reduced Maxwell-Dirac equations, we will retain all of the $k^{\mu}$ dependent functions with the implicit understanding that two of them can be eliminated. 

\subsubsection{Vector potential} As in the spherical case, we will look at each term in the numerator of in first rational term in (\ref{Gauge invariant vector potential}) separately, then take the sum. Unlike in the spherical case, we do not show any of the calculation steps, as they are too lengthy, but not difficult. The four components of the vector potential are
\begin{subequations}
\begin{align}
&B^{0}=\{[\rho(j_{c,\rho}k_{d}+j_{c}k_{d,\rho}-j_{d,\rho}k_{c}-j_{d}k_{c,\rho})+2(j_{c}k_{d}-j_{d}k_{c})-2m\sigma j_{a}](\sigma^{2}-\omega^{2}) \nonumber \\
&\qquad+\rho(j_{d}k_{c}-j_{c}k_{d})(\sigma\sigma_{\rho}-\omega\omega_{\rho})+\rmi\rho(j_{a}k_{b}-j_{b}k_{a})(\sigma_{\rho}\omega-\sigma\omega_{\rho})\} \nonumber \\
&\qquad\cdot[2q(\sigma^{2}-\omega^{2})^{2}]^{-1}, \\
&B^{1}=x[(j_{d,t}k_{c}+j_{d}k_{c,t}-j_{c,t}k_{d}-j_{c}k_{d,t}-2m\sigma j_{b})(\sigma^{2}-\omega^{2}) \nonumber \\
&\qquad+(j_{c}k_{d}-j_{d}k_{c})(\sigma\sigma_{t}-\omega\omega_{t})+\rmi(j_{b}k_{a}-j_{a}k_{b})(\sigma_{t}\omega-\sigma\omega_{t})][2q(\sigma^{2}-\omega^{2})^{2}]^{-1} \nonumber \\
&\qquad-y\{[j_{b,t}k_{d}+j_{b}k_{d,t}-j_{d,t}k_{b}-j_{d}k_{b,t}+(1/\rho)(j_{a,\rho}k_{d}+j_{a}k_{d,\rho}-j_{d,\rho}k_{a} \nonumber \\
&\qquad-j_{d}k_{a,\rho})-2m\sigma j_{c}](\sigma^{2}-\omega^{2})+(j_{d}k_{b}-j_{b}k_{d})(\sigma\sigma_{t}-\omega\omega_{t}) \nonumber \\
&\qquad+(1/\rho)(j_{d}k_{a}-j_{a}k_{d})(\sigma\sigma_{\rho}-\omega\omega_{\rho})+\rmi(j_{c}k_{a}-j_{a}k_{c})(\sigma_{t}\omega-\sigma\omega_{t}) \nonumber \\
&\qquad+\rmi\rho(j_{c}k_{b}-j_{b}k_{c})(\sigma_{\rho}\omega-\sigma\omega_{\rho})\}[2q(\sigma^{2}-\omega^{2})^{2}]^{-1}, \\
&B^{2}=y[(j_{d,t}k_{c}+j_{d}k_{c,t}-j_{c,t}k_{d}-j_{c}k_{d,t}-2m\sigma j_{b})(\sigma^{2}-\omega^{2}) \nonumber \\
&\qquad+(j_{c}k_{d}-j_{d}k_{c})(\sigma\sigma_{t}-\omega\omega_{t})+\rmi(j_{b}k_{a}-j_{a}k_{b})(\sigma_{t}\omega-\sigma\omega_{t})][2q(\sigma^{2}-\omega^{2})^{2}]^{-1} \nonumber \\
&\qquad+x\{[j_{b,t}k_{d}+j_{b}k_{d,t}-j_{d,t}k_{b}-j_{d}k_{b,t}+(1/\rho)(j_{a,\rho}k_{d}+j_{a}k_{d,\rho}-j_{d,\rho}k_{a} \nonumber \\
&\qquad-j_{d}k_{a,\rho})-2m\sigma j_{c}](\sigma^{2}-\omega^{2})+(j_{d}k_{b}-j_{b}k_{d})(\sigma\sigma_{t}-\omega\omega_{t}) \nonumber \\
&\qquad+(1/\rho)(j_{d}k_{a}-j_{a}k_{d})(\sigma\sigma_{\rho}-\omega\omega_{\rho})+\rmi(j_{c}k_{a}-j_{a}k_{c})(\sigma_{t}\omega-\sigma\omega_{t}) \nonumber \\
&\qquad+\rmi\rho(j_{c}k_{b}-j_{b}k_{c})(\sigma_{\rho}\omega-\sigma\omega_{\rho})\}[2q(\sigma^{2}-\omega^{2})^{2}]^{-1}, \\
&B^{3}=\{[\rho^{2}(j_{c,t}k_{b}+j_{c}k_{b,t}-j_{b,t}k_{c}-j_{b}k_{c,t})+\rho(j_{c,\rho}k_{a}+j_{c}k_{a,\rho}-j_{a,\rho}k_{c}-j_{a}k_{c,\rho}) \nonumber \\
&\qquad+2(j_{c}k_{a}-j_{a}k_{c})-2m\sigma j_{d}](\sigma^{2}-\omega^{2})+\rho^{2}(j_{b}k_{c}-j_{c}k_{b})(\sigma\sigma_{t}-\omega\omega_{t}) \nonumber \\
&\qquad+\rho(j_{a}k_{c}-j_{c}k_{a})(\sigma\sigma_{\rho}-\omega\omega_{\rho})+\rmi(j_{d}k_{a}-j_{a}k_{d})(\sigma_{t}\omega-\sigma\omega_{t}) \nonumber \\
&\qquad+\rmi\rho(j_{d}k_{b}-j_{b}k_{d})(\sigma_{\rho}\omega-\sigma\omega_{\rho})\}[2q(\sigma^{2}-\omega^{2})^{2}]^{-1}.
\end{align}
\end{subequations}
From inspection, we can see that the gauge invariant vector potential assumes the correct form for a cylindrically symmetric four vector field
\begin{equation}\label{Cylindrically symmetric vector potential}
B^{\mu}=\left(\begin{array}{c}B_{a} \\
xB_{b}-yB_{c} \\
yB_{b}+xB_{c} \\
B_{d}\end{array}\right),
\end{equation}
where $B_{a}$, etc. are functions of the invariants $t$ and $\rho$, and are defined as
\begin{subequations}
\begin{align}
& B_{a}=\{[\rho(j_{c,\rho}k_{d}+j_{c}k_{d,\rho}-j_{d,\rho}k_{c}-j_{d}k_{c,\rho})+2(j_{c}k_{d}-j_{d}k_{c})-2m\sigma j_{a}](\sigma^{2}-\omega^{2}) \nonumber \\
&\qquad+\rho(j_{d}k_{c}-j_{c}k_{d})(\sigma\sigma_{\rho}-\omega\omega_{\rho})+\rmi\rho(j_{a}k_{b}-j_{b}k_{a})(\sigma_{\rho}\omega-\sigma\omega_{\rho})\} \nonumber \\
&\qquad\cdot[2q(\sigma^{2}-\omega^{2})^{2}]^{-1},\label{Cylindrical symmetry, Ba} \\
& B_{b}=[(j_{d,t}k_{c}+j_{d}k_{c,t}-j_{c,t}k_{d}-j_{c}k_{d,t}-2m\sigma j_{b})(\sigma^{2}-\omega^{2}) \nonumber \\
&\qquad+(j_{c}k_{d}-j_{d}k_{c})(\sigma\sigma_{t}-\omega\omega_{t})+\rmi(j_{b}k_{a}-j_{a}k_{b})(\sigma_{t}\omega-\sigma\omega_{t})][2q(\sigma^{2}-\omega^{2})^{2}]^{-1},\label{Cylindrical symmetry, Bb} \\
& B_{c}=\{[j_{b,t}k_{d}+j_{b}k_{d,t}-j_{d,t}k_{b}-j_{d}k_{b,t}+(1/\rho)(j_{a,\rho}k_{d}+j_{a}k_{d,\rho}-j_{d,\rho}k_{a}-j_{d}k_{a,\rho}) \nonumber \\
&\qquad-2m\sigma j_{c}](\sigma^{2}-\omega^{2})+(j_{d}k_{b}-j_{b}k_{d})(\sigma\sigma_{t}-\omega\omega_{t})+(1/\rho)(j_{d}k_{a} \nonumber \\
&\qquad-j_{a}k_{d})(\sigma\sigma_{\rho}-\omega\omega_{\rho})+\rmi(j_{c}k_{a}-j_{a}k_{c})(\sigma_{t}\omega-\sigma\omega_{t})+\rmi\rho(j_{c}k_{b}-j_{b}k_{c})(\sigma_{\rho}\omega \nonumber \\
&\qquad-\sigma\omega_{\rho})\}[2q(\sigma^{2}-\omega^{2})^{2}]^{-1},\label{Cylindrical symmetry, Bc} \\
& B_{d}=\{[\rho^{2}(j_{c,t}k_{b}+j_{c}k_{b,t}-j_{b,t}k_{c}-j_{b}k_{c,t})+\rho(j_{c,\rho}k_{a}+j_{c}k_{a,\rho}-j_{a,\rho}k_{c}-j_{a}k_{c,\rho}) \nonumber \\
&\qquad+2(j_{c}k_{a}-j_{a}k_{c})-2m\sigma j_{d}](\sigma^{2}-\omega^{2})+\rho^{2}(j_{b}k_{c}-j_{c}k_{b})(\sigma\sigma_{t}-\omega\omega_{t}) \nonumber \\
&\qquad+\rho(j_{a}k_{c}-j_{c}k_{a})(\sigma\sigma_{\rho}-\omega\omega_{\rho})+\rmi(j_{d}k_{a}-j_{a}k_{d})(\sigma_{t}\omega-\sigma\omega_{t}) \nonumber \\
&\qquad+\rmi\rho(j_{d}k_{b}-j_{b}k_{d})(\sigma_{\rho}\omega-\sigma\omega_{\rho})\}[2q(\sigma^{2}-\omega^{2})^{2}]^{-1}.\label{Cylindrical symmetry, Bd}
\end{align}
\end{subequations}

\subsubsection{Field strength tensor} We approach the calculation of $F_{\mu\nu}$ as we did the vector potential, in that we calculate the four-curl of $B^{\mu}$ and the rational term in (\ref{Manifestly gauge invariant field strength tensor}) separately, then sum them together. Obviously, these calculations would be very time consuming to do by hand, so we use Mathematica to carry out the expansions, and manually factorizing. Due to the size of these expressions, their explicit form in terms of $j^{\mu}$, $k^{\mu}$, $\sigma$ and $\omega$ are relegated to appendix D. We can calculate the form of the four curl of $B^{\mu}$ by substituting (\ref{Cylindrically symmetric vector potential}) into (\ref{Manifestly gauge invariant field strength tensor}), which can in turn be expressed in terms of $j^{\mu}$, etc. by applying (\ref{Cylindrical symmetry, Ba})-(\ref{Cylindrical symmetry, Bd}). The $\mu=0$, $\nu=1$ four curl term is
\begin{equation}
\partial_{0}B_{1}-\partial_{1}B_{0}=\partial_{t}(-xB_{b}+yB_{c})-\partial_{x}B_{a}=-x[\partial_{t}B_{b}+(1/\rho)\partial_{\rho}B_{a}]+y\partial_{t}B_{c},
\end{equation}
and if we take account of the rational term, we find that
\begin{equation}
F_{01}=-xF_{a}+yF_{b},
\end{equation}
where $F_{a}$, $F_{b}$, etc., are functions of $t$ and $\rho$. The $\mu=0$, $\nu=2$ four curl term is
\begin{equation}
\partial_{0}B_{2}-\partial_{2}B_{0}=\partial_{t}(-yB_{b}-xB_{c})-\partial_{y}B_{a}=-y[\partial_{t}B_{b}+(1/\rho)\partial_{\rho}B_{a}]-x\partial_{t}B_{c},
\end{equation}
which when including the rational term, gives
\begin{equation}
F_{02}=-yF_{a}-xF_{b}.
\end{equation}
Now, considering the form of the rational term in $F_{03}$, we can see that $\partial_{3}j_{\sigma}$ and $\partial_{3}k_{\sigma}$ cause the entire term to vanish, due to $z$-translation invariance. The $\partial_{3}B_{0}$ part of the four curl vanishes for the same reason, so we are left with
\begin{equation}
F_{03}=\partial_{0}B_{3}=\partial_{t}B_{d}=-F_{c}.
\end{equation}
The $\mu=1$, $\nu=2$ four curl term is
\begin{equation}
\partial_{1}B_{2}-\partial_{2}B_{1}=\partial_{x}(-yB_{b}-xB_{c})-\partial_{y}(-xB_{b}+yB_{c})=-2B_{c}-\rho\partial_{\rho}B_{c},
\end{equation}
which when including the rational term, is of the form
\begin{equation}
F_{12}=F_{d}.
\end{equation}
The rational terms in $F_{13}$ and $F_{23}$ are both zero for the same reason as in $F_{03}$, so we are just left with the four curl term in both cases. In the $\mu=1$, $\nu=3$ case we have
\begin{equation}
F_{13}=\partial_{1}B_{3}=(x/\rho)\partial_{\rho}B_{d}=xF_{e},
\end{equation}
and when $\mu=2$, $\nu=3$, we have
\begin{equation}
F_{23}=\partial_{2}B_{3}=(y/\rho)\partial_{\rho}B_{d}=yF_{e}.
\end{equation}
So our field strength tensor form for cylindrical symmetry is
\begin{equation}\label{Cylindrical symmetry, field strength tensor}
F_{\mu\nu}=\left(\begin{array}{cccc}0 & -xF_{a}+yF_{b} & -yF_{a}-xF_{b} & -F_{c} \\
xF_{a}-yF_{b} & 0 & F_{d} & xF_{e} \\
yF_{a}+xF_{b} & -F_{d} & 0 & yF_{e} \\
F_{c} & -xF_{e} & -yF_{e} & 0\end{array}\right),
\end{equation}
where the forms of $F_{a}$, etc. in terms of $j^{\mu}$, $k^{\mu}$, $\sigma$ and $\omega$ are given in (\ref{Cylindrical symmetry appendix, Fa})-(\ref{Cylindrical symmetry appendix, Fe}).

\subsubsection{Maxwell equations}
Substituting (\ref{Cylindrically symmetric j-mu}) and (\ref{Cylindrical symmetry, field strength tensor}) into (\ref{Maxwell equations}), it is easy to obtain the four equations
\begin{subequations}
\begin{align}
&qj_{a}=2F_{a}+\rho\partial_{\rho}F_{a},\label{Cylindrical symmetry first Maxwell equation} \\
&q(xj_{b}-yj_{c})=x(-\partial_{t}F_{a})-y[-\partial_{t}F_{b}+(1/\rho)\partial_{\rho}F_{d}],\label{Cylindrical mu=1 Maxwell equation before algebraic manipulation} \\
&q(yj_{b}+xj_{c})=y(-\partial_{t}F_{a})+x[-\partial_{t}F_{b}+(1/\rho)\partial_{\rho}F_{d}],\label{Cylindrical mu=2 Maxwell equation before algebraic manipulation} \\
&qj_{d}=2F_{e}-\partial_{t}F_{c}+\rho\partial_{\rho}F_{e}.\label{Cylindrical symmetry second Maxwell equation}
\end{align}
\end{subequations}
If we multiply (\ref{Cylindrical mu=1 Maxwell equation before algebraic manipulation}) by $x$ and add (\ref{Cylindrical mu=2 Maxwell equation before algebraic manipulation}) multiplied by $y$, we obtain
\begin{equation}\label{Cylindrical symmetry third Maxwell equation}
qj_{b}=-\partial_{t}F_{a}.
\end{equation}
Likewise, if we take the combination $x$ times (\ref{Cylindrical mu=2 Maxwell equation before algebraic manipulation}) and subtract $y$ times (\ref{Cylindrical mu=1 Maxwell equation before algebraic manipulation}), we get
\begin{equation}\label{Cylindrical symmetry fourth Maxwell equation}
qj_{c}=-\partial_{t}F_{b}+(1/\rho)\partial_{\rho}F_{d}.
\end{equation}
The Maxwell equations for cylindrical symmetry therefore reduces to the set (\ref{Cylindrical symmetry first Maxwell equation}), (\ref{Cylindrical symmetry second Maxwell equation}), (\ref{Cylindrical symmetry third Maxwell equation}) and (\ref{Cylindrical symmetry fourth Maxwell equation}), equations dependent only on functions of $t$ and $\rho$. These expressions are far too long to write explicitly, even in the appendix, but the full Maxwell-Dirac equations can be obtained simply by substituting in the expressions (\ref{Cylindrical symmetry appendix, Fa})-(\ref{Cylindrical symmetry appendix, Fe}) from appendix D. In addition to the four Maxwell-Dirac equations, we have the three equations provided by the Fierz identities (\ref{Cylindrical inner product Fierz identity}) and (\ref{Cylindrical orthogonality Fierz identity}), as well as the continuity equations
\begin{subequations}
\begin{align}
&2j_{b}+\partial_{t}j_{a}+\rho\partial_{\rho}j_{b}=0,\label{Cylindrical symmetry, continuity equation} \\
&2k_{b}+\partial_{t}k_{a}+\rho\partial_{\rho}k_{b}=-2\rmi m\omega,\label{Cylindrical symmetry, partial conservation of axial current equation}
\end{align}
\end{subequations}
obtained by applying the cylindrical four vector forms to (\ref{Continuity equation}) and (\ref{Partial conservation of axial current}).

\subsection{$P_{11,2}$ symmetry (``screw'' subgroup)}
\subsubsection{Fierz identities}
From (\ref{P-11,2 invariant four vector field}), $j^{\mu}$ and $k^{\mu}$ have the form
\begin{equation}
j^{\mu}=\left(\begin{array}{c}l_{+}j_{a} \\
0 \\
0 \\
l_{+}j_{a}\end{array}\right),\ \ \ k^{\mu}=\left(\begin{array}{c}l_{+}k_{a} \\
0 \\
0 \\
l_{+}k_{a}\end{array}\right),
\end{equation}
where $j_{a}$ and $k_{a}$ are constants. Scalar fields $\sigma$ and $\omega$ are also constants. Applying these forms to the inner product Fierz identity (\ref{Fierz inner product identity}) results in the expression
\begin{equation}\label{P_11,2 inner product Fierz identity}
\sigma^{2}-\omega^{2}=0,
\end{equation}
because our $P_{11,2}$ invariant four vectors are null. The orthogonality condition (\ref{Fierz orthogonality identity}) results in $0=0$, providing no further information.

\subsubsection{Vector potential}
Ignoring the fact that $\sigma^{2}-\omega^{2}=0$ for the moment, we shall carry out the Maxwell-Dirac reduction to check what happens. Consider the gauge invariant vector potential (\ref{Gauge invariant vector potential}). Since $\sigma$ and $\omega$ are constants, all of the derivative terms involving them vanish. The $\epsilon^{\mu\nu\rho\sigma}\partial_{\nu}(j_{\rho}k_{\sigma})$ term also vanishes due to antisymmetry because the only non-zero derivatives are for $\nu=0$ when $\rho=0$, $\sigma=3$ and vice-versa. We are left with
\begin{equation}
B^{\mu}=-\frac{1}{q}\frac{m\sigma j^{\mu}}{\sigma^{2}-\omega^{2}},
\end{equation}
which in explicit component form is
\begin{equation}
B^{\mu}=\left(\begin{array}{c}-l_{+}m\sigma j_{a}/q(\sigma^{2}-\omega^{2}) \\
0 \\
0 \\
-l_{+}m\sigma j_{a}/q(\sigma^{2}-\omega^{2})\end{array}\right).
\end{equation}

\subsubsection{Field strength tensor}
The rational term in (\ref{Manifestly gauge invariant field strength tensor}) vanishes due to the antisymmetry of $\epsilon^{\sigma\rho\kappa\tau}$ and the fact that there are only two non-zero $B^{\mu}$ components, so we are just left with the four-curl term. The only non-zero component is $F_{03}=-F_{30}$, which is
\begin{equation}
F_{03}=\partial_{t}B_{3}-\partial_{z}B_{0}=2m\sigma j_{a}/q(\sigma^{2}-\omega^{2}),
\end{equation}
a constant term. Since the field strength tensor is constant the left-hand side of the Maxwell equations (\ref{Maxwell equations}) vanishes, leaving us with the result
\begin{equation}
j_{a}=0.
\end{equation}

\subsubsection{Continuity equations}
Now consider the two continuity equations (\ref{Continuity equation}) and (\ref{Partial conservation of axial current}). Applying our $P_{11,2}$ invariant forms, we find that
\begin{subequations}
\begin{align}
&j_{a}=0, \\
&k_{a}=-\rmi m\omega,
\end{align}
\end{subequations}
the first equation confirming our Maxwell-Dirac result. Let us consider the result $\sigma^{2}-\omega^{2}=0$ more closely. This can be rearranged to give $\sigma=\pm\omega$, but since $\omega$ is pure imaginary and $\sigma$ is real, the only case in which they can be equal is when they are both zero. This in turn means that $k_{a}=0$ also. We have thus obtained a closed form solution to the Maxwell-Dirac equations under $P_{11,2}$ symmetry, which unfortunately is constrained to be the trivial solution
\begin{align}
&\sigma=\omega=0, \nonumber \\
&j^{\mu}=k^{\mu}=\underline{0}.\label{P_11,2 Maxwell-Dirac solution}
\end{align}
Note that this solution was obtained using only the Fierz identities and continuity equations; it is unnecessary to deal with the full Maxwell-Dirac equations in this case.

\subsection{$\widetilde{P}_{13,10}$ symmetry}
\subsubsection{Fierz identities}
From (\ref{P_13,10 invariant four vector field}), the $\widetilde{P}_{13,10}$ invariant form of $j^{\mu}$ is
\begin{equation}
j^{\mu}=\left(\begin{array}{c}(l_{+}/|l_{+}|)\rme^{-2y/\lambda}j_{a}+(l_{-}/|l_{-}|)\rme^{2y/\lambda}j_{b} \\
j_{c} \\
j_{d} \\
(l_{+}/|l_{+}|)\rme^{-2y/\lambda}j_{a}-(l_{-}/|l_{-}|)\rme^{2y/\lambda}j_{b}\end{array}\right),
\end{equation}
where $j_{a}$, $j_{b}$, etc. are constants. The axial four vector $k^{\mu}$ has the same form, but with $k_{a}$ replacing $j_{a}$, and so on. Remember that $\lambda>0$ is a continuous parameter associated with the $\widetilde{B}_{\lambda}$ generator, with each value representing a different symmetry. Applying the symmetric forms of $j^{\mu}$ and $k^{\mu}$ to the Fierz identities (\ref{Fierz inner product identity}) and (\ref{Fierz orthogonality identity}) gives
\begin{align}
&4(L/|L|)j_{a}j_{b}-j_{c}^{2}-j_{d}^{2}=-4(L/|L|)k_{a}k_{b}+k_{c}^{2}+k_{d}^{2}=\sigma^{2}-\omega^{2},\label{P-13,10 inner product Fierz identity} \\
&2(L/|L|)(j_{a}k_{b}+j_{b}k_{a})-j_{c}k_{c}-j_{d}k_{d}=0,\label{P-13,10 orthogonality Fierz identity}
\end{align}
where $L\equiv l_{+}l_{-}$ Unlike previous examples, we will not apply these immediately to replace elements of $k^{\mu}$, but wait until the Maxwell-Dirac system has been obtained.

\subsubsection{Vector potential}
Since $\sigma$ and $\omega$ are constants, the derivatives of these objects in (\ref{Gauge invariant vector potential}) vanish, leaving us with
\begin{equation}\label{Vector potential, scalar field derivatives zero}
B^{\mu}=\frac{\epsilon^{\mu\nu\rho\sigma}\partial_{\nu}(j_{\rho}k_{\sigma})-2m\sigma j^{\mu}}{2q(\sigma^{2}-\omega^{2})}.
\end{equation}
The four-vectors only vary in the $y-$direction, so the only non-zero derivatives are for $\nu=2$ when $\rho=0$, $\sigma=3$ or $\rho=3$, $\sigma=0$. Setting $\mu=0$, the first term in the numerator is
\begin{align}
&\epsilon^{02\rho\sigma}\partial_{2}(j_{\rho}k_{\sigma})=\epsilon^{0213}\partial_{2}(j_{1}k_{3})+\epsilon^{0231}\partial_{2}(j_{3}k_{1}) \nonumber \\
&\qquad=(l_{+}/|l_{+}|)(2/\lambda)\rme^{-2y/\lambda}(j_{c}k_{a}-j_{a}k_{c})+(l_{-}/|l_{-}|)(2/\lambda)\rme^{2y/\lambda}(j_{c}k_{b}-j_{b}k_{c}).
\end{align}
Substituting into (\ref{Vector potential, scalar field derivatives zero}) for $\mu=0$, we get
\begin{equation}
B^{0}=(l_{+}/|l_{+}|)\rme^{-2y/\lambda}B_{a}+(l_{-}/|l_{-}|)\rme^{2y/\lambda}B_{b},
\end{equation}
where $B_{a}$ and $B_{b}$ are the constants
\begin{align}
&B_{a}=\frac{(1/\lambda)(j_{c}k_{a}-j_{a}k_{c})-m\sigma j_{a}}{q(\sigma^{2}-\omega^{2})}, \\
&B_{b}=\frac{(1/\lambda)(j_{c}k_{b}-j_{b}k_{c})-m\sigma j_{b}}{q(\sigma^{2}-\omega^{2})}.
\end{align}
Setting $\mu=1$ and $\mu=2$, we find that the first numerator term in (\ref{Vector potential, scalar field derivatives zero}) vanishes in both cases, so these components are the constants
\begin{align}
&B^{1}=B_{c}=-\frac{m\sigma j_{c}}{q(\sigma^{2}-\omega^{2})}, \\
&B^{2}=B_{d}=-\frac{m\sigma j_{d}}{q(\sigma^{2}-\omega^{2})}.
\end{align}
Lastly, we have $\mu=3$. The first numerator term is
\begin{align}
&\epsilon^{32\rho\sigma}\partial_{2}(j_{\rho}k_{\sigma})=\epsilon^{3201}\partial_{2}(j_{0}k_{1})+\epsilon^{3210}\partial_{2}(j_{1}k_{0}) \nonumber \\
&\qquad=(l_{+}/|l_{+}|)(2/\lambda)\rme^{-2y/\lambda}(j_{c}k_{a}-j_{a}k_{c})-(l_{-}/|l_{-}|)(2/\lambda)\rme^{2y/\lambda}(j_{c}k_{b}-j_{b}k_{c}),
\end{align}
which when substituting into $B^{3}$ gives the final component
\begin{equation}
B^{3}=(l_{+}/|l_{+}|)\rme^{-2y/\lambda}B_{a}-(l_{-}/|l_{-}|)\rme^{2y/\lambda}B_{b}.
\end{equation}
We have determined that $B^{\mu}$ has the correct form for a $\widetilde{P}_{13,10}$ invariant four vector field
\begin{equation}
B^{\mu}=\left(\begin{array}{c}(l_{+}/|l_{+}|)\rme^{-2y/\lambda}B_{a}+(l_{-}/|l_{-}|)\rme^{2y/\lambda}B_{b} \\
B_{c} \\
B_{d} \\
(l_{+}/|l_{+}|)\rme^{-2y/\lambda}B_{a}-(l_{-}/|l_{-}|)\rme^{2y/\lambda}B_{b}\end{array}\right),
\end{equation}
where $B_{a}$, etc. are constants, as required.

\subsubsection{Field strength tensor}
Consider the rational term in (\ref{Manifestly gauge invariant field strength tensor}). Since the only non-constants in $j_{\sigma}$ and $k_{\sigma}$ are $\rme^{\pm2y/\lambda}$, the only non-vanishing derivative is $\partial_{2}\equiv\partial_{y}$. In general, the rational term must vanish, because $\mu\neq\nu$, so they cannot both be 2, and either derivative vanishing leads to the entire rational term vanishing. So the field strength tensor reduces to the four curl of $B^{\mu}$ for this symmetry group
\begin{equation}
F_{\mu\nu}=\partial_{\mu}B_{\nu}-\partial_{\nu}B_{\mu}.
\end{equation}
For similar reasons as above, the only non-vanishing elements of $F_{\mu\nu}$ are those where one of the indices is 2, and the other is either 0 or 3. Setting $\mu=0$, $\nu=2$, we get
\begin{equation}
F_{02}=\partial_{0}B_{2}-\partial_{2}B_{0}=(l_{+}/|l_{+}|)\rme^{-2y/\lambda}F_{a}-(l_{-}/|l_{-}|)\rme^{2y/\lambda}F_{b},
\end{equation}
where $F_{a}$ and $F_{b}$ are the constants
\begin{align}
&F_{a}=\frac{(2/\lambda)(j_{c}k_{a}-j_{a}k_{c})-2m\sigma j_{a}}{\lambda q(\sigma^{2}-\omega^{2})}, \\
&F_{b}=\frac{(2/\lambda)(j_{c}k_{b}-j_{b}k_{c})-2m\sigma j_{b}}{\lambda q(\sigma^{2}-\omega^{2})}.
\end{align}
Setting $\mu=2$, $\nu=3$, we get
\begin{equation}
F_{23}=\partial_{2}B_{3}-\partial_{3}B_{2}=(l_{+}/|l_{+}|)\rme^{-2y/\lambda}F_{a}+(l_{-}/|l_{-}|)\rme^{2y/\lambda}F_{b},
\end{equation}
giving us the only other independent non-zero component of $F_{\mu\nu}$.

\subsubsection{Maxwell equations}
Setting $\mu=0$ in (\ref{Maxwell equations}), we get
\begin{equation}
\partial_{1}F^{10}+\partial_{2}F^{20}+\partial_{3}F^{30}=q\left(\frac{l_{+}}{|l_{+}|}\rme^{-2y/\lambda}j_{a}+\frac{l_{-}}{|l_{-}|}\rme^{2y/\lambda}j_{b}\right).
\end{equation}
The only non-zero term on the left hand side is $\partial_{2}F^{20}=\partial_{y}F_{02}$, so carrying out the derivative and rearranging gives us
\begin{align}\label{P-13,10 mu=0 Maxwell equation}
&\frac{l_{+}}{|l_{+}|}\rme^{-2y/\lambda}\left\{\frac{4}{\lambda^{2}q(\sigma^{2}-\omega^{2})}\left[\frac{1}{\lambda}(j_{c}k_{a}-j_{a}k_{c})-m\sigma j_{a}\right]+qj_{a}\right\} \nonumber \\
&+\frac{l_{-}}{|l_{-}|}\rme^{2y/\lambda}\left\{\frac{4}{\lambda^{2}q(\sigma^{2}-\omega^{2})}\left[\frac{1}{\lambda}(j_{c}k_{b}-j_{b}k_{c})-m\sigma j_{b}\right]+qj_{b}\right\}=0.
\end{align}
Setting $\mu=1$ gives
\begin{equation}
\partial_{0}F^{01}+\partial_{2}F^{21}+\partial_{3}F^{31}=qj_{c}.
\end{equation}
Since the entire left hand side vanishes, we are left with
\begin{equation}\label{P-13,10 mu=1 Maxwell equation}
j_{c}=0.
\end{equation}
Setting $\mu=2$, we find for the same reason that
\begin{equation}\label{P-13,10 mu=2 Maxwell equation}
j_{d}=0.
\end{equation}
Lastly, setting $\mu=3$ gives
\begin{equation}
\partial_{0}F^{03}+\partial_{1}F^{13}+\partial_{2}F^{23}=q\left(\frac{l_{+}}{|l_{+}|}\rme^{-2y/\lambda}j_{a}-\frac{l_{-}}{|l_{-}|}\rme^{2y/\lambda}j_{b}\right).
\end{equation}
Recognizing that the only non-zero term on the left is $\partial_{2}F^{23}=\partial_{y}F_{23}$, we end up with
\begin{align}\label{P-13,10 mu=3 Maxwell equation}
&\frac{l_{+}}{|l_{+}|}\rme^{-2y/\lambda}\left\{\frac{4}{\lambda^{2}q(\sigma^{2}-\omega^{2})}\left[\frac{1}{\lambda}(j_{c}k_{a}-j_{a}k_{c})-m\sigma j_{a}\right]+qj_{a}\right\} \nonumber \\
&-\frac{l_{-}}{|l_{-}|}\rme^{2y/\lambda}\left\{\frac{4}{\lambda^{2}q(\sigma^{2}-\omega^{2})}\left[\frac{1}{\lambda}(j_{c}k_{b}-j_{b}k_{c})-m\sigma j_{b}\right]+qj_{b}\right\}=0.
\end{align}
Adding (\ref{P-13,10 mu=0 Maxwell equation}) and (\ref{P-13,10 mu=3 Maxwell equation}), discarding the non-zero coefficient and applying (\ref{P-13,10 mu=1 Maxwell equation}), we get
\begin{equation}
\frac{4}{\lambda^{2}q(\sigma^{2}-\omega^{2})}\left(-\frac{j_{a}k_{c}}{\lambda}-m\sigma j_{a}\right)+qj_{a}=0.
\end{equation}
Canceling the common factor $j_{a}$ and rearranging to solve for $k_{c}$ gives
\begin{equation}\label{P-13,10 kc Maxwell-Dirac equation}
k_{c}=\frac{\lambda^{3}q^{2}(\sigma^{2}-\omega^{2})}{4}-\lambda m\sigma.
\end{equation}
Subtracting (\ref{P-13,10 mu=3 Maxwell equation}) from (\ref{P-13,10 mu=0 Maxwell equation}), applying (\ref{P-13,10 mu=1 Maxwell equation}) then discarding the non-zero coefficient and the common factor $j_{b}$ gives exactly (\ref{P-13,10 kc Maxwell-Dirac equation}). Therefore, the three equations (\ref{P-13,10 mu=1 Maxwell equation}), (\ref{P-13,10 mu=2 Maxwell equation}) and (\ref{P-13,10 kc Maxwell-Dirac equation}) constitute the Maxwell-Dirac equations for the $\widetilde{P}_{13,10}$ subalgebra.

\subsubsection{Fierz-Maxwell-Dirac reduction}
From this point on, for simplicity we shall set all of the discontinuous factors $L/|L|=1$, choosing the positive sign. We can further simplify the $k_{c}$ expression if we take into account the partial conservation of the axial four vector (\ref{Partial conservation of axial current}). Since the only non-zero derivatives of $k^{\mu}$ are $\partial_{2}k^{0}$ and $\partial_{2}k^{3}$, the left hand side vanishes, so the consistency condition
\begin{equation}
\omega=0
\end{equation}
must hold. So (\ref{P-13,10 kc Maxwell-Dirac equation}) simplifies to an expression quadratic in $\sigma$
\begin{equation}\label{P-13,10 kc after setting omega=0}
k_{c}=\frac{\lambda^{3}q^{2}\sigma^{2}}{4}-\lambda m\sigma.
\end{equation}
Note that the continuity equation (\ref{Continuity equation}) gives the redundant expression $0=0$, as the left side vanishes for the same reasons as $\partial_{\mu}k^{\mu}$. Now consider the outer parts of the Fierz identity (\ref{P-13,10 inner product Fierz identity}). Setting $j_{c}=j_{d}=\omega=0$, this becomes
\begin{equation}
\sigma^{2}=4j_{a}j_{b},
\end{equation}
which when substituted into (\ref{P-13,10 kc after setting omega=0}) gives
\begin{equation}\label{P-13,10 kc after sigma replaced}
k_{c}=\lambda^{3}q^{2}j_{a}j_{b}\mp2\lambda m\sqrt{j_{a}j_{b}}.
\end{equation}
Additionally, looking at the left hand parts of (\ref{P-13,10 inner product Fierz identity}), we can see that setting $j_{c}=j_{d}=0$ gives us
\begin{equation}\label{P-13,10 Fierz inner product identity, left side after jc=jd=0}
-4j_{a}j_{b}-4k_{a}k_{b}+k_{c}^{2}+k_{d}^{2}=0.
\end{equation}
Setting $j_{c}=j_{d}=0$ in the orthogonality Fierz identity (\ref{P-13,10 orthogonality Fierz identity}) gives the additional relationship
\begin{equation}\label{P-13,10 reduced orthogonality Fierz identity}
j_{a}k_{b}=-j_{b}k_{a}.
\end{equation}
Substituting (\ref{P-13,10 kc after sigma replaced}) into (\ref{P-13,10 Fierz inner product identity, left side after jc=jd=0}) gives an algebraic equation of the form
\begin{equation}\label{P-13,10 reduced Maxwell-Dirac equation}
f(j_{a},j_{b},k_{a},k_{b},k_{d};\lambda)=0,
\end{equation}
where the function on the left is
\begin{equation}\label{P-13,10 reduced MD-Fierz equation}
f(j_{a},j_{b},k_{a},k_{b},k_{d};\lambda)=\lambda^{6}q^{4}(j_{a}j_{b})^{2}\mp4\lambda^{4}q^{2}m(j_{a}j_{b})^{3/2}+4(\lambda^{2}m^{2}-1)j_{a}j_{b}-4k_{a}k_{b}+k_{d}^{2}.
\end{equation}
Values of $j_{a}$, $j_{b}$, $k_{a}$, $k_{b}$ and $k_{d}$ which solve (\ref{P-13,10 reduced Maxwell-Dirac equation}) for a given $\lambda$ constitute solutions to the Fierz-Maxwell-Dirac equations, symmetric under the $\widetilde{P}_{13,10}$ Poincar\'{e} subalgebra. Note that we can still eliminate one of the constants by imposing (\ref{P-13,10 reduced orthogonality Fierz identity}).

\section{Conclusions}
In this study, we developed a manifestly gauge invariant tensor formalism for the Maxwell-Dirac system, then explored the reduction of this system under four different symmetry subalgebras of the Poincar\'{e} group.

Initially, a brief recap of the Dirac equation inversion for an Abelian gauge field $A^{\mu}$ was given, followed by a short discussion on Fierz identities and bilinear products of spinors. Next, the inverted Dirac equation was recast into a purely tensorial form, and a gauge-independent vector potential $B^{\mu}$ was defined by subtracting the gauge-dependent parts from $A^{\mu}$. Following a brief discussion on the tetrad of mutually orthogonal vector fields, the electromagnetic field strength tensor $F_{\mu\nu}$ was recast into a manifestly gauge invariant form, involving only $B^{\mu}$ and gauge independent tensors. The resulting set of tensorial, manifestly gauge invariant Maxwell-Dirac equations, Fierz identities and consistency conditions was summarized in the list (\ref{Fierz inner product identity})-(\ref{Partial conservation of axial current}). In the next section, the invariants and the forms of four vector fields for a given set of generators were calculated by finding solutions of the vanishing Lie derivative PDEs via the method of characteristics. Four example Poincar\'{e} subalgebras were studied in particular: the standard spherical and cylindrical symmetries, as well as the more unusual splitting $P_{11,2}$ and non-splitting $\widetilde{P}_{13,10}$ subalgebras of PWZ \cite{Patera-Winternitz-Zassenhaus-1975}.

The first two subalgebras reduced dependent variables to functions of two independent variables, whereas the last two reduced dependent variables to constants, because of transitive action on Minkowski space. The number of independent variables for the different symmetry subalgebras had a large impact on the complexity of the reduced Maxwell-Dirac system. For each subalgebra, the invariant forms were applied to the bilinear tensor fields $\sigma$, $\omega$, $j^{\mu}$ and $k^{\mu}$, then the reduced forms of the Fierz identities, gauge invariant vector potential, field strength tensor, and Maxwell equations were calculated, with the reduced set of Maxwell-Dirac equations and consistency conditions presented at the end of each subsection.

In the case of spherical symmetry, the Maxwell-Dirac equations took the form of two coupled PDEs (\ref{Spherical symmetry, first MD equation}) and (\ref{Spherical symmetry, second MD equation}) in terms of the dependent functions $j_{a}$, $j_{b}$, $\sigma$ and $\omega$, as well as their $t$ and $r$ derivatives up to third order. Additional information is provided in the spherically reduced Fierz identity (\ref{Spherical symmetry, unused Fierz identity}) and the continuity equations, (\ref{Spherical symmetry, continuity equation}) and (\ref{Spherical symmetry, partial conservation of axial current equation}). Interestingly, magnetic monopoles appear in the field strength tensor, but ultimately play no part in the coupled Maxwell-Dirac system.

Cylindrical symmetry resulted in a more complicated system, with the Maxwell-Dirac equations given implicitly by the set (\ref{Cylindrical symmetry first Maxwell equation}), (\ref{Cylindrical symmetry second Maxwell equation}), (\ref{Cylindrical symmetry third Maxwell equation}) and (\ref{Cylindrical symmetry fourth Maxwell equation}), where the field strength tensor dependent functions of $t$ and $\rho$ are given explicitly in appendix D. Five more equations are provided by the three Fierz identities (\ref{Cylindrical inner product Fierz identity}) and (\ref{Cylindrical orthogonality Fierz identity}), as well as the two continuity equations (\ref{Cylindrical symmetry, continuity equation}) and (\ref{Cylindrical symmetry, partial conservation of axial current equation}).

The first of the PWZ subalgebras, the splitting $P_{11,2}$, resulted in a strong reduction of the Maxwell-Dirac system, to the point where the only allowed solution was the trivial one (\ref{P_11,2 Maxwell-Dirac solution}), where all of the tensor fields are zero. This provided a good demonstration of the fact that if a symmetry is too restrictive, the only solutions are vanishing fields.

The last example was the non-splitting subalgebra $\widetilde{P}_{13,10}$, where the Maxwell-Dirac system, combined with the Fierz identities and continuity equations boiled down to the two algebraic equations (\ref{P-13,10 reduced orthogonality Fierz identity}) and (\ref{P-13,10 reduced Maxwell-Dirac equation}); solutions correspond to finding sets of constants $j_{a}$, $j_{b}$, $k_{a}$, $k_{b}$ and $k_{d}$ (where one other than $k_{d}$ can be eliminated) that solve this equation for a given type of symmetry, defined by the continuous parameter $\lambda>0$. Finding such a family of solutions to this equation is a matter for further study, and is an interesting problem, as it would potentially extend the types of known solutions beyond massless plane waves and solitons.

Finding numerical solutions to the spherical and cylindrical systems of equations presented in this paper would also be an interesting exercise, and could be compared to the numerical solutions of Radford and Booth \cite{Radford-1996}, \cite{Booth-Radford-1997}, \cite{Radford-Booth-1999}, which were performed in the two-spinor formalism. Maxwell-Dirac symmetry under the hyperbolic group $SO(2,1)$ would make a good comparison case in further studies of the $SO(3)$ spherical symmetry reduction, due to their algebraic similarity. In the longer term, an exhaustive study of the Maxwell-Dirac symmetry reductions, under the entire list of Poincar\'{e} subalgebras given by PWZ \cite{Patera-Winternitz-Zassenhaus-1975} could be done, as well as subsequent studies of their solutions.

\section*{Acknowledgements}
The authors wish to acknowledge the work of Graham Legg, whose Master's thesis \cite{Legg-2007} provided the inspiration for much of this work. This study was financially supported by the Australian Postgraduate Awards (APA) program.

\appendix

\section{Algebraic Identities}
Throughout this paper we use the Levi-Civita symbol, defined as
\begin{equation}\label{cases}
\epsilon^{\mu\nu\rho\sigma}=\begin{cases}+1&$if $\{\mu,\nu,\rho,\sigma\}$ is an even permutation of $\{0,1,2,3\} \\
-1&$if is an odd permutation$ \\
0&$otherwise$,\end{cases}
\end{equation}
with the additional property
\begin{equation}
\epsilon_{\mu\nu\rho\sigma}=\rmdet(\eta)\epsilon^{\mu\nu\rho\sigma}=-\epsilon^{\mu\nu\rho\sigma}.
\end{equation}
For convenience, we another rank-4 antisymmetric symbol, the shorthand for which is
\begin{equation}
\delta^{\mu\nu\rho\sigma}=\rmi(\eta^{\mu\rho}\eta^{\nu\sigma}-\eta^{\mu\sigma}\eta^{\nu\rho}).
\end{equation}

\subsection{Dirac Identities}
\begin{align}
&\{\gamma^{\mu},\gamma^{\nu}\}=2\eta^{\mu\nu} \\
&[\gamma^{\mu},\gamma^{\nu}]=-2\rmi\sigma^{\mu\nu} \\
&\gamma^{5}=\gamma_{5}=-(\rmi/4!)\epsilon_{\mu\nu\rho\sigma}\gamma^{\mu}\gamma^{\nu}\gamma^{\rho}\gamma^{\sigma}=\rmi\gamma^{0}\gamma^{1}\gamma^{2}\gamma^{3}=-\rmi\gamma_{0}\gamma_{1}\gamma_{2}\gamma_{3} \\
&\gamma_{5}^{2}=I \\
&\{\gamma_{5},\gamma^{\mu}\}=0 \\
&[\gamma_{5},\sigma^{\mu\nu}]=0 \\
&\gamma^{\mu}\gamma^{\nu}=\eta^{\mu\nu}-\rmi\sigma^{\mu\nu}\label{gamma-gamma Dirac identity} \\
&\gamma^{\mu}\gamma_{\mu}=4 \\
&\gamma^{\mu}\gamma_{5}\gamma_{\mu}=-4\gamma_{5} \\
&\gamma^{\mu}\gamma^{\nu}\gamma^{\lambda}=\eta^{\mu\nu}\gamma^{\lambda}+\eta^{\nu\lambda}\gamma^{\mu}-\eta^{\mu\lambda}\gamma^{\nu}-\rmi\epsilon^{\mu\nu\lambda\sigma}\gamma_{5}\gamma_{\sigma} \\
&\gamma^{\nu}\gamma^{\mu}\gamma_{\nu}=-2\gamma^{\mu} \\
&\gamma^{\nu}\gamma_{5}\gamma^{\mu}\gamma_{\nu}=2\gamma_{5}\gamma^{\mu} \\
&\gamma^{\mu}\gamma^{\nu}\gamma^{\sigma}\gamma^{\epsilon}=\eta^{\mu\nu}\eta^{\sigma\epsilon}+\eta^{\nu\sigma}\eta^{\mu\epsilon}-\eta^{\mu\sigma}\eta^{\nu\epsilon}-\rmi\eta^{\mu\nu}\sigma^{\sigma\epsilon}-\rmi\eta^{\nu\sigma}\sigma^{\mu\epsilon}+\rmi\eta^{\mu\sigma}\sigma^{\nu\epsilon}+\rmi\eta^{\mu\epsilon}\sigma^{\sigma\nu} \nonumber \\
&\qquad+\rmi\eta^{\nu\epsilon}\sigma^{\mu\sigma}+\rmi\eta^{\sigma\epsilon}\sigma^{\nu\mu}-\rmi\epsilon^{\mu\nu\sigma\epsilon}\gamma_{5} \\
&\gamma^{\epsilon}\sigma^{\mu\nu}=\rmi\eta^{\epsilon\mu}\gamma^{\nu}-\rmi\eta^{\epsilon\nu}\gamma^{\mu}+\epsilon^{\mu\nu\epsilon\sigma}\gamma_{5}\gamma_{\sigma} \\
&\sigma^{\mu\nu}\gamma^{\epsilon}=\rmi\eta^{\nu\epsilon}\gamma^{\mu}-\rmi\eta^{\mu\epsilon}\gamma^{\nu}+\epsilon^{\mu\nu\epsilon\sigma}\gamma_{5}\gamma_{\sigma} \\
&\gamma^{\mu}\sigma^{\sigma\epsilon}\gamma^{\nu}=\rmi\eta^{\epsilon\nu}\eta^{\mu\sigma}-\rmi\eta^{\sigma\nu}\eta^{\mu\epsilon}+\eta^{\epsilon\nu}\sigma^{\mu\sigma}-\eta^{\sigma\nu}\sigma^{\mu\epsilon}-\epsilon^{\sigma\epsilon\nu\mu}\gamma_{5}+\rmi\epsilon^{\sigma\epsilon\nu\lambda}\gamma_{5}\sigma^{\mu}{}_{\lambda} \\
&\gamma^{\sigma}\sigma^{\mu\nu}\gamma_{\sigma}=0 \\
&-\epsilon^{\lambda\rho\sigma\epsilon}\epsilon_{\lambda}{}^{\mu\nu\tau}=\eta^{\rho\mu}\eta^{\sigma\nu}\eta^{\epsilon\tau}-\eta^{\rho\mu}\eta^{\epsilon\nu}\eta^{\sigma\tau}+\eta^{\rho\nu}\eta^{\sigma\tau}\eta^{\epsilon\mu}-\eta^{\rho\nu}\eta^{\epsilon\tau}\eta^{\sigma\mu}+\eta^{\rho\tau}\eta^{\sigma\mu}\eta^{\epsilon\nu} \nonumber \\
&\qquad-\eta^{\rho\tau}\eta^{\epsilon\mu}\eta^{\sigma\nu}
\end{align}

\subsection{Charge conjugation identities}
\begin{align}
&C^{-1}\gamma_{\mu}^{\rmT}C=-\gamma_{\mu} \\
&C^{-1}\gamma_{5}^{\rmT}C=\gamma_{5} \\
&C^{-1}\sigma_{\mu\nu}^{\rmT}C=\sigma_{\mu\nu} \\
&C^{-1}(\gamma_{\mu}\gamma_{\nu})^{\rmT}C=\gamma_{\nu}\gamma_{\mu} \\
&C^{-1}(\gamma_{5}\gamma_{\mu})^{\rmT}C=\gamma_{5}\gamma_{\mu}
\end{align}
The relationship between a charge conjugate bilinear and a regular bilinear, for commuting spinor fields is
\begin{equation}\label{Charge conjugation bilinear identity}
\psibar{}^{\rmc}\Gamma\chi^{\rmc}=-\chibar C^{-1}\Gamma^{\rmT}C\psi,
\end{equation}
where $\Gamma$ is an element of the Dirac algebra. Some particular examples, using the above charge conjugation identities are
\begin{align}
&\psibar{}^{\rmc}\psi^{\rmc}=-\psibar\psi \\
&\psibar{}^{\rmc}\gamma_{\mu}\psi^{\rmc}=\psibar\gamma_{\mu}\psi \\
&\psibar{}^{\rmc}\sigma_{\mu\nu}\psi{}^{\rmc}=\psibar\sigma_{\mu\nu}\psi \\
&\psibar{}^{\rmc}\gamma_{\mu}\gamma^{\nu}(\partial_{\nu}\psi^{\rmc})=-(\partial_{\nu}\psibar)\gamma^{\nu}\gamma_{\mu}\psi \\
&\psibar{}^{\rmc}\psi=\psibar{}^{\rmc}\gamma_{5}\gamma_{\mu}\psi=\psibar{}^{\rmc}\gamma_{5}\psi=0 \\
\end{align}
The last identity is due to the fact these expressions equal their own negatives.

\subsection{Dirac bilinear notation}
The shorthand for gauge-independent Dirac bilinear tensors is as follows:
\begin{align}
&\sigma=\psibar\psi \\
&j^{\mu}=\psibar\gamma^{\mu}\psi \\
&s^{\mu\nu}=\psibar\sigma^{\mu\nu}\psi \\
&\sdual^{\mu\nu}=\psibar\gamma_{5}\sigma^{\mu\nu}\psi \\
&k^{\mu}=\psibar\gamma_{5}\gamma^{\mu}\psi \\
&\omega=\psibar\gamma_{5}\psi.
\end{align}
With regards to convention, note that some authors define the pseudoscalar bilinear to be $\omega\equiv\psibar\rmi\gamma_{5}\psi$. The gauge-dependent bilinear tensors are
\begin{align}
&m^{\mu}+\rmi n^{\mu}=\psibar{}^{\rmc}\gamma^{\mu}\psi \\
&m^{\mu}=\rmRe[\psibar{}^{\rmc}\gamma^{\mu}\psi]=(1/2)(\psibar{}^{\rmc}\gamma^{\mu}\psi+\psibar\gamma^{\mu}\psi^{\rmc}) \\
&n^{\mu}=\rmIm[\psibar{}^{\rmc}\gamma^{\mu}\psi]=(\rmi/2)(\psibar\gamma^{\mu}\psi^{\rmc}-\psibar{}^{\rmc}\gamma^{\mu}\psi).
\end{align}
The last two equations follow from the bilinear complex conjugation identity
\begin{equation}\label{Complex conjugation bilinear identity}
(\chibar\Gamma\psi)^{*}=\psibar(\gamma_{0}\Gamma^{\dagger}\gamma_{0})\chi,
\end{equation}
which implies, for $\Gamma=\gamma^{\mu}$
\begin{equation}
(\psibar{}^{\rmc}\gamma^{\mu}\psi)^{*}=\psibar{}\gamma^{\mu}\psi^{\rmc}.
\end{equation}

\section{Expressions from bilinearization of the Dirac equation}
We list here the various expressions resulting from left-multiplying the Dirac equation and its charge conjugate by $\psibar\Gamma$ and $\psibar{}^{\rmc}\Gamma$ respectively (where $\Gamma$ is an element of the Dirac algebra), then 1) subtracting the charge conjugate equation from the regular equation and 2) adding the charge conjugate equation to the regular equation.

\vspace{5mm}

\noindent $\Gamma=\psibar$:
\begin{align}
&j^{\nu}A_{\nu}=\frac{\rmi}{2q}[\psibar\gamma^{\nu}(\partial_{\nu}\psi)-(\partial_{\nu}\psibar)\gamma^{\nu}\psi]-\frac{m\sigma}{q} \\
&\partial_{\nu}j^{\nu}=0
\end{align}
$\Gamma=\psibar\gamma_{5}$:
\begin{align}
&\partial_{\nu}k^{\nu}=-2\rmi m\omega \\
&k^{\nu}A_{\nu}=\frac{\rmi}{2q}[\psibar\gamma_{5}\gamma^{\nu}(\partial_{\nu}\psi)-(\partial_{\nu}\psibar)\gamma_{5}\gamma^{\nu}\psi]
\end{align}
$\Gamma=\psibar\gamma_{\mu}$:
\begin{align}
&s_{\mu}{}^{\nu}A_{\nu}=\frac{\rmi}{2q}[\psibar\sigma_{\mu}{}^{\nu}(\partial_{\nu}\psi)-(\partial_{\nu}\psibar)\sigma_{\mu}{}^{\nu}\psi]-\frac{\partial_{\mu}\sigma}{2q} \\
&A_{\mu}=\frac{1}{2q}\frac{\rmi[\psibar(\partial_{\mu}\psi)-(\partial_{\mu}\psibar)\psi]+\partial_{\nu}s_{\mu}{}^{\nu}-2mj_{\mu}}{\sigma}
\end{align}
$\Gamma=\psibar\gamma_{5}\gamma_{\mu}$:
\begin{align}
&\sdual_{\mu}{}^{\nu}A_{\nu}=\frac{\rmi}{2q}[\psibar\gamma_{5}\sigma_{\mu}{}^{\nu}(\partial_{\nu}\psi)-(\partial_{\nu}\psibar)\gamma_{5}\sigma_{\mu}{}^{\nu}\psi]-\frac{\partial_{\mu}\omega}{2q}-\frac{\rmi mk_{\mu}}{q} \\
&A_{\mu}=\frac{1}{2q}\frac{\rmi[\psibar\gamma_{5}(\partial_{\mu}\psi)-(\partial_{\mu}\psibar)\gamma_{5}\psi]+\partial_{\nu}\sdual_{\mu}{}^{\nu}}{\omega}
\end{align}
$\Gamma=\psibar\sigma_{\mu\nu}$:
\begin{align}
&\delta_{\mu\nu}{}^{\rho\sigma}A_{\rho}j_{\sigma}=\frac{1}{2q}\left\{\rmi\delta_{\mu\nu}{}^{\rho\sigma}[\psibar\gamma_{\sigma}(\partial_{\rho}\psi)-(\partial_{\rho}\psibar)\gamma_{\sigma}\psi]-\rmi\epsilon_{\mu\nu}{}^{\rho\sigma}\partial_{\rho}k_{\sigma}\right\} \\
&\epsilon_{\mu\nu}{}^{\rho\sigma}A_{\rho}k_{\sigma}=\frac{1}{2q}\left\{\rmi\epsilon_{\mu\nu}{}^{\rho\sigma}[\psibar\gamma_{5}\gamma_{\sigma}(\partial_{\rho}\psi)-(\partial_{\rho}\psibar)\gamma_{5}\gamma_{\sigma}\psi]-\rmi\delta_{\mu\nu}{}^{\rho\sigma}\partial_{\rho}j_{\sigma}-2ms_{\mu\nu}\right\}
\end{align}
$\Gamma=\psibar\gamma_{5}\sigma_{\mu\nu}$:
\begin{align}
&\epsilon_{\mu\nu}{}^{\rho\sigma}A_{\rho}j_{\sigma}=\frac{1}{2q}\left\{\rmi\epsilon_{\mu\nu}{}^{\rho\sigma}[\psibar\gamma_{\sigma}(\partial_{\rho}\psi)-(\partial_{\rho}\psibar)\gamma_{\sigma}\psi]-\rmi\delta_{\mu\nu}{}^{\rho\sigma}\partial_{\rho}k_{\sigma}\right\} \\
&\delta_{\mu\nu}{}^{\rho\sigma}A_{\rho}k_{\sigma}=\frac{1}{2q}\left\{\rmi\delta_{\mu\nu}{}^{\rho\sigma}[\psibar\gamma_{5}\gamma_{\sigma}(\partial_{\rho}\psi)-(\partial_{\rho}\psibar)\gamma_{5}\gamma_{\sigma}\psi]-\rmi\epsilon_{\mu\nu}{}^{\rho\sigma}\partial_{\rho}j_{\sigma}+2m\sdual_{\mu\nu}\right\}
\end{align}

\section{Vector potential in tensor form}
This appendix contains a more detailed derivation of the inverted Dirac equation in terms of bilinear tensors only, to supplement the brief outline contained in section 3.1. Throughout, we will make heavy use of the identities contained within appendix A.

Given the two different forms of the inverted Abelian Dirac equation
\begin{align}
A_{\mu}&=\frac{1}{2q}\frac{\rmi[\psibar(\partial_{\mu}\psi)-(\partial_{\mu}\psibar)\psi]+\partial_{\nu}s_{\mu}{}^{\nu}-2mj_{\mu}}{\sigma} \label{Inverted regular Dirac Equation with spinors} \\
A_{\mu}&=\frac{1}{2q}\frac{\rmi[\psibar\gamma_{5}(\partial_{\mu}\psi)-(\partial_{\mu}\psibar)\gamma_{5}\psi]+\partial_{\nu}\sdual_{\mu}{}^{\nu}}{\omega}, \label{Inverted alternate Dirac Equation with spinors}
\end{align}
we can combine these into a single equation by adding them together and dividing by 2
\begin{align}\label{Inverted combined Dirac Equation with spinors}
&A_{\mu}=\frac{1}{4q}\left\{\frac{\rmi[\psibar(\partial_{\mu}\psi)-(\partial_{\mu}\psibar)\psi]\omega+\rmi[\psibar\gamma_{5}(\partial_{\mu}\psi)-(\partial_{\mu}\psibar)\gamma_{5}\psi]\sigma}{\sigma\omega}+\frac{\partial_{\nu}s_{\mu}{}^{\nu}}{\sigma}+\frac{\partial_{\nu}\sdual_{\mu}{}^{\nu}}{\omega}\right. \nonumber \\
&\qquad\qquad\left.-\frac{2mj_{\mu}}{\sigma}\right\}.
\end{align}
The appropriate tensor forms needed to replace the spinor terms in (\ref{Inverted combined Dirac Equation with spinors}) are $j^{\nu}(\partial_{\mu}k_{\nu}\!)$ and $m^{\nu}(\partial_{\mu}n_{\nu}\!)$. Consider the first
\begin{equation}
j^{\nu}(\partial_{\mu}k_{\nu})=\psibar\gamma^{\nu}\psi\cdot(\partial_{\mu}\psibar)\gamma_{5}\gamma_{\nu}\psi+\psibar\gamma^{\nu}\psi\cdot\psibar\gamma_{5}\gamma_{\nu}(\partial_{\mu}\psi).
\end{equation}
Fierz expanding the first term gives
\begin{align}
&\psibar\gamma^{\nu}\psi\cdot(\partial_{\mu}\psibar)\gamma_{5}\gamma_{\nu}\psi=-(\partial_{\mu}\psibar)\psi\cdot\omega-(1/2)(\partial_{\mu}\psibar)\gamma_{\sigma}\psi\cdot k^{\sigma} \nonumber \\
&\qquad-(1/2)(\partial_{\mu}\psibar)\gamma_{5}\gamma_{\sigma}\psi\cdot j^{\sigma}+(\partial_{\mu}\psibar)\gamma_{5}\psi\cdot\sigma,
\end{align}
and the second term gives
\begin{align}
&\psibar\gamma^{\nu}\psi\cdot\psibar\gamma_{5}\gamma_{\nu}(\partial_{\mu}\psi)=-\psibar\gamma_{5}(\partial_{\mu}\psi)\cdot\sigma-(1/2)\psibar\gamma_{5}\gamma_{\sigma}(\partial_{\mu}\psi)\cdot j^{\sigma} \nonumber \\
&\qquad-(1/2)\psibar\gamma_{\sigma}(\partial_{\mu}\psi)\cdot k^{\sigma}+\psibar(\partial_{\mu}\psi)\cdot\omega.
\end{align}
Combining these, and rearranging the equation gives
\begin{equation}
 j^{\nu}(\partial_{\mu}k_{\nu})=(2/3)[\psibar(\partial_{\mu}\psi)-(\partial_{\mu}\psibar)\psi]\omega-(2/3)[\psibar\gamma_{5}(\partial_{\mu}\psi)-(\partial_{\mu}\psibar)\gamma_{5}\psi]\sigma-(1/3)k^{\nu}(\partial_{\mu}j_{\nu}).
\end{equation}
We can see that in order to completely derive this Fierz identity, we need to calculate $k^{\nu}(\partial_{\mu}j_{\nu})$, and eliminate it by substitution. This is the same situation that arises when calculating the Fierz identities for $j^{\nu}j_{\nu}$ and $k^{\nu}k_{\nu}$.
\begin{equation}
k^{\nu}(\partial_{\mu}j_{\nu})=\psibar\gamma_{5}\gamma^{\nu}\psi\cdot(\partial_{\mu}\psibar)\gamma_{\nu}\psi+\psibar\gamma_{5}\gamma^{\nu}\psi\cdot\psibar\gamma_{\nu}(\partial_{\mu}\psi),
\end{equation}
now Fierz expanding the two terms respectively
\begin{align}
&\psibar\gamma_{5}\gamma^{\nu}\psi\cdot(\partial_{\mu}\psibar)\gamma_{\nu}\psi=(\partial_{\mu}\psibar)\psi\cdot\omega-(1/2)(\partial_{\mu}\psibar)\gamma_{\sigma}\psi\cdot k^{\sigma} \nonumber \\
&\qquad-(1/2)(\partial_{\mu}\psibar)\gamma_{5}\gamma_{\sigma}\psi\cdot j^{\sigma}-(\partial_{\mu}\psibar)\gamma_{5}\psi\cdot\sigma, \\
&\psibar\gamma_{5}\gamma^{\nu}\psi\cdot\psibar\gamma_{\nu}(\partial_{\mu}\psi)=-\psibar(\partial_{\mu}\psi)\cdot\omega-(1/2)\psibar\gamma_{\sigma}(\partial_{\mu}\psi)\cdot k^{\sigma} \nonumber \\
&\qquad-(1/2)\psibar\gamma_{5}\gamma_{\sigma}(\partial_{\mu}\psi)\cdot j^{\sigma}+\psibar\gamma_{5}(\partial_{\mu}\psi)\cdot\sigma.
\end{align}
Adding the terms gives
\begin{align}
&k^{\nu}(\partial_{\mu}j_{\nu})=(2/3)[\psibar\gamma_{5}(\partial_{\mu}\psi)-(\partial_{\mu}\psibar)\gamma_{5}\psi]\sigma-(2/3)[\psibar(\partial_{\mu}\psi)-(\partial_{\mu}\psibar)\psi]\omega \nonumber \\
&\qquad-(1/3)j^{\nu}(\partial_{\mu}k_{\nu}),
\end{align}
and substituting this into the $j^{\nu}(\partial_{\mu}k_{\nu})$ identity gives
\begin{equation}\label{j dk Fierz identity}
j^{\nu}(\partial_{\mu}k_{\nu})=[\psibar(\partial_{\mu}\psi)-(\partial_{\mu}\psibar)\psi]\omega-[\psibar\gamma_{5}(\partial_{\mu}\psi)-(\partial_{\mu}\psibar)\gamma_{5}\psi]\sigma.
\end{equation}
Likewise, substituting this into the $k^{\nu}(\partial_{\mu}j_{\nu})$ identity gives
\begin{equation}
k^{\nu}(\partial_{\mu}j_{\nu})=[\psibar\gamma_{5}(\partial_{\mu}\psi)-(\partial_{\mu}\psibar)\gamma_{5}\psi]\sigma-[\psibar(\partial_{\mu}\psi)-(\partial_{\mu}\psibar)\psi]\omega,
\end{equation}
which implies the new Fierz identity
\begin{equation}\label{j times partial k Fierz identity}
j^{\nu}(\partial_{\mu}k_{\nu})=-k^{\nu}(\partial_{\mu}j_{\nu})=[\psibar(\partial_{\mu}\psi)-(\partial_{\mu}\psibar)\psi]\omega-[\psibar\gamma_{5}(\partial_{\mu}\psi)-(\partial_{\mu}\psibar)\gamma_{5}\psi]\sigma.
\end{equation}
Now we consider $m^{\nu}(\partial_{\mu}n_{\nu})$, which, after applying the derivative and expanding is
\begin{align}
&m^{\nu}(\partial_{\mu}n_{\nu})=(\rmi/4)[\psibar{}^{\rmc}\gamma^{\nu}\psi\cdot(\partial_{\mu}\psibar)\gamma_{\nu}\psi^{\rmc}+\psibar{}^{\rmc}\gamma^{\nu}\psi\cdot\psibar\gamma_{\nu}(\partial_{\mu}\psi^{\rmc}) \nonumber \\
&\quad-\psibar{}^{\rmc}\gamma^{\nu}\psi\cdot(\partial_{\mu}\psibar{}^{\rmc})\gamma_{\nu}\psi-\psibar{}^{\rmc}\gamma^{\nu}\psi\cdot\psibar{}^{\rmc}\gamma_{\nu}(\partial_{\mu}\psi)+\psibar\gamma^{\nu}\psi^{\rmc}\cdot(\partial_{\mu}\psibar)\gamma_{\nu}\psi^{\rmc} \nonumber \\
&\quad+\psibar\gamma^{\nu}\psi^{\rmc}\cdot\psibar\gamma_{\nu}(\partial_{\mu}\psi^{\rmc})-\psibar\gamma^{\nu}\psi^{\rmc}\cdot(\partial_{\mu}\psibar{}^{\rmc})\gamma_{\nu}\psi-\psibar\gamma^{\nu}\psi^{\rmc}\cdot\psibar{}^{\rmc}\gamma_{\nu}(\partial_{\mu}\psi)].
\end{align}
After Fierz expanding and applying the charge conjugation identity (\ref{Charge conjugation bilinear identity}), the eight terms respectively are
\begin{align}
&\psibar{}^{\rmc}\gamma^{\nu}\psi\cdot(\partial_{\mu}\psibar)\gamma_{\nu}\psi^{\rmc}=-(\partial_{\mu}\psibar)\psi\cdot\sigma-(1/2)(\partial_{\mu}\psibar)\gamma_{\sigma}\psi\cdot j^{\sigma} \nonumber \\
&\qquad+(1/2)(\partial_{\mu}\psibar)\gamma_{5}\gamma_{\sigma}\psi\cdot k^{\sigma}+(\partial_{\mu}\psibar)\gamma_{5}\psi\cdot\omega, \\
&\psibar{}^{\rmc}\gamma^{\nu}\psi\cdot\psibar\gamma_{\nu}(\partial_{\mu}\psi^{\rmc})=-(\partial_{\mu}\psibar)\psi\cdot\sigma-(1/2)(\partial_{\mu}\psibar)\gamma_{\sigma}\psi\cdot j^{\sigma} \nonumber \\
&\qquad+(1/2)(\partial_{\mu}\psibar)\gamma_{5}\gamma_{\sigma}\psi\cdot k^{\sigma}+(\partial_{\mu}\psibar)\gamma_{5}\psi\cdot\omega, \\
&\psibar{}^{\rmc}\gamma^{\nu}\psi\cdot(\partial_{\mu}\psibar{}^{\rmc})\gamma_{\nu}\psi=(\partial_{\mu}\psibar{}^{\rmc})\psi\cdot\psibar{}^{\rmc}\psi-(1/2)(\partial_{\mu}\psibar{}^{\rmc})\gamma_{\sigma}\psi\cdot\psibar{}^{\rmc}\gamma^{\sigma}\psi \nonumber \\
&\qquad-(1/2)(\partial_{\mu}\psibar{}^{\rmc})\gamma_{5}\gamma_{\sigma}\psi\cdot\psibar{}^{\rmc}\gamma_{5}\gamma^{\sigma}\psi-(\partial_{\mu}\psibar{}^{\rmc})\gamma_{5}\psi\cdot\psibar{}^{\rmc}\gamma_{5}\psi, \\
&\psibar{}^{\rmc}\gamma^{\nu}\psi\cdot\psibar{}^{\rmc}\gamma_{\nu}(\partial_{\mu}\psi)=\psibar{}^{\rmc}\psi\cdot\psibar{}^{\rmc}(\partial_{\mu}\psi)-(1/2)\psibar{}^{\rmc}\gamma_{\sigma}\psi\cdot\psibar{}^{\rmc}\gamma^{\sigma}(\partial_{\mu}\psi) \nonumber \\
&\qquad-(1/2)\psibar{}^{\rmc}\gamma_{5}\gamma_{\sigma}\psi\cdot\psibar{}^{\rmc}\gamma_{5}\gamma^{\sigma}(\partial_{\mu}\psi)-\psibar{}^{\rmc}\gamma_{5}\psi\cdot\psibar{}^{\rmc}\gamma_{5}(\partial_{\mu}\psi), \\
&\psibar\gamma^{\nu}\psi^{\rmc}\cdot(\partial_{\mu}\psibar)\gamma_{\nu}\psi^{\rmc}=(\partial_{\mu}\psibar)\psi^{\rmc}\cdot\psibar\psi^{\rmc}-(1/2)(\partial_{\mu}\psibar)\gamma_{\sigma}\psi^{\rmc}\cdot\psibar\gamma^{\sigma}\psi^{\rmc} \nonumber \\
&\qquad-(1/2)(\partial_{\mu}\psibar)\gamma_{5}\gamma_{\sigma}\psi^{\rmc}\cdot\psibar\gamma_{5}\gamma^{\sigma}\psi^{\rmc}-(\partial_{\mu}\psibar)\gamma_{5}\psi^{\rmc}\cdot\psibar\gamma_{5}\psi^{\rmc}, \\
&\psibar\gamma^{\nu}\psi^{\rmc}\cdot\psibar\gamma_{\nu}(\partial_{\mu}\psi^{\rmc})=\psibar\psi^{\rmc}\cdot\psibar(\partial_{\mu}\psi^{\rmc})-(1/2)\psibar\gamma_{\sigma}\psi^{\rmc}\cdot\psibar\gamma^{\sigma}(\partial_{\mu}\psi^{\rmc}) \nonumber \\
&\qquad-(1/2)\psibar\gamma_{5}\gamma_{\sigma}\psi^{\rmc}\cdot\psibar\gamma_{5}\gamma^{\sigma}(\partial_{\mu}\psi^{\rmc})-\psibar\gamma_{5}\psi^{\rmc}\cdot\psibar\gamma_{5}(\partial_{\mu}\psi^{\rmc}), \\
&\psibar\gamma^{\nu}\psi^{\rmc}\cdot(\partial_{\mu}\psibar{}^{\rmc})\gamma_{\nu}\psi=-\psibar(\partial_{\mu}\psi)\sigma-(1/2)\psibar\gamma_{\sigma}(\partial_{\mu}\psi)\cdot j^{\sigma} \nonumber \\
&\qquad+(1/2)\psibar\gamma_{5}\gamma_{\sigma}(\partial_{\mu}\psi)\cdot k^{\sigma}+\psibar\gamma_{5}(\partial_{\mu}\psi)\cdot\omega, \\
&\psibar\gamma^{\nu}\psi^{\rmc}\cdot\psibar{}^{\rmc}\gamma_{\nu}(\partial_{\mu}\psi)=-\psibar(\partial_{\mu}\psi)\cdot\sigma-(1/2)\psibar\gamma_{\sigma}(\partial_{\mu}\psi)\cdot j^{\sigma} \nonumber \\
&\qquad+(1/2)\psibar\gamma_{5}\gamma_{\sigma}(\partial_{\mu}\psi)\cdot k^{\sigma}+\psibar\gamma_{5}(\partial_{\mu}\psi)\cdot\omega.
\end{align}
Adding these together and gathering terms gives
\begin{align}
&m^{\nu}(\partial_{\mu}n_{\nu})=(\rmi/4)\{2[\psibar(\partial_{\mu}\psi)-(\partial_{\mu}\psibar)\psi]\sigma-2[\psibar\gamma_{5}(\partial_{\mu}\psi)-(\partial_{\mu}\psibar)\gamma_{5}\psi]\omega \nonumber \\
&\qquad+[\psibar\gamma_{\sigma}(\partial_{\mu}\psi)-(\partial_{\mu}\psibar)\gamma_{\sigma}\psi]j^{\sigma}-[\psibar\gamma_{5}\gamma_{\sigma}(\partial_{\mu}\psi)-(\partial_{\mu}\psibar)\gamma_{5}\gamma_{\sigma}\psi]k^{\sigma} \nonumber \\
&\qquad-\partial_{\mu}(\psibar{}^{\rmc}\psi)\cdot\psibar{}^{\rmc}\psi+\partial_{\mu}(\psibar\psi^{\rmc})\cdot\psibar\psi^{\rmc}+\partial_{\mu}(\psibar{}^{\rmc}\gamma_{5}\psi)\cdot\psibar{}^{\rmc}\gamma_{5}\psi \nonumber \\
&\qquad-\partial_{\mu}(\psibar\gamma_{5}\psi^{\rmc})\cdot\psibar\gamma_{5}\psi^{\rmc}+(1/2)\partial_{\mu}(\psibar{}^{\rmc}\gamma_{\sigma}\psi)\cdot\psibar{}^{\rmc}\gamma^{\sigma}\psi \nonumber \\
&\qquad-(1/2)\partial_{\mu}(\psibar\gamma_{\sigma}\psi^{\rmc})\cdot\psibar\gamma^{\sigma}\psi^{\rmc}+(1/2)\partial_{\mu}(\psibar{}^{\rmc}\gamma_{5}\gamma_{\sigma}\psi)\cdot\psibar{}^{\rmc}\gamma_{5}\gamma^{\sigma}\psi \nonumber \\
&\qquad-(1/2)\partial_{\mu}(\psibar\gamma_{5}\gamma_{\sigma}\psi^{\rmc})\cdot\psibar\gamma_{5}\gamma^{\sigma}\psi^{\rmc}\}.
\end{align}
Now, using (\ref{Charge conjugation bilinear identity}), and setting $\chi^{\rmc}=\psi$, which implies $\chibar=\psibar{}^{\rmc}$, we can show that $\psibar{}^{\rmc}\psi=0$, $\psibar{}^{\rmc}\gamma_{5}\gamma_{\sigma}\psi=0$ and $\psibar{}^{\rmc}\gamma_{5}\psi=0$, because they equal their own negatives. The same goes for the corresponding bilinears with the charge conjugation index switched. Our equation now whittles down to
\begin{align}
&m^{\nu}(\partial_{\mu}n_{\nu})=(\rmi/4)\{2[\psibar(\partial_{\mu}\psi)-(\partial_{\mu}\psibar)\psi]\sigma-2[\psibar\gamma_{5}(\partial_{\mu}\psi)-(\partial_{\mu}\psibar)\gamma_{5}\psi]\omega \nonumber \\
&\qquad+[\psibar\gamma_{\sigma}(\partial_{\mu}\psi)-(\partial_{\mu}\psibar)\gamma_{\sigma}\psi]j^{\sigma}-[\psibar\gamma_{5}\gamma_{\sigma}(\partial_{\mu}\psi)-(\partial_{\mu}\psibar)\gamma_{5}\gamma_{\sigma}\psi]k^{\sigma} \nonumber \\
&\qquad+(1/2)\partial_{\mu}(\psibar{}^{\rmc}\gamma_{\sigma}\psi)\cdot\psibar{}^{\rmc}\gamma^{\sigma}\psi-(1/2)\partial_{\mu}(\psibar\gamma_{\sigma}\psi^{\rmc})\cdot\psibar\gamma^{\sigma}\psi^{\rmc}\}.
\end{align}
We can use the result of the complex conjugation bilinear identity (\ref{Complex conjugation bilinear identity}), $(\psibar{}^{\rmc}\gamma_{\mu}\psi)^{*}=\psibar\gamma_{\mu}\psi^{\rmc}$, which implies that $\psibar\gamma_{\mu}\psi^{\rmc}=m_{\mu}-\rmi n_{\mu}$. So taking the last two terms from the equation for $m^{\nu}(\partial_{\mu}n_{\nu})$
\begin{align}
&(1/2)\partial_{\mu}(\psibar{}^{\rmc}\gamma_{\sigma}\psi)\cdot\psibar{}^{\rmc}\gamma^{\sigma}\psi-(1/2)\partial_{\mu}(\psibar\gamma_{\sigma}\psi^{\rmc})\cdot\psibar\gamma^{\sigma}\psi^{\rmc} \nonumber \\
&\qquad=(1/2)\partial_{\mu}(m_{\sigma}+\rmi n_{\sigma})(m^{\sigma}+\rmi n^{\sigma})-(1/2)\partial_{\mu}(m_{\sigma}-\rmi n_{\sigma})(m^{\sigma}-\rmi n^{\sigma}) \nonumber \\
&\qquad=2\rmi\partial_{\mu}(m_{\sigma}n^{\sigma}) \nonumber \\
&\qquad=0,
\end{align}
by the orthogonality of $m_{\mu}$ and $n_{\nu}$. Our equation now becomes
\begin{align}\label{Second-last expression for m-partial-n}
&m^{\nu}(\partial_{\mu}n_{\nu})=(\rmi/4)\{2[\psibar(\partial_{\mu}\psi)-(\partial_{\mu}\psibar)\psi]\sigma-2[\psibar\gamma_{5}(\partial_{\mu}\psi)-(\partial_{\mu}\psibar)\gamma_{5}\psi]\omega \nonumber \\
&\qquad+[\psibar\gamma_{\nu}(\partial_{\mu}\psi)-(\partial_{\mu}\psibar)\gamma_{\nu}\psi]j^{\nu}-[\psibar\gamma_{5}\gamma_{\nu}(\partial_{\mu}\psi)-(\partial_{\mu}\psibar)\gamma_{5}\gamma_{\nu}\psi]k^{\nu}\}.
\end{align}
We can eliminate the last two terms via a Fierz expansion process analogous to that of $j^{\nu}(\partial_{\mu}k_{\nu})$ and $k^{\nu}(\partial_{\mu}j_{\nu})$. The expanded terms on the left are
\begin{align}
&\psibar\gamma_{\nu}(\partial_{\mu}\psi)\cdot\psibar\gamma^{\nu}\psi=\psibar(\partial_{\mu}\psi)\cdot\sigma-(1/2)\psibar\gamma_{\nu}(\partial_{\mu}\psi)\cdot j^{\nu}-(1/2)\psibar\gamma_{5}\gamma_{\nu}(\partial_{\mu}\psi)\cdot k^{\nu} \nonumber \\
&\qquad-\psibar\gamma_{5}(\partial_{\mu}\psi)\cdot\omega, \\
&(\partial_{\mu}\psibar)\gamma_{\nu}\psi\cdot\psibar\gamma^{\nu}\psi=(\partial_{\mu}\psibar)\psi\cdot\sigma-(1/2)(\partial_{\mu}\psibar)\gamma_{\nu}\psi\cdot j^{\nu}-(1/2)(\partial_{\mu}\psibar)\gamma_{5}\gamma_{\nu}\psi\cdot k^{\sigma} \nonumber \\
&\qquad-(\partial_{\mu}\psibar)\gamma_{5}\psi\cdot\omega.
\end{align}
Subtracting the second term from the first and rearranging gives
\begin{align}\label{Proto [...]j-nu identity}
&[\psibar\gamma_{\nu}(\partial_{\mu}\psi)-(\partial_{\mu}\psibar)\gamma_{\nu}\psi]j^{\nu}=(2/3)[\psibar(\partial_{\mu}\psi)-(\partial_{\mu}\psibar)\psi]\sigma \nonumber \\
&\qquad-(2/3)[\psibar\gamma_{5}(\partial_{\mu}\psi)-(\partial_{\mu}\psibar)\gamma_{5}\psi]\omega-(1/3)[\psibar\gamma_{5}\gamma_{\nu}(\partial_{\mu}\psi)-(\partial_{\mu}\psibar)\gamma_{5}\gamma_{\nu}\psi]k^{\nu},
\end{align}
where we can see that as mentioned before, we must also consider the Fierz expansion of $[\psibar\gamma_{5}\gamma_{\nu}(\partial_{\mu}\psi)$ $-(\partial_{\mu}\psibar)\gamma_{5}\gamma_{\nu}\psi]k^{\nu}$. The two Fierz expanded terms are
\begin{align}
&\psibar\gamma_{5}\gamma_{\nu}(\partial_{\mu}\psi)\cdot\psibar\gamma_{5}\gamma^{\nu}\psi=-\psibar(\partial_{\mu}\psi)\cdot\sigma-(1/2)\psibar\gamma_{\nu}(\partial_{\mu}\psi)\cdot j^{\nu} \nonumber \\
&\qquad-(1/2)\psibar\gamma_{5}\gamma_{\nu}(\partial_{\mu}\psi)\cdot k^{\nu}+\psibar\gamma_{5}(\partial_{\mu}\psi)\cdot\omega, \\
&(\partial_{\mu}\psibar)\gamma_{5}\gamma_{\nu}\psi\cdot\psibar\gamma_{5}\gamma^{\nu}\psi=-(\partial_{\mu}\psibar)\psi\cdot\sigma-(1/2)(\partial_{\mu}\psibar)\gamma_{\nu}\psi\cdot j^{\nu} \nonumber \\
&\qquad-(1/2)(\partial_{\mu}\psibar)\gamma_{5}\gamma_{\nu}\psi\cdot k^{\nu}+(\partial_{\mu}\psibar)\gamma_{5}\psi\cdot\omega.
\end{align}
Subtracting the second term from the first and rearranging gives
\begin{align}\label{Proto [...]k-nu identity}
&[\psibar\gamma_{5}\gamma_{\nu}(\partial_{\mu}\psi)-(\partial_{\mu}\psibar)\gamma_{5}\gamma_{\nu}\psi]k^{\nu}=-(2/3)[\psibar(\partial_{\mu}\psi)-(\partial_{\mu}\psibar)\psi]\sigma \nonumber \\
&\qquad+(2/3)[\psibar\gamma_{5}(\partial_{\mu}\psi)-(\partial_{\mu}\psibar)\gamma_{5}\psi]\omega-(1/3)[\psibar\gamma_{\nu}(\partial_{\mu}\psi)-(\partial_{\mu}\psibar)\gamma_{\nu}\psi]j^{\nu}.
\end{align}
Substituting this into (\ref{Proto [...]j-nu identity}) gives the identity
\begin{equation}
[\psibar\gamma_{\nu}(\partial_{\mu}\psi)-(\partial_{\mu}\psibar)\gamma_{\nu}\psi]j^{\nu}=[\psibar(\partial_{\mu}\psi)-(\partial_{\mu}\psibar)\psi]\sigma-[\psibar\gamma_{5}(\partial_{\mu}\psi)-(\partial_{\mu}\psibar)\gamma_{5}\psi]\omega,
\end{equation}
then subsequently substituting this into (\ref{Proto [...]k-nu identity}) gives
\begin{equation}
[\psibar\gamma_{5}\gamma_{\nu}(\partial_{\mu}\psi)-(\partial_{\mu}\psibar)\gamma_{5}\gamma_{\nu}\psi]k^{\nu}=-[\psibar(\partial_{\mu}\psi)-(\partial_{\mu}\psibar)\psi]\sigma+[\psibar\gamma_{5}(\partial_{\mu}\psi)-(\partial_{\mu}\psibar)\gamma_{5}\psi]\omega.
\end{equation}
This provides us with another Fierz identity
\begin{align}
&[\psibar\gamma_{\nu}(\partial_{\mu}\psi)-(\partial_{\mu}\psibar)\gamma_{\nu}\psi]j^{\nu}=-[\psibar\gamma_{5}\gamma_{\nu}(\partial_{\mu}\psi)-(\partial_{\mu}\psibar)\gamma_{5}\gamma_{\nu}\psi]k^{\nu} \nonumber \\
&\qquad=[\psibar(\partial_{\mu}\psi)-(\partial_{\mu}\psibar)\psi]\sigma-[\psibar\gamma_{5}(\partial_{\mu}\psi)-(\partial_{\mu}\psibar)\gamma_{5}\psi]\omega,
\end{align}
which when substituted into (\ref{Second-last expression for m-partial-n}) gives its final form
\begin{equation}\label{m dn Fierz identity}
m^{\nu}(\partial_{\mu}n_{\nu})=\rmi[\psibar(\partial_{\mu}\psi)-(\partial_{\mu}\psibar)\psi]\sigma-\rmi[\psibar\gamma_{5}(\partial_{\mu}\psi)-(\partial_{\mu}\psibar)\gamma_{5}\psi]\omega.
\end{equation}
It can be shown via a similar process, that
\begin{equation}\label{m times partial n Fierz identity}
m^{\nu}(\partial_{\mu}n_{\nu})=-n^{\nu}(\partial_{\mu}m_{\nu}).
\end{equation}
We can now use the identities (\ref{j dk Fierz identity}) and (\ref{m dn Fierz identity}) to describe the spinorial objects in the inverted Dirac equation in terms of tensors alone. From (\ref{j dk Fierz identity}), we get
\begin{equation}
[\psibar(\partial_{\mu}\psi)-(\partial_{\mu}\psibar)\psi]=j^{\nu}(\partial_{\mu}k_{\nu})\omega^{-1}+[\psibar\gamma_{5}(\partial_{\mu}\psi)-(\partial_{\mu}\psibar)\gamma_{5}\psi]\sigma\omega^{-1},
\end{equation}
and from (\ref{m dn Fierz identity}) we get
\begin{equation}
[\psibar\gamma_{5}(\partial_{\mu}\psi)-(\partial_{\mu}\psibar)\gamma_{5}\psi]=\rmi m^{\nu}(\partial_{\mu}n_{\nu})\omega^{-1}+[\psibar(\partial_{\mu}\psi)-(\partial_{\mu}\psibar)\psi]\sigma\omega^{-1}.
\end{equation}
After substitution and rearrangement, we get the required spinor replacement identities
\begin{align}
&[\psibar(\partial_{\mu}\psi)-(\partial_{\mu}\psibar)\psi]=-(\sigma^{2}-\omega^{2})^{-1}[j^{\nu}(\partial_{\mu}k_{\nu})\omega+\rmi m^{\nu}(\partial_{\mu}n_{\nu})\sigma],\label{Pseudo-product rule on sigma tensor form} \\
&[\psibar\gamma_{5}(\partial_{\mu}\psi)-(\partial_{\mu}\psibar)\gamma_{5}\psi]=-(\sigma^{2}-\omega^{2})^{-1}[j^{\nu}(\partial_{\mu}k_{\nu})\sigma+\rmi m^{\nu}(\partial_{\mu}n_{\nu})\omega].\label{Pseudo-product rule on omega tensor form}
\end{align}
Combining these two identities in the form they appear in (\ref{Inverted combined Dirac Equation with spinors})
\begin{align}
&\{\rmi[\psibar(\partial_{\mu}\psi)-(\partial_{\mu}\psibar)\psi]\omega+\rmi[\psibar\gamma_{5}(\partial_{\mu}\psi)-(\partial_{\mu}\psibar)\gamma_{5}\psi]\sigma\}(\sigma\omega)^{-1} \nonumber \\
&\qquad=\{\omega[m^{\nu}(\partial_{\mu}n_{\nu})\sigma-\rmi j^{\nu}(\partial_{\mu}k_{\nu})\omega]+\sigma[m^{\nu}(\partial_{\mu}n_{\nu})\omega-\rmi j^{\nu}(\partial_{\mu}k_{\nu})\sigma]\} \nonumber \\
&\qquad\qquad\cdot[\sigma\omega(\sigma^{2}-\omega^{2})]^{-1} \nonumber \\
&\qquad=\frac{2m^{\nu}(\partial_{\mu}n_{\nu})}{\sigma^{2}-\omega^{2}}-\frac{\rmi j^{\nu}(\partial_{\mu}k_{\nu})}{\sigma^{2}-\omega^{2}}\left(\frac{\sigma^{2}+\omega^{2}}{\sigma\omega}\right).
\end{align}
Substitute into (\ref{Inverted combined Dirac Equation with spinors}):
\begin{equation}\label{Inverted combined Dirac Equation tensor form}
A_{\mu}=\frac{1}{4q}\left\{\frac{2m^{\nu}(\partial_{\mu}n_{\nu})}{\sigma^{2}-\omega^{2}}-\rmi j^{\nu}(\partial_{\mu}k_{\nu})\left[\frac{\sigma^{2}+\omega^{2}}{\sigma\omega(\sigma^{2}-\omega^{2})}\right]+\frac{\partial_{\nu}s_{\mu}{}^{\nu}}{\sigma}+\frac{\partial_{\nu}\sdual_{\mu}{}^{\nu}}{\omega}-\frac{2mj_{\mu}}{\sigma}\right\},
\end{equation}
which technically, is an expression for $A_{\mu}$ exclusively in tensor form. However, we can simplify this. Substituting (\ref{Pseudo-product rule on sigma tensor form}) into the original inverted Dirac equation (\ref{Inverted regular Dirac Equation with spinors}) gives
\begin{equation}\label{Inverted regular Dirac Equation tensor form}
A_{\mu}=\frac{1}{2q}\left\{\frac{m^{\nu}(\partial_{\mu}n_{\nu})}{\sigma^{2}-\omega^{2}}-\rmi j^{\nu}(\partial_{\mu}k_{\nu})\left[\frac{\omega}{\sigma(\sigma^{2}-\omega^{2})}\right]+\frac{\partial_{\nu}s_{\mu}{}^{\nu}}{\sigma}-\frac{2mj_{\mu}}{\sigma}\right\},
\end{equation}
and substituting (\ref{Pseudo-product rule on omega tensor form}) into the alternative inverted Dirac equation (\ref{Inverted alternate Dirac Equation with spinors}) gives
\begin{equation}\label{Inverted alternate Dirac Equation tensor form}
A_{\mu}=\frac{1}{2q}\left\{\frac{m^{\nu}(\partial_{\mu}n_{\nu})}{\sigma^{2}-\omega^{2}}-\rmi j^{\nu}(\partial_{\mu}k_{\nu})\left[\frac{\sigma}{\omega(\sigma^{2}-\omega^{2})}\right]+\frac{\partial_{\nu}\sdual_{\mu}{}^{\nu}}{\omega}\right\}.
\end{equation}
Adding (\ref{Inverted regular Dirac Equation tensor form}) and (\ref{Inverted alternate Dirac Equation tensor form}) and dividing by 2 results in (\ref{Inverted combined Dirac Equation tensor form}), the combined form of $A_{\mu}$ already obtained. But if we \emph{subtract} these two and rearrange, we obtain a new identity that we can use to eliminate $\rmi j^{\nu}(\partial_{\mu}k_{\nu})$
\begin{equation}
\rmi j^{\nu}(\partial_{\mu}k_{\nu})=2m\omega j_{\mu}+\sigma\partial_{\nu}\sdual_{\mu}{}^{\nu}-\omega\partial_{\nu}s_{\mu}{}^{\nu}.
\end{equation}
Take the $\rmi j^{\nu}(\partial_{\mu}k_{\nu})$ term from (\ref{Inverted combined Dirac Equation tensor form}) and substitute the above identity
\begin{align}
&-\rmi j^{\nu}(\partial_{\mu}k_{\nu})\left[\frac{\sigma^{2}+\omega^{2}}{\sigma\omega(\sigma^{2}-\omega^{2})}\right]=-(2m\omega j_{\mu}+\sigma\partial_{\nu}\sdual_{\mu}{}^{\nu}-\omega\partial_{\nu}s_{\mu}{}^{\nu})\left[\frac{\sigma^{2}+\omega^{2}}{\sigma\omega(\sigma^{2}-\omega^{2})}\right] \nonumber \\
&\qquad=\frac{-2m\sigma^{2}\omega j_{\mu}-2m\omega^{3}j_{\mu}-\sigma^{3}\partial_{\nu}\sdual_{\mu}{}^{\nu}-\sigma\omega^{2}\partial_{\nu}\sdual_{\mu}{}^{\nu}+\sigma^{2}\omega\partial_{\nu}s_{\mu}{}^{\nu}+\omega^{3}\partial_{\nu}s_{\mu}{}^{\nu}}{\sigma\omega(\sigma^{2}-\omega^{2})}.
\end{align}
Rearrange the last three terms in (\ref{Inverted combined Dirac Equation tensor form})
\begin{align}
&\frac{\partial_{\nu}s_{\mu}{}^{\nu}}{\sigma}+\frac{\partial_{\nu}\sdual_{\mu}{}^{\nu}}{\omega}-\frac{2mj_{\mu}}{\sigma}=\left(\frac{\omega\partial_{\nu}s_{\mu}{}^{\nu}+\sigma\partial_{\nu}\sdual_{\mu}{}^{\nu}-2m\omega j_{\mu}}{\sigma\omega}\right)\left(\frac{\sigma^{2}-\omega^{2}}{\sigma^{2}-\omega^{2}}\right) \nonumber \\
&\qquad=\frac{-2m\sigma^{2}\omega j_{\mu}+2m\omega^{3}j_{\mu}+\sigma^{3}\partial_{\nu}\sdual_{\mu}{}^{\nu}-\sigma\omega^{2}\partial_{\nu}\sdual_{\mu}{}^{\nu}+\sigma^{2}\omega\partial_{\nu}s_{\mu}{}^{\nu}-\omega^{3}\partial_{\nu}s_{\mu}{}^{\nu}}{\sigma\omega(\sigma^{2}-\omega^{2})},
\end{align}
and adding the previous two equations together gives
\begin{align}
&-\rmi j^{\nu}(\partial_{\mu}k_{\nu})\left[\frac{\sigma^{2}+\omega^{2}}{\sigma\omega(\sigma^{2}-\omega^{2})}\right]+\frac{\partial_{\nu}s_{\mu}{}^{\nu}}{\sigma}+\frac{\partial_{\nu}\sdual_{\mu}{}^{\nu}}{\omega}-\frac{2mj_{\mu}}{\sigma} \nonumber \\
&\qquad=\frac{2\sigma^{2}\omega\partial_{\nu}s_{\mu}{}^{\nu}-2\sigma\omega^{2}\partial_{\nu}\sdual_{\mu}{}^{\nu}-4m\sigma^{2}\omega j_{\mu}}{\sigma\omega(\sigma^{2}-\omega^{2})} \nonumber \\
&\qquad=\frac{2(\sigma\partial_{\nu}s_{\mu}{}^{\nu}-\omega\partial_{\nu}\sdual_{\mu}{}^{\nu}-2m\sigma j_{\mu})}{(\sigma^{2}-\omega^{2})}.
\end{align}
Therefore, our final form of $A_{\mu}$ in tensor form is
\begin{equation}
A_{\mu}=\frac{1}{2q}\frac{m^{\nu}(\partial_{\mu}n_{\nu})+\sigma\partial_{\nu}s_{\mu}{}^{\nu}-\omega\partial_{\nu}\sdual_{\mu}{}^{\nu}-2m\sigma j_{\mu}}{\sigma^{2}-\omega^{2}}.
\end{equation}

\section{Cylindrically symmetric field strength tensor}
In this appendix, we present the dependent functions of $t$ and $\rho$ in the cylindrically symmetric field strength tensor in terms of the $j^{\mu}$, $k^{\mu}$, $\sigma$ and $\omega$ tensor fields. The functions appearing in $F_{01}=-xF_{a}+yF_{b}$ and $F_{02}=-yF_{a}-xF_{b}$ are
\begin{align}\label{Cylindrical symmetry appendix, Fa}
&F_{a}=-2[q(\sigma^{2}-\omega^{2})^{3}]^{-1}\{(j_{d}k_{c,t}-j_{c}k_{d,t}-k_{d}j_{c,t}+k_{c}j_{d,t}-2m\sigma j_{b})(\sigma\sigma_{t}-\omega\omega_{t}) \nonumber \\
&\qquad\cdot(\sigma^{2}-\omega^{2})+(j_{c}k_{d,\rho}-j_{d}k_{c,\rho}-k_{c}j_{d,\rho}+k_{d}j_{c,\rho})(\sigma\sigma_{\rho}-\omega\omega_{\rho})(\sigma^{2}-\omega^{2}) \nonumber \\
&\qquad+(j_{c}k_{d}-j_{d}k_{c})(\sigma\sigma_{t}-\omega\omega_{t})^{2}+(j_{d}k_{c}-j_{c}k_{d})(\sigma\sigma_{\rho}-\omega\omega_{\rho})^{2}+(2/\rho)(j_{c}k_{d} \nonumber \\
&\qquad-j_{d}k_{c}-m\sigma j_{a})(\sigma\sigma_{\rho}-\omega\omega_{\rho})(\sigma^{2}-\omega^{2})+\rmi[(j_{b}k_{a}-j_{a}k_{b})(\sigma_{t}\omega-\sigma\omega_{t}) \nonumber \\
&\qquad\cdot(\sigma\sigma_{t}-\omega\omega_{t})+(j_{a}k_{b}-j_{b}k_{a})(\sigma_{\rho}\omega-\sigma\omega_{\rho})(\sigma\sigma_{\rho}-\omega\omega_{\rho})]\} \nonumber \\
&\qquad+[2q(\sigma^{2}-\omega^{2})^{2}]^{-1}\{(j_{d}k_{c,tt}+2j_{d,t}k_{c,t}+j_{d,tt}k_{c}-j_{c}k_{d,tt}-2j_{c,t}k_{d,t} \nonumber \\
&\qquad-j_{c,tt}k_{d}+j_{c}k_{d,\rho\rho}+2j_{c,\rho}k_{d,\rho}+j_{c,\rho\rho}k_{d}-j_{d}k_{c,\rho\rho}-2j_{d,\rho}k_{c,\rho}-j_{d,\rho\rho}k_{c} \nonumber \\
&\qquad-2m\sigma_{t}j_{b}-2m\sigma j_{b,t})(\sigma^{2}-\omega^{2})+(j_{d}k_{c,t}+j_{d,t}k_{c}-j_{c}k_{d,t}-j_{c,t}k_{d} \nonumber \\
&\qquad-4m\sigma j_{b})(\sigma\sigma_{t}-\omega\omega_{t})+(j_{c}k_{d,\rho}+j_{c,\rho}k_{d}-j_{d}k_{c,\rho}-j_{d,\rho}k_{c})(\sigma\sigma_{\rho}-\omega\omega_{\rho}) \nonumber \\
&\qquad+(j_{c}k_{d}-j_{d}k_{c})(\sigma_{t}^{2}+\sigma\sigma_{tt}-\omega_{t}^{2}-\omega\omega_{tt})+(j_{d}k_{c}-j_{c}k_{d})(\sigma_{\rho}^{2}+\sigma\sigma_{\rho\rho} \nonumber \\
&\qquad-\omega_{\rho}^{2}-\omega\omega_{\rho\rho})+(1/\rho)(3j_{c}k_{d,\rho}+3j_{c,\rho}k_{d}-3j_{d}k_{c,\rho}-3j_{d,\rho}k_{c}-2m\sigma_{\rho}j_{a} \nonumber \\
&\qquad-2m\sigma j_{a,\rho})(\sigma^{2}-\omega^{2})+(1/\rho)(3j_{c}k_{d}-3j_{d}k_{c}-4m\sigma j_{a})(\sigma\sigma_{\rho}-\omega\omega_{\rho}) \nonumber \\
&\qquad+\rmi[(j_{b}k_{a,t}+j_{b,t}k_{a}-j_{a}k_{b,t}-j_{a,t}k_{b})(\sigma_{t}\omega-\sigma\omega_{t})+(j_{a}k_{b,\rho}+j_{a,\rho}k_{b} \nonumber \\
&\qquad-j_{b}k_{a,\rho}-j_{b,\rho}k_{a})(\sigma_{\rho}\omega-\sigma\omega_{\rho})+(j_{b}k_{a}-j_{a}k_{b})(\sigma_{tt}\omega-\sigma\omega_{tt})+(j_{a}k_{b} \nonumber \\
&\qquad-j_{b}k_{a})(\sigma_{\rho\rho}\omega-\sigma\omega_{\rho\rho})+(1/\rho)(j_{a}k_{b}-j_{b}k_{a})(\sigma_{\rho}\omega-\sigma\omega_{\rho})]+(j_{d}k_{a} \nonumber \\
&\qquad-j_{a}k_{d})(j_{b}j_{c,t}-j_{c}j_{b,t}-k_{b}k_{c,t}+k_{c}k_{b,t})+(j_{c}k_{b}-j_{b}k_{c})(j_{d}j_{a,t}-j_{a}j_{d,t} \nonumber \\
&\qquad-k_{d}k_{a,t}+k_{a}k_{d,t})-\rho[(j_{a}k_{b}-j_{b}k_{a})(j_{d,t}j_{c,\rho}-j_{c,t}j_{d,\rho}-k_{d,t}k_{c,\rho}+k_{c,t}k_{d,\rho}) \nonumber \\
&\qquad+(j_{a}k_{c}-j_{c}k_{a})(j_{b,t}j_{d,\rho}-j_{d,t}j_{b,\rho}-k_{b,t}k_{d,\rho}+k_{d,t}k_{b,\rho})+(j_{a}k_{d}-j_{d}k_{a}) \nonumber \\
&\qquad\cdot(j_{c,t}j_{b,\rho}-j_{b,t}j_{c,\rho}-k_{c,t}k_{b,\rho}+k_{b,t}k_{c,\rho})+(j_{b}k_{c}-j_{c}k_{b})(j_{d,t}j_{a,\rho}-j_{a,t}j_{d,\rho} \nonumber \\
&\qquad-k_{d,t}k_{a,\rho}+k_{a,t}k_{d,\rho})+(j_{b}k_{d}-j_{d}k_{b})(j_{a,t}j_{c,\rho}-j_{c,t}j_{a,\rho}-k_{a,t}k_{c,\rho}+k_{c,t}k_{a,\rho}) \nonumber \\
&\qquad+(j_{c}k_{d}-j_{d}k_{c})(j_{b,t}j_{a,\rho}-j_{a,t}j_{b,\rho}-k_{b,t}k_{a,\rho}+k_{a,t}k_{b,\rho})]\},
\end{align}
as well as
\begin{align}
&F_{b}=-2[q(\sigma^{2}-\omega^{2})^{3}]^{-1}(\sigma\sigma_{t}-\omega\omega_{t})\{(j_{b}k_{d,t}+j_{b,t}k_{d}-j_{d}k_{b,t}-j_{d,t}k_{b}-2m\sigma j_{c}) \nonumber \\
&\qquad\cdot(\sigma^{2}-\omega^{2})+(j_{d}k_{b}-j_{b}k_{d})(\sigma\sigma_{t}-\omega\omega_{t})+(1/\rho)(j_{a}k_{d,\rho}+j_{a,\rho}k_{d}-j_{d}k_{a,\rho} \nonumber \\
&\qquad-j_{d,\rho}k_{a})(\sigma^{2}-\omega^{2})+(1/\rho)(j_{d}k_{a}-j_{a}k_{d})(\sigma\sigma_{\rho}-\omega\omega_{\rho})+\rmi[(j_{c}k_{a}-j_{a}k_{c}) \nonumber \\
&\qquad\cdot(\sigma_{t}\omega-\sigma\omega_{t})+\rho(j_{c}k_{b}-j_{b}k_{c})(\sigma_{\rho}\omega-\sigma\omega_{\rho})]\}+[2q(\sigma^{2}-\omega^{2})^{2}]^{-1}\{(j_{b}k_{d,tt} \nonumber \\
&\qquad+2j_{b,t}k_{d,t}+j_{b,tt}k_{d}-j_{d}k_{b,tt}-2j_{d,t}k_{b,t}-j_{d,tt}k_{b}-2m\sigma_{t}j_{c}-2m\sigma j_{c,t}) \nonumber \\
&\qquad\cdot(\sigma^{2}-\omega^{2})+(j_{b}k_{d,t}+j_{b,t}k_{d}-j_{d}k_{b,t}-j_{d,t}k_{b}-4m\sigma j_{c})(\sigma\sigma_{t}-\omega\omega_{t}) \nonumber \\
&\qquad+(j_{d}k_{b}-j_{b}k_{d})(\sigma_{t}^{2}+\sigma\sigma_{tt}-\omega_{t}^{2}-\omega\omega_{tt})+(1/\rho)(j_{a}k_{d,t\rho}+j_{a,t}k_{d,\rho} \nonumber \\
&\qquad+j_{a,\rho}k_{d,t}+j_{a,t\rho}k_{d}-j_{d}k_{a,t\rho}-j_{d,t}k_{a,\rho}-j_{d,\rho}k_{a,t}-j_{d,t\rho}k_{a})(\sigma^{2}-\omega^{2}) \nonumber \\
&\qquad+(2/\rho)(j_{a}k_{d,\rho}+j_{a,\rho}k_{d}-j_{d}k_{a,\rho}-j_{d,\rho}k_{a})(\sigma\sigma_{t}-\omega\omega_{t})+(1/\rho)(j_{d}k_{a,t} \nonumber \\
&\qquad+j_{d,t}k_{a}-j_{a}k_{d,t}-j_{a,t}k_{d})(\sigma\sigma_{\rho}-\omega\omega_{\rho})+(1/\rho)(j_{d}k_{a}-j_{a}k_{d})(\sigma_{t}\sigma_{\rho}+\sigma\sigma_{t\rho} \nonumber \\
&\qquad-\omega_{t}\omega_{\rho}-\omega\omega_{t\rho})+\rmi[(j_{c}k_{a,t}+j_{c,t}k_{a}-j_{a}k_{c,t}-j_{a,t}k_{c})(\sigma_{t}\omega-\sigma\omega_{t}) \nonumber \\
&\qquad+(j_{c}k_{a}-j_{a}k_{c})(\sigma_{tt}\omega-\sigma\omega_{tt})+\rho(j_{c}k_{b,t}+j_{c,t}k_{b}-j_{b}k_{c,t}-j_{b,t}k_{c})(\sigma_{\rho}\omega \nonumber \\
&\qquad-\sigma\omega_{\rho})+\rho(j_{c}k_{b}-j_{b}k_{c})(\sigma_{\rho}\omega_{t}+\sigma_{t\rho}\omega-\sigma_{t}\omega_{\rho}-\sigma\omega_{t\rho})+(j_{a}k_{d}-j_{d}k_{a}) \nonumber \\
&\qquad\cdot(j_{b}j_{b,t}+j_{c}j_{c,t}-k_{b}k_{b,t}-k_{c}k_{c,t})+(j_{b}k_{b}+j_{c}k_{c})(j_{a,t}j_{d}-j_{a}j_{d,t}+k_{a,t}k_{d} \nonumber \\
&\qquad-k_{a}k_{d,t})+(j_{b}^{2}+j_{c}^{2})(j_{d,t}k_{a}-j_{a,t}k_{d})+(k_{b}^{2}+k_{c}^{2})(j_{a}k_{d,t}-j_{d}k_{a,t})\}.
\end{align}
The function in $F_{03}=-F_{c}$ is
\begin{align}
&F_{c}=2[q(\sigma^{2}-\omega^{2})^{3}]^{-1}(\sigma\sigma_{t}-\omega\omega_{t})\{2(j_{c}k_{a}-j_{a}k_{c}-m\sigma j_{d})(\sigma^{2}-\omega^{2}) \nonumber \\
&\qquad+\rho(j_{c}k_{a,\rho}+j_{c,\rho}k_{a}-j_{a}k_{c,\rho}-j_{a,\rho}k_{c})(\sigma^{2}-\omega^{2})+\rho(j_{a}k_{c}-j_{c}k_{a})(\sigma\sigma_{\rho} \nonumber \\
&\qquad-\omega\omega_{\rho})+\rho^{2}(j_{c}k_{b,t}+j_{c,t}k_{b}-j_{b}k_{c,t}-j_{b,t}k_{c})(\sigma^{2}-\omega^{2})+\rho^{2}(j_{b}k_{c}-j_{c}k_{b}) \nonumber \\
&\qquad\cdot(\sigma\sigma_{t}-\omega\omega_{t})+\rmi[(j_{d}k_{a}-j_{a}k_{d})(\sigma_{t}\omega-\sigma\omega_{t})+\rho(j_{d}k_{b}-j_{b}k_{d})(\sigma_{\rho}\omega-\sigma\omega_{\rho})]\} \nonumber \\
&\qquad-[2q(\sigma^{2}-\omega^{2})^{2}]^{-1}\{2(j_{c}k_{a,t}+j_{c,t}k_{a}-j_{a}k_{c,t}-j_{a,t}k_{c}-m\sigma_{t}j_{d}-m\sigma j_{d,t}) \nonumber \\
&\qquad\cdot(\sigma^{2}-\omega^{2})+4(j_{c}k_{a}-j_{a}k_{c}-m\sigma j_{d})(\sigma\sigma_{t}-\omega\omega_{t})+\rho(j_{c}k_{a,t\rho}+j_{c,t}k_{a,\rho} \nonumber \\
&\qquad+j_{c,\rho}k_{a,t}+j_{c,t\rho}k_{a}-j_{a}k_{c,t\rho}-j_{a,t}k_{c,\rho}-j_{a,\rho}k_{c,t}-j_{a,t\rho}k_{c})(\sigma^{2}-\omega^{2}) \nonumber \\
&\qquad+2\rho(j_{c}k_{a,\rho}+j_{c,\rho}k_{a}-j_{a}k_{c,\rho}-j_{a,\rho}k_{c})(\sigma\sigma_{t}-\omega\omega_{t})+\rho(j_{a}k_{c,t}+j_{a,t}k_{c} \nonumber \\
&\qquad-j_{c}k_{a,t}-j_{c,t}k_{a})(\sigma\sigma_{\rho}-\omega\omega_{\rho})+\rho(j_{a}k_{c}-j_{c}k_{a})(\sigma_{t}\sigma_{\rho}+\sigma\sigma_{t\rho}-\omega_{t}\omega_{\rho} \nonumber \\
&\qquad-\omega\omega_{t\rho})+\rho^{2}(j_{c}k_{b,tt}+2j_{c,t}k_{b,t}+j_{c,tt}k_{b}-j_{b}k_{c,tt}-2j_{b,t}k_{c,t}-j_{b,tt}k_{c}) \nonumber \\
&\qquad\cdot(\sigma^{2}-\omega^{2})+\rho^{2}(j_{c}k_{b,t}+j_{c,t}k_{b}-j_{b}k_{c,t}-j_{b,t}k_{c})(\sigma\sigma_{t}-\omega\omega_{t})+\rho^{2}(j_{b}k_{c} \nonumber \\
&\qquad-j_{c}k_{b})(\sigma_{t}^{2}+\sigma\sigma_{tt}-\omega_{t}^{2}-\omega\omega_{tt})+\rmi[(j_{d}k_{a}-j_{a}k_{d})(\sigma_{tt}\omega-\sigma\omega_{tt})+(j_{d}k_{a,t} \nonumber \\
&\qquad+j_{d,t}k_{a}-j_{a}k_{d,t}-j_{a,t}k_{d})(\sigma_{t}\omega-\sigma\omega_{t})+\rho(j_{d}k_{b,t}+j_{d,t}k_{b}-j_{b}k_{d,t}-j_{b,t}k_{d}) \nonumber \\
&\qquad\cdot(\sigma_{\rho}\omega-\sigma\omega_{\rho})+\rho(j_{d}k_{b}-j_{b}k_{d})(\sigma_{\rho}\omega_{t}+\sigma_{t\rho}\omega-\sigma_{t}\omega_{\rho}-\sigma\omega_{t\rho})]\}
\end{align}
The function $F_{12}=F_{d}$ is
\begin{align}
&F_{d}=-[2q(\sigma^{2}-\omega^{2})^{2}]^{-1}\{(2j_{b}k_{d,t}+2j_{b,t}k_{d}-2j_{d}k_{b,t}-2j_{d,t}k_{b}+j_{a}k_{d,\rho\rho} \nonumber \\
&\qquad+2j_{a,\rho}k_{d,\rho}+j_{a,\rho\rho}k_{d}-j_{d}k_{a,\rho\rho}-2j_{d,\rho}k_{a,\rho}-j_{d,\rho\rho}k_{a}-4m\sigma j_{c})(\sigma^{2}-\omega^{2}) \nonumber \\
&\qquad+2(j_{d}k_{b}-j_{b}k_{d})(\sigma\sigma_{t}-\omega\omega_{t})+(j_{a}k_{d,\rho}+j_{a,\rho}k_{d}-j_{d}k_{a,\rho}-j_{d,\rho}k_{a})(\sigma\sigma_{\rho} \nonumber \\
&\qquad-\omega\omega_{\rho})+(j_{d}k_{a}-j_{a}k_{d})(\sigma_{\rho}^{2}+\sigma\sigma_{\rho\rho}-\omega_{\rho}^{2}-\omega\omega_{\rho\rho})+(1/\rho)(j_{a}k_{d,\rho}+j_{a,\rho}k_{d} \nonumber \\
&\qquad-j_{d}k_{a,\rho}-j_{d,\rho}k_{a})(\sigma^{2}-\omega^{2})+(1/\rho)(j_{d}k_{a}-j_{a}k_{d})(\sigma\sigma_{\rho}-\omega\omega_{\rho})+\rho(j_{b}k_{d,t\rho} \nonumber \\
&\qquad+j_{b,t}k_{d,\rho}+j_{b,\rho}k_{d,t}+j_{b,t\rho}k_{d}-j_{d}k_{b,t\rho}-j_{d,t}k_{b,\rho}-j_{d,\rho}k_{b,t}-j_{d,t\rho}k_{b}-2m\sigma_{\rho}j_{c} \nonumber \\
&\qquad-2m\sigma j_{c,\rho})(\sigma^{2}-\omega^{2})+\rho(j_{d}k_{b,\rho}+j_{d,\rho}k_{b}-j_{b}k_{d,\rho}-j_{b,\rho}k_{d})(\sigma\sigma_{t}-\omega\omega_{t}) \nonumber \\
&\qquad+2\rho(j_{b}k_{d,t}+j_{b,t}k_{d}-j_{d}k_{b,t}-j_{d,t}k_{b}-2m\sigma j_{c})(\sigma\sigma_{\rho}-\omega\omega_{\rho})+\rho(j_{d}k_{b} \nonumber \\
&\qquad-j_{b}k_{d})(\sigma_{t}\sigma_{\rho}+\sigma\sigma_{t\rho}-\omega_{t}\omega_{\rho}-\omega\omega_{t\rho})+\rmi[2(j_{c}k_{a}-j_{a}k_{c})(\sigma_{t}\omega-\sigma\omega_{t}) \nonumber \\
&\qquad+\rho(j_{c}k_{a,\rho}+j_{c,\rho}k_{a}-j_{a}k_{c,\rho}-j_{a,\rho}k_{c})(\sigma_{t}\omega-\sigma\omega_{t})+3\rho(j_{c}k_{b}-j_{b}k_{c})(\sigma_{\rho}\omega \nonumber \\
&\qquad-\sigma\omega_{\rho})+\rho(j_{c}k_{a}-j_{a}k_{c})(\sigma_{t}\omega_{\rho}+\sigma_{t\rho}\omega-\sigma_{\rho}\omega_{t}-\sigma\omega_{t\rho})+\rho^{2}(j_{c}k_{b}-j_{b}k_{c}) \nonumber \\
&\qquad\cdot(\sigma_{\rho\rho}\omega-\sigma\omega_{\rho\rho})+\rho^{2}(\sigma_{\rho}\omega-\sigma\omega_{\rho})(j_{c}k_{b,\rho}+j_{c,\rho}k_{b}-j_{b}k_{c,\rho}-j_{b,\rho}k_{c})] \nonumber \\
&\qquad\cdot+(j_{a}k_{d}-j_{d}k_{a})(j_{b}^{2}+j_{c}^{2}-k_{b}^{2}-k_{c}^{2})-\rho[(j_{a}k_{b}-j_{b}k_{a})(j_{b}j_{d,\rho}-k_{b}k_{d,\rho}) \nonumber \\
&\qquad+(j_{a}k_{c}-j_{c}k_{a})(j_{c}j_{d,\rho}-k_{c}k_{d,\rho})+(j_{b}k_{d}-j_{d}k_{b})(j_{b}j_{a,\rho}-k_{b}k_{a,\rho})+(j_{c}k_{d} \nonumber \\
&\qquad-j_{d}k_{c})(j_{c}j_{a,\rho}-k_{c}k_{a,\rho})+(j_{d}k_{a}-j_{a}k_{d})(j_{b}j_{b,\rho}+j_{c}j_{c,\rho}-k_{b}k_{b,\rho}-k_{c}k_{c,\rho})]\} \nonumber \\
&\qquad+2[q(\sigma^{2}-\omega^{2})^{3}]^{-1}(\sigma\sigma_{\rho}-\omega\omega_{\rho})\{(j_{a,\rho}k_{d}+j_{a}k_{d,\rho}-j_{d}k_{a,\rho}-j_{d,\rho}k_{a})(\sigma^{2}-\omega^{2}) \nonumber \\
&\qquad+(j_{d}k_{a}-j_{a}k_{d})(\sigma\sigma_{\rho}-\omega\omega_{\rho})+\rho(j_{d}k_{b}-j_{b}k_{d})(\sigma\sigma_{t}-\omega\omega_{t})+\rho(j_{b}k_{d,t} \nonumber \\
&\qquad+j_{b,t}k_{d}-j_{d}k_{b,t}-j_{d,t}k_{b}-2m\sigma j_{c})(\sigma^{2}-\omega^{2})+\rmi[\rho(j_{c}k_{a}-j_{a}k_{c})(\sigma_{t}\omega-\sigma\omega_{t}) \nonumber \\
&\qquad+\rho^{2}(j_{c}k_{b}-j_{b}k_{c})(\sigma_{\rho}\omega-\sigma\omega_{\rho})]\}.
\end{align}
Lastly, the function in $F_{13}=xF_{e}$ and $F_{23}=yF_{e}$ is
\begin{align}\label{Cylindrical symmetry appendix, Fe}
&F_{e}=-2[q(\sigma^{2}-\omega^{2})^{3}]^{-1}(\sigma\sigma_{\rho}-\omega\omega_{\rho})\{(j_{c}k_{a,\rho}+j_{c,\rho}k_{a}-j_{a}k_{c,\rho}-j_{a,\rho}k_{c}) \nonumber \\
&\qquad\cdot(\sigma^{2}-\omega^{2})+(j_{a}k_{c}-j_{c}k_{a})(\sigma\sigma_{\rho}-\omega\omega_{\rho})+(2/\rho)(j_{c}k_{a}-j_{a}k_{c}-m\sigma j_{d}) \nonumber \\
&\qquad\cdot(\sigma^{2}-\omega^{2})+\rho(j_{c}k_{b,t}+j_{c,t}k_{b}-j_{b}k_{c,t}-j_{b,t}k_{c})(\sigma^{2}-\omega^{2})+\rho(j_{b}k_{c}-j_{c}k_{b}) \nonumber \\
&\qquad\cdot(\sigma\sigma_{t}-\omega\omega_{t})+\rmi[(j_{d}k_{b}-j_{b}k_{d})(\sigma_{\rho}\omega-\sigma\omega_{\rho})+(1/\rho)(j_{d}k_{a}-j_{a}k_{d})(\sigma_{t}\omega \nonumber \\
&\qquad-\sigma\omega_{t})]\}+[2q(\sigma^{2}-\omega^{2})^{2}]^{-1}\{(2j_{c}k_{b,t}+2j_{c,t}k_{b}-2j_{b}k_{c,t}-2j_{b,t}k_{c}+j_{c}k_{a,\rho\rho} \nonumber \\
&\qquad+2j_{c,\rho}k_{a,\rho}+j_{c,\rho\rho}k_{a}-j_{a}k_{c,\rho\rho}-2j_{a,\rho}k_{c,\rho}-j_{a,\rho\rho}k_{c})(\sigma^{2}-\omega^{2})+2(j_{b}k_{c} \nonumber \\
&\qquad-j_{c}k_{b})(\sigma\sigma_{t}-\omega\omega_{t})+(j_{c}k_{a,\rho}+j_{c,\rho}k_{a}-j_{a}k_{c,\rho}-j_{a,\rho}k_{c})(\sigma\sigma_{\rho}-\omega\omega_{\rho}) \nonumber \\
&\qquad+(j_{a}k_{c}-j_{c}k_{a})(\sigma_{\rho}^{2}+\sigma\sigma_{\rho\rho}-\omega_{\rho}^{2}-\omega\omega_{\rho\rho})+(1/\rho)(3j_{c}k_{a,\rho}+3j_{c,\rho}k_{a} \nonumber \\
&\qquad-3j_{a}k_{c,\rho}-3j_{a,\rho}k_{c}-2m\sigma_{\rho}j_{d}-2m\sigma j_{d,\rho})(\sigma^{2}-\omega^{2})+(1/\rho)(3j_{c}k_{a} \nonumber \\
&\qquad-3j_{a}k_{c}-4m\sigma j_{d})(\sigma\sigma_{\rho}-\omega\omega_{\rho})+\rho(j_{c}k_{b,t\rho}+j_{c,t}k_{b,\rho}+j_{c,\rho}k_{b,t}+j_{c,t\rho}k_{b} \nonumber \\
&\qquad-j_{b}k_{c,t\rho}-j_{b,t}k_{c,\rho}-j_{b,\rho}k_{c,t}-j_{b,t\rho}k_{c})(\sigma^{2}-\omega^{2})+\rho(j_{b}k_{c,\rho}+j_{b,\rho}k_{c}-j_{c}k_{b,\rho} \nonumber \\
&\qquad-j_{c,\rho}k_{b})(\sigma\sigma_{t}-\omega\omega_{t})+2\rho(j_{c}k_{b,t}+j_{c,t}k_{b}-j_{b}k_{c,t}-j_{b,t}k_{c})(\sigma\sigma_{\rho}-\omega\omega_{\rho}) \nonumber \\
&\qquad+\rho(j_{b}k_{c}-j_{c}k_{b})(\sigma_{t}\sigma_{\rho}+\sigma\sigma_{t\rho}-\omega_{t}\omega_{\rho}-\omega\omega_{t\rho})+\rmi[(j_{d}k_{b,\rho}+j_{d,\rho}k_{b}-j_{b}k_{d,\rho} \nonumber \\
&\qquad-j_{b,\rho}k_{d})(\sigma_{\rho}\omega-\sigma\omega_{\rho})+(j_{d}k_{b}-j_{b}k_{d})(\sigma_{\rho\rho}\omega-\sigma\omega_{\rho\rho})+(1/\rho)(j_{d}k_{a,\rho}+j_{d,\rho}k_{a} \nonumber \\
&\qquad-j_{a}k_{d,\rho}-j_{a,\rho}k_{d})(\sigma_{t}\omega-\sigma\omega_{t})+(1/\rho)(j_{d}k_{b}-j_{b}k_{d})(\sigma_{\rho}\omega-\sigma\omega_{\rho}) \nonumber \\
&\qquad+(1/\rho)(j_{d}k_{a}-j_{a}k_{d})(\sigma_{t\rho}\omega+\sigma_{t}\omega_{\rho}-\sigma_{\rho}\omega_{t}-\sigma\omega_{t\rho})]\}.
\end{align}



\end{document}